# Structure Prediction of Ionic Epitaxial Interfaces with Ogre Demonstrated for Colloidal Heterostructures of Lead Halide Perovskites

*Stefano Toso[1,†], Derek Dardzinski[2,†], Liberato Manna[1]*, Noa Marom,[2,3,4]**

[1] Nanochemistry Department, Istituto Italiano di Tecnologia, Genova, 16163, Italy

[2] Department of Materials Science and Engineering, Carnegie Mellon University, Pittsburgh, PA, 15213, United States

[3] Department of Physics, Carnegie Mellon University, Pittsburgh, PA, 15213, United States

[4] Department of Chemistry, Carnegie Mellon University, Pittsburgh, PA, 15213, United States

[†] These authors contributed equally

E-mail: liberato.manna@iit.it, nmarom@andrew.cmu.edu

**Table of Contents.**

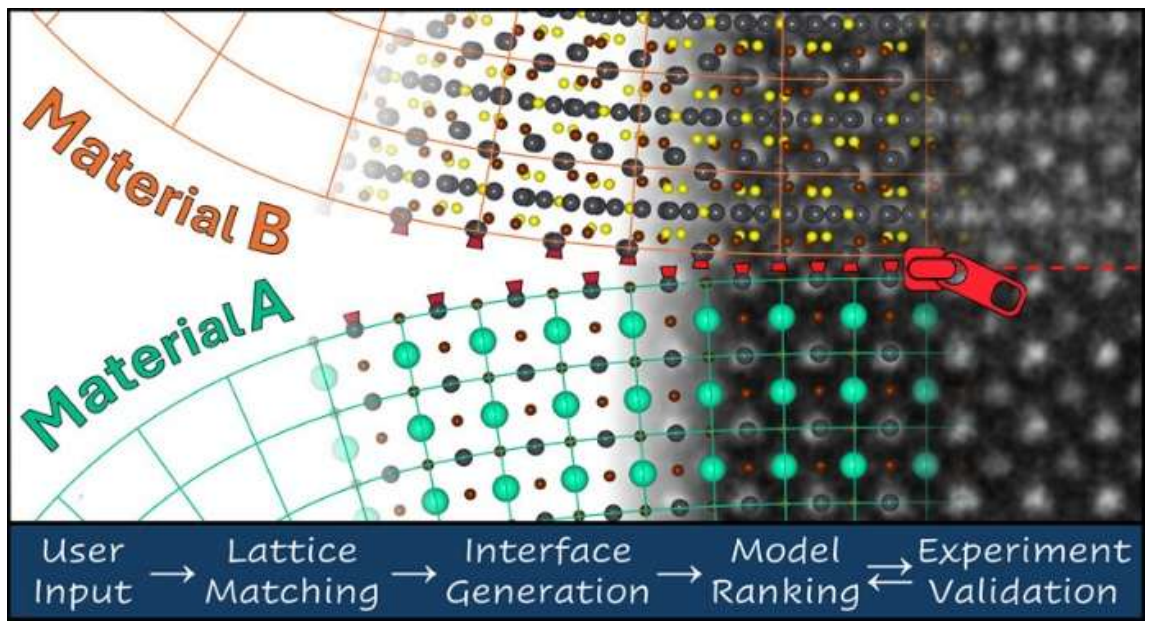

**Abstract.**

Colloidal epitaxial heterostructures are nanoparticles composed of two different materials connected at an interface, which can exhibit properties different from those of their individual components. Combining dissimilar materials offers opportunities to create several functional heterostructures. Yet, assessing structural compatibility (the main prerequisite for epitaxial growth) is challenging when pairing complex materials with different lattice parameters/crystal structures. This complicates both the selection of target heterostructures for synthesis and the assignment of interface models when new heterostructures are obtained. Here, we demonstrate Ogre as a powerful tool to accelerate the design and characterization of colloidal heterostructures. To this end we implemented developments tailored for the efficient prediction of epitaxial interfaces between ionic/polar materials, which encompass most colloidal semiconductors. These include pre-screening candidate models based on charge balance at the interface and using a classical potential for fast energy evaluations, with parameters automatically extracted from the input bulk structures. These developments are validated for $CsPbBr_3/Pb_4S_3Br_2$ heterostructures, where Ogre produces interface models in agreement with density functional theory and experiments. Furthermore, we use Ogre to rationalize the templating effect of $CsPbCl_3$ on the growth of lead sulfochlorides, where perovskite seeds induce the formation of $Pb_4S_3Cl_2$ rather than $Pb_3S_2Cl_2$ due to better epitaxial compatibility. Combining Ogre simulations with experimental data enables us to unravel the structure and composition of the hitherto unsolved $CsPbBr_3/Bi_xPb_yS_z$ interface and assign a structure to many other reported metal halide/oxide based interfaces. The Ogre package is available on GitHub or via the OgreInterface desktop application, available for Windows, Linux and Mac.

## 1. Introduction

Colloidal heterostructures are nanoparticles composed of two materials connected at an interface. Such architectures can profitably combine the properties of their components, and may exhibit unique functionalities emerging from their interaction. Prominent examples are core-shell quantum dots[1–3] and photocatalytic nanocomposites,[4–7] whose properties stem directly from the electronic structure of the junction. The increasing availability of colloidal materials has encouraged a property-driven approach to the design of new heterostructures, where materials are paired based on desired features like band alignment, plasmon resonance, or other functionalities.[8–12] However, materials selected based on their properties are not always structurally compatible,[13] and the absence of suitable structural relations may lead to the formation of amorphous or highly defective interfaces. Such heterostructures might still be useful if the target functionality is insensitive to the nature of the interface, as is the case for protective oxide shells.[14,15] However, heterostructures expressing electronic, magnetic, and optical properties that arise from quantum interactions at the junction may be impaired.[16–25]

To express their full potential, such heterostructures require epitaxial interfaces that ensure a precise matching between the two materials, and can grow virtually defect-free. Indeed, the most successful colloidal heterostructures are fully epitaxial architectures between isostructural materials (e.g., CdSe/CdS, CdSe/ZnSe, InP/ZnS),[2,26–29] which combine outstanding optoelectronic properties with a simple and predictable growth process. However, isostructural interfaces cover only a small fraction of the wide design space offered by colloidal chemistry, and the increasing number of reported heterostructures between non-isostructural materials, often obtained by chance, suggests that there is much more to explore.[30–43]

Nevertheless, synthesizing new heterostructures is a painstaking process of trial and error, as even compatible materials often tend to crystallize separately, and the co-presence of many elements in the reaction medium can lead to competing byproducts. Hence, failing to couple a specific pair of materials raises the question of whether such a heterostructure is intrinsically impossible to grow, or if the right conditions have not been found yet. Moreover, the morphology of nanocrystals may evolve during post-synthetic treatments, complicating the identification of facets involved in the interface growth.[44–49] Finally, the instability of some colloidal nanomaterials under an electron beam can hinder the collection of atomic resolution images, which are needed to verify if a heterostructure is truly epitaxial and determine the interface structure.[50,51]

These challenges can be addressed through computer simulations, which help direct synthetic efforts toward materials that are likely to match epitaxially, and can provide models to assist the interpretation of experiments. Several tools for the prediction of interface structures have been developed to this end.[48,52–59] These often employ hierarchical workflows, in which fast methods like classical force fields, score functions, or machine learning are used for the initial screening of models, followed by density functional theory (DFT) calculations to predict the structure and properties of the most promising candidates.[58–65] A similar approach is implemented in the Ogre open-source Python package for the prediction of organic and inorganic epitaxial interfaces.[66–69] Like other tools of its kind, Ogre was initially developed with the goal of predicting the structure of large-area epitaxial interfaces grown by thin film deposition methods, and its first applications focused on semiconductor/metal interfaces.[66] Such tools are not well-suited to identify prospective interfaces that could be grown by colloidal chemistry, as this requires a quick evaluation of numerous interfaces rather than the detailed simulation of a few. The task is further complicated by the complex compositions and structures adopted by many colloidal materials, like the lead halide perovskites studied here, which make approaches that rely heavily on DFT less suitable. Lastly, most structure prediction codes are designed for specialists, while synthetic chemists would benefit from tools that are easy to use, and do not require coding experience or access to advanced computing facilities.

To address these needs, we have implemented in Ogre a fast workflow for the prediction of epitaxial interfaces between polar compounds, which encompass most colloidal nanomaterials (e.g. $CsPbBr_3$, CdS, ZnO, etc.). This includes two new features that take advantage of the predominantly ionic nature of these materials to dramatically accelerate predictions. The first is a preliminary screening of interface models based on charge balance, which reduces the computational load by eliminating unreasonable candidates. The second feature is a classical potential consisting of a Coulomb term[70,71] and a repulsive Born term,[66,72] whose parameters are determined automatically based on the bulk structures of the input materials. This potential allows to quickly evaluate the energy of candidate structures, and can produce interface models and energy rankings in good agreement with DFT at a fraction of its computational cost. These strategies enable predicting the structure of epitaxial interfaces between polar materials in just few minutes on a simple laptop, making Ogre an excellent screening tool. We note that this approach is not suitable when the dominant interactions are not electrostatic, as is the case for metallic, covalent, and van der Waals dispersion bonding. To cover these cases, DFT, machine-learned potentials, and alternative score functions are available in Ogre.[66–69]

In what follows, the new Ogre workflow for the prediction of polar epitaxial interfaces is presented and validated extensively for several case studies. In **Section 2.1**, we illustrate the full workflow in detail, using the $CsPbBr_3/Pb_4S_3Br_2$ colloidal heterostructures reported by some of us as an example.[73,74] For this system, Ogre correctly identifies all the known epitaxial relations, and for the case study of the (100)⫽(010) orientation, it produces an interface model that matches both DFT-based predictions and high-resolution scanning transmission electron microscopy (HR-STEM) images of the heterostructures. In **Section 2.2**, we demonstrate how Ogre can help elucidate the outcome of a synthesis by revisiting a previous study of perovskite/lead sulfochloride heterostructures, conducted by some of us.[75] Specifically, the empirical observation that $CsPbCl_3$ nanocrystals template the selective nucleation of $Pb_4S_3Cl_2$ while suppressing the growth of the competing phase $Pb_3S_2Cl_2$ is rationalized based on Ogre's predictions of lattice matching and interface stability. In **Section 2.3**, we revisit several heterostructures involving $CsPbBr_3$ to demonstrate how Ogre can help assign structural models to newly synthesized interfaces, taking advantage of characterization results as a part of the prediction workflow.[34,36,76,77] A prominent example is the $CsPbBr_3$/Pb-Bi-S heterostructures recently reported by some of us,[39] whose hitherto unknown composition and interface connectivity are unraveled by combining information from HR-TEM experiments with Ogre predictions. Finally, in **Section 2.4** we demonstrate Ogre's broad applicability beyond colloidal metal halides by reproducing the structure of known heteroepitaxial interfaces between pairs of oxides, which are among the most studied polar materials for thin film growth.[78–81]

In addition to being efficient and versatile, Ogre is accessible to non-specialized users via the *OgreInterface* desktop application, available for Windows, Linux, and Mac. All scripts and reference bulk structures needed to reproduce the results presented here are available as Supplementary Material, and serve as examples for newly approaching users. By being directly accessible to synthetic chemists, we expect Ogre to become a versatile and powerful tool for the discovery and investigation of new epitaxial heterostructures between polar materials.

## 2. Results and Discussion

### 2.1. Ogre workflow demonstrated for $CsPbBr_3/Pb_4S_3Br_2$

Given the crystal structure of two materials, *A* and *B*, Ogre first identifies all favorable commensurate epitaxial relations between them, and then attempts to predict the structure of the resulting interfaces. The output is a set of atomistic models with optimized epitaxial registry (i.e., lateral offset) and interface distance, ranked by energy from the most to the least stable. The prediction workflow proceeds through three steps, illustrated in **Figure 1**: *lattice matching*, *interface generation*, and *surface matching and ranking*. **Sections 2.1.1-2.1.3** demonstrate each step of the algorithm for the (100)⫽(010) – $CsPbBr_3/Pb_4S_3Br_2$ epitaxial interface, which some of us have previously characterized in the form of colloidal heterostructures.[73] In **Section 2.1.4**, the performance of the Ogre classical potential is validated against DFT for the same interface.

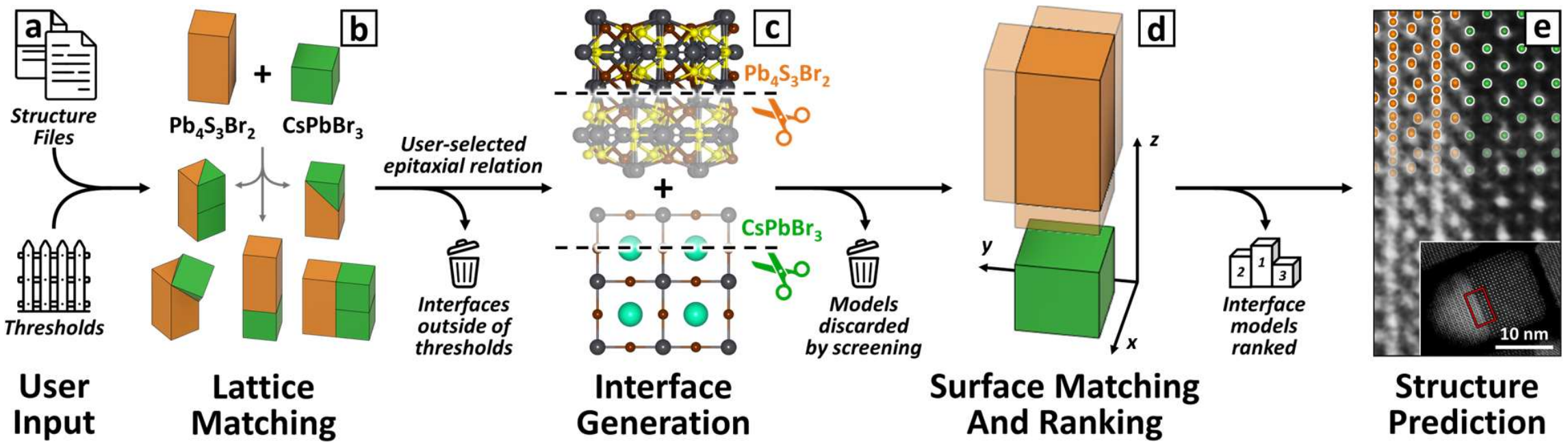

**Figure 1. Ogre interface prediction workflow.** a) Ogre takes as an input the bulk structures of two materials, along with user-defined thresholds for strain, supercell area, and the Miller indices of lattice planes to consider. b) The *lattice matching* step identifies all the commensurate supercells that yield domain-matched interfaces within the specified thresholds. c) For selected epitaxial orientations, the *interface generation* step creates all possible surface terminations for both materials, and combines them to construct atomistic models of the interface. d) For each model, the *surface matching and ranking* step identifies the optimal epitaxial registry (i.e., lateral offset) and distance between the two domains, using a fast electrostatic potential to evaluate energies. e) The output is a set of atomistic models ranked by stability. Top-ranked models can be compared with available experimental data, as demonstrated here for a $Pb_4S_3Br_2/CsPbBr_3$ heterostructure (only heavy atoms shown for clarity). HR-STEM images adapted with permission.[73] Copyright 2023, American Chemical Society.

We note that in this work $CsPbBr_3$ is treated as cubic despite adopting a slightly distorted orthorhombic structure.[82–85] This approximation greatly simplifies the discussion of results, and does not adversely affect the outcome of predictions (see Figures S1-7 and Tables S1-4).

### 2.1.1 Lattice matching

The *lattice matching* step uses the Zur-McGill algorithm[52] to identify commensurate epitaxial relations between the two materials, denoted here as $(hkl)_A /\!/ (h'k'l')_B$ based on the lattice planes being matched. In short, the algorithm attempts to combine multiple 2D-cells describing the lateral periodicity of $(hkl)_A$ and $(h'k'l')_B$ planes to construct a common 2D-supercell that can represent both materials at the interface. An epitaxial match is found if such supercell(s) exist within the user-defined constraints for strain and area. We note that one $(hkl)_A /\!/ (h'k'l')_B$ combination may produce several non-equivalent 2D-supercells that differ by a relative rotation of the two materials around an axis perpendicular to their contact plane.[66] In that case, the smallest 2D-supercell is selected (see Figure S8 and related discussion). To avoid any ambiguities, Table S5 specifies for all interfaces discussed in this work a pair of lattice vectors that are parallel to each other in the plane of the interface, denoted here as $[hkl]_A \Uparrow [h'k'l']_B$.

At this stage, the user can impose constraints on strain and supercell area to exclude implausible matches. In general, high strain is unfavorable, as it can lead to the formation of defects at the interface or even hinder its growth. However, a certain degree of mismatch is tolerable, depending on the type of interface being studied. For example, the strain limit for high-quality epitaxial semiconductor films is considered to be around 2%.[86] Conversely, nanoscale interfaces tend to be more tolerant because strain can be effectively absorbed by lattice deformations if the contact surface is limited to few $nm^2$. For instance, the $CsPbBr_3/CsPb_2Br_5$ heterostructures reported by Zheng et al. accommodate a 2.7% mismatch by bending into rings where the material with the largest lattice step faces outward (see **Section 2.2**).[36] Other examples of high-strain colloidal interfaces are CdS/CdSe (4.2%)[87] and the rather extreme InAs/ZnS (12.0%).[88] For the colloidal heterostructures studied here, we set the strain threshold to 10%.

For selecting the area threshold there are conflicting considerations. On the one hand, larger supercells enable finding commensurate domains with lower strain, approaching the limit where an infinite area produces supercells without any strain. On the other hand, smaller supercells increase the chance of good atom-to-atom correspondence at the interface because the lattice sites of the two materials are more likely to coincide. By default, Ogre sets the surface area threshold, $S_T$, to:

$$S_T = 2 \times \max[S_A(hkl)\ ;\ S_B(h'k'l')] \tag{1}$$

In short, for each $(hkl)_A /\!/ (h'k'l')_B$ pair, $S_T$ is set to twice the area of the largest 2D-cell among the two planes. This enables finding supercells that can encompass up to two single-material cells positioned side-by-side, thus allowing for a more extended interface repeating unit. The user can select a numerical hard threshold if needed (see Equation S7 and related discussion).

The user may also impose constraints on the Miller indices to be searched, based on their pre-existing knowledge of the system of interest. For example, the planes exposed by one or both materials might be known, as is the case for well-faceted nanocrystal seeds, or for the epitaxial growth of a thin film on a single-crystalline substrate. In the absence of such information, or if multiple hypothetical orientations are being considered, it is still reasonable to limit the search to low Miller indices because high-index crystal terminations are usually less stable, and therefore rarely exposed.[44–47,49,89]

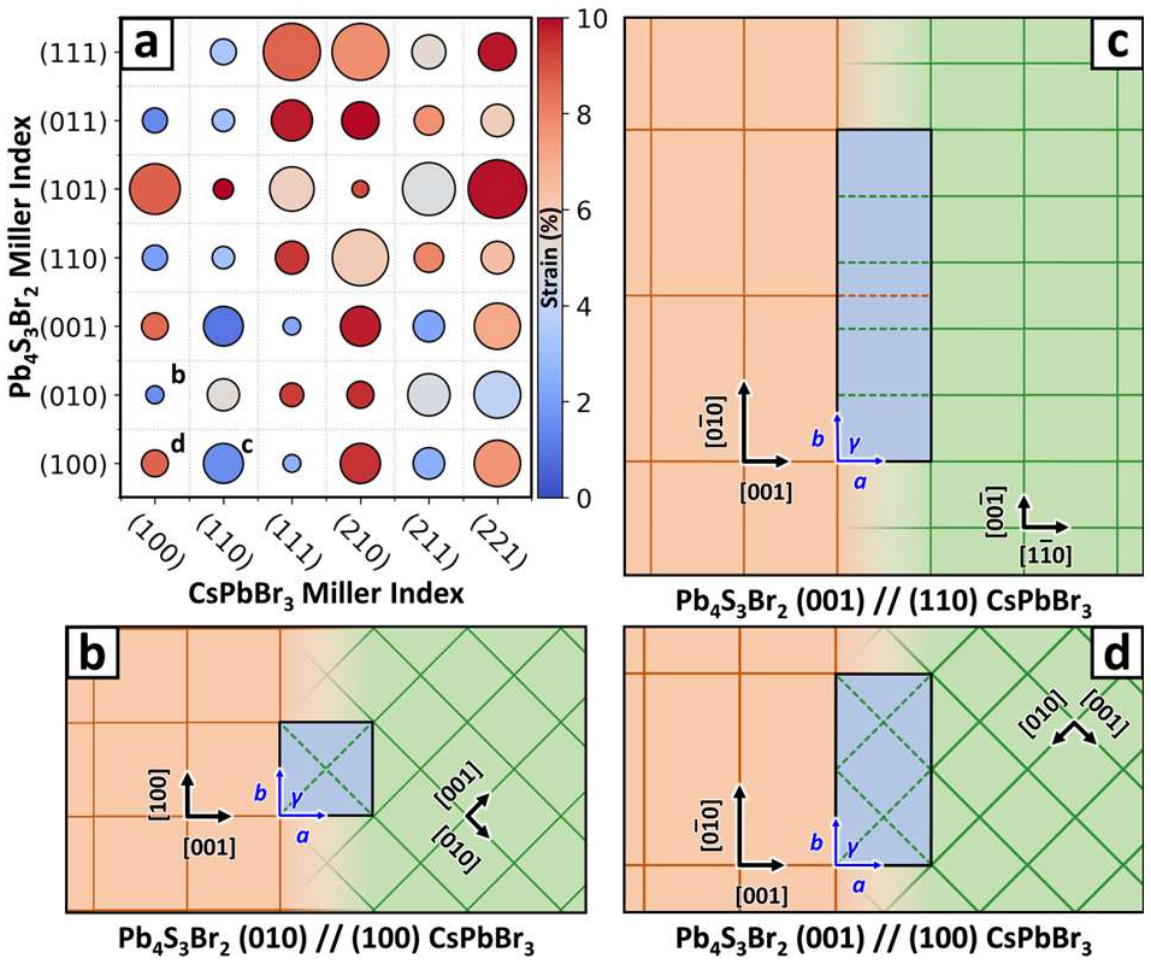

**Figure 2. *Lattice matching* results for $CsPbBr_3$/$Pb_4S_3Br_2$.** a) Domain-matched interfaces identified within the specified constraints. Each circle indicates a match, with the color corresponding to the strain and the size representing the interface area. The interfaces marked with letters have been reported experimentally, and their 2D-supercell is shown in the corresponding panels. b-d) Supercells of the $CsPbBr_3$/$Pb_4S_3Br_2$ interfaces reported in (b) Ref. 73 and (c,d) Ref. 90. The $CsPbBr_3$ 2D-lattice is colored in green, the $Pb_4S_3Br_2$ 2D-lattice in orange, and the domain-matched commensurate 2D-supercell is shown in blue.

**Figure 2** summarizes the lattice matching results for the $CsPbBr_3/Pb_4S_3Br_2$ system. For ease of visualization $h,k,l \leq 1$ for $Pb_4S_3Br_2$ and $h,k,l \leq 2$ for $CsPbBr_3$ are shown here (see Figure S9 for a more extended search). Each circle in **Figure 2a** corresponds to a domain-matched epitaxial relation identified within the constraints: the color indicates the strain, while the size represents the supercell area. Interfaces reported experimentally are marked with a letter, pointing to the corresponding 2D-supercell (**Figures 2b-d**).[73,90] The (100)∥(010) – $CsPbBr_3/Pb_4S_3Br_2$ interface (**Figure 2b**) stands out for having both a low strain of 1.6% and a small area of 68 Å$^2$. Indeed, this interface was the first to be reported and fully characterized by some of us for the $CsPbBr_3/Pb_4S_3Br_2$ system.[73]

Ogre also identifies two additional epitaxial relations that were recently reported by Das et al:[90] the (110)∥(001) interface, with 0.9% mismatch and an area of 241 Å$^2$ (**Figure 2c**), and the (100)∥(001) interface, with 8.6% mismatch and an area of 136 Å$^2$ (**Figure 2d**). Because these interfaces were observed in nanoscale heterostructures, the contact surface between the two materials was likely small enough to tolerate what would otherwise be a high strain value. It is worth noting that the *a* and *c* lattice parameters of $Pb_4S_3Br_2$ are very close, leading to (*hkl*) ≈ (*lkh*) for this material. This causes similarities between the matches found by Ogre for the (100) and (001) planes of $Pb_4S_3Br_2$, and would also complicate the experimental distinction between these orientations. In the absence of atomic-resolution images of the interface, we cannot exclude that some of the heterostructures observed by Das et al.[90] might have adopted alternative epitaxial relations (see Figures S10-14 and Tables S6-10 for further discussion).

Based on the results shown in **Figure 2**, there are several additional plausible matches that have not been reported experimentally. For example, the (111)∥(100) interface (2.5% strain and 118 Å$^2$ area) and the (110)∥(110) interface (3.1% strain and 145 Å$^2$ area) appear comparable to the (100)∥(010) interface reported by some of us, and more favorable than those reported by Das et al.[90] It is likely that these interfaces have not been observed owing to the known tendency of $CsPbBr_3$ nanocrystals to express the (100) facets.[44,91]

### 2.1.2. Interface generation

The *lattice matching* procedure relies solely on lattice parameters, with no consideration of interatomic interactions and bonding at the interface. To assess if any of the identified epitaxial relations can produce a chemically stable interface, an exhaustive set of atomistic models must be constructed and ranked by relative energy. To this end, Ogre cleaves the two materials at planes parallel to the interface to generate surface slabs with different terminations,[66,92] which are then combined to construct interface models. At this stage, Ogre can be set to distribute the strain on the two materials equally, as is the case in most colloidal heterostructures, or instead let the lattice of one material conform to the other, which might be better suited to describe the growth of a thin film on a bulk substrate.

**Figure 3** illustrates the construction of interface models for the (100)⫽(010) – $CsPbBr_3/Pb_4S_3Br_2$ interface.[73] Depending on the complexity of the materials involved, this process can result in a large number of candidate structures. Therefore, multiple strategies are implemented to minimize the number of models that proceed to the *surface matching and ranking* step of the algorithm, which is the most computationally intensive. Specifically, while generating slab models, Ogre clusters together nearly co-planar atoms to construct only chemically sound surface terminations. The resulting slabs are then screened to eliminate those that are equivalent by symmetry, thus avoiding redundancies (see Figures S15-16).

To further reduce the number of interface models, we have developed a fast screening algorithm based on the charge balance at the interface. To this end, each slab is assigned a surface charge $Q$, defined as:

$$Q = -\frac{\sum_i q_i \cdot d_i}{D} \qquad (2)$$

where $i$ runs over all ions in the slab, $q_i$ is the ion charge, $d_i$ is its distance from the interface, and $D$ is the total thickness of the slab (see Equations S8-S9 and Figure S17 for derivation). Conceptually, $Q$ represents the uncompensated charge found at the surface of each material, which will interact with the other material to construct the interface. Because charges of the same sign repel each other, interfaces formed by [+/+] or [–/–] slab pairs can be immediately excluded based on the sign of $Q$.

In principle, the magnitude of $Q$ could also be used to identify [+/–], [+/0], and [– /0] models that do not achieve full charge balance at the interface, which could serve as a further screening criterion. However, plenty of examples for unbalanced and yet stable interfaces have been reported, such as $Sm_2CuO_4/LaFeO_3$,[93] $LaAlO_3/SrTiO_3$,[94] and many Si/oxide[95] or metal/oxide[96] interfaces. This is possible because charge imbalance can be locally compensated by the formation of defects (ionic compensation) or a change of oxidation states (electronic compensation), which are often considered as the origin of 2D-conductivity in epitaxial interfaces between insulators.[94,97,98]

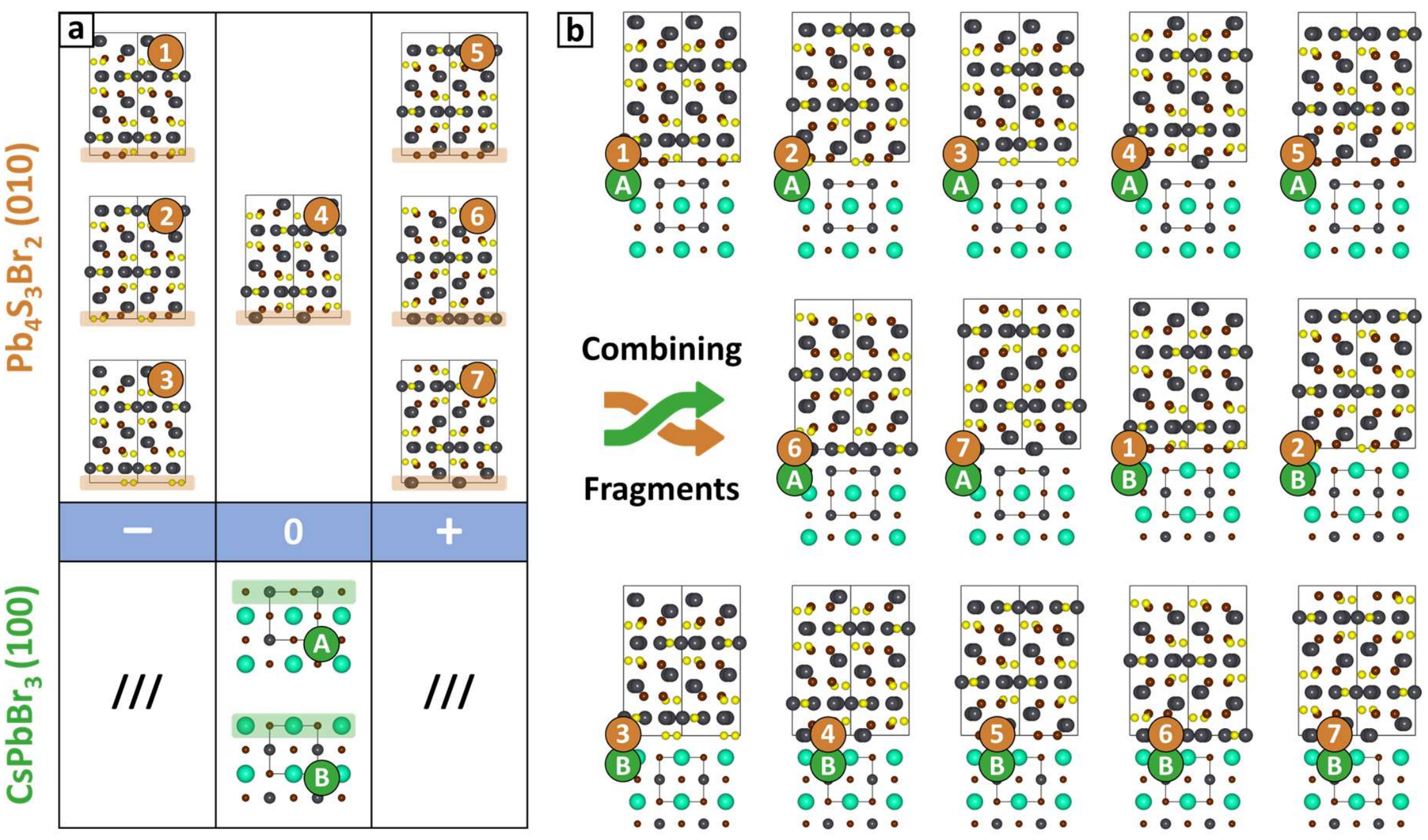

**Figure 3. Interface generation.** a) Possible terminations for the (010) $Pb_4S_3Br_2$ and (100) $CsPbBr_3$ surfaces. The +, 0, and – symbols refer to the surface charge of each slab. Colored shadings indicate the side of the slab that will contact the other material to form the interface, which is where the charge is calculated. b) Models of all the (100)⫽(010) – $CsPbBr_3/Pb_4S_3Br_2$ non-equivalent interfaces obtained by combining the slabs in panel (a), shown prior to the *surface matching* step. Cs atoms are colored in cyan, Pb in gray, S in yellow, and Br in brown.

---

The effectiveness of such compensation mechanisms depends on the materials involved, as not all elements can assume multiple oxidation states, and different structures might tolerate defects to a lesser extent than others. Furthermore, considering charge compensation mechanisms requires simulations that are too elaborate to be performed at a preliminary screening stage. For example, achieving a realistic concentration of defects may require large supercells, and multiple configurations must be considered to account for their random distribution. Conversely, variable oxidation states can be implicitly handled by DFT without affecting the size of the simulation, but pose challenges for classical force fields that explicitly assign a charge to atoms. Nevertheless, **Section 2.1.3** shows that the classical potential used for *surface matching and ranking* can handle unbalanced interfaces even in the absence of charge compensation mechanisms. Therefore, we opted to retain the [+/–], [+/0], and [– /0] cases for further scrutiny. The user can decide to exclude such interface models based on their knowledge of the system of interest.

Overall, the effectiveness of the preliminary screening strategies employed by Ogre depends on the materials under scrutiny. For example, **Figure 3** shows all the possible slab combinations considered further for the (100)∥(010) – $CsPbBr_3$/$Pb_4S_3Br_2$ interface. In this case, the number of $Pb_4S_3Br_2$ slabs is decreased from 14 to the 7 shown in **Figure 3a** by excluding symmetry-equivalent terminations, whereas the only two possible terminations for cubic $CsPbBr_3$ (i.e., CsBr and $PbBr_2$ planes formed by coplanar ions) cannot be reduced further. The advantage of clustering becomes more evident when considering the orthorhombic structure of $CsPbBr_3$, where ions are slightly shifted out of planarity. In that case, some of the possible cleavage planes would separate those ions, producing defective slabs with missing atoms at their surface. By clustering nearly co-planar ions together, Ogre identifies only the two chemically sound terminations also found for cubic $CsPbBr_3$ (see Figure S16). Moreover, as both $CsPbBr_3$ terminations yield slabs of the $Q = 0$ type, none of the 14 resulting interface models in **Figure 3b** are excluded based on charge balance. The $CsPbBr_3$/$Bi_2PbS_4$ interface discussed in **Section 2.3** provides an example of effective screening based on charge balance, where the number of models is reduced from 20 to 12 by excluding the [+/+] and [-/-] cases (see Figures S17-18).

### 2.1.3. Surface matching and ranking

When pairing two surface slabs, their relative position must be optimized to identify the best possible bonding configuration across the interface. To this end, Ogre employs particle swarm optimization[99,100] to efficiently explore the 3D space of parameters formed by the relative epitaxial registry (i.e., lateral *xy*-offset) and *z*-distance between the two slabs. The optimal configuration is the one that minimizes the energy of the interface model, $E_{AB}$. To visualize the outcome of *surface matching*, Ogre produces a 2D energy map by shifting one material on top of the other within the interface supercell at their optimal *z*-distance (**Figure 4a**), and computes an energy *vs* interfacial distance curve at their optimal *xy*-offset (**Figure 4b**). For visualization purposes, these plots display the adhesion energy $E_{adh}$, which is obtained by subtracting from $E_{AB}$ the energies of the two isolated slabs ($E_A$ and $E_B$). The sign of $E_{adh}$ indicates whether or not the resulting interface model is more stable than the two constituent surface slabs. In these maps, the global minimum of $E_{adh}$ corresponds to the relative slabs position that produces the most stable configuration, while local minima identify metastable configurations. As a part of its output, Ogre produces an atomistic model of the optimal interface configuration (**Figure 4c**), which can be opened with structure visualization programs like Vesta.[101]

Next, all surface-matched interface models are compared to identify the most stable interface structure. As a ranking metric, Ogre adopts the interface energy $E_{int}$, defined as:[68]

$$E_{int} = \frac{E_{AB} - (E_{AA'} + E_{BB'})/2}{S} \tag{3}$$

where $E_{AA'}$ and $E_{BB'}$ are the cleavage energies of the two materials $A$ and $B$ (i.e., the energy obtained when two slabs are re-assembled into a bulk-like slab with twice the thickness at a given cleavage plane), and $S$ is the supercell area. Essentially, $E_{int}$ compares the energy of the interface against that of the parent bulk materials, indicating how favorable it is to interrupt the growth of one domain and switch to the other (See Figures S19-22 for further discussion).

As mentioned above, evaluating the energy terms ($E_{AB}$, $E_{AA'}$, etc.) is the most computationally intensive task of the algorithm. To make this step computationally efficient, we have developed a classical potential that enables estimating the energy of interfaces at a fraction of the computational cost of DFT. In short, all energy terms are calculated using pair-wise interatomic potentials, consisting of an electrostatic Coulomb term computed via the damped-shifted force potential,[70,71,102] and a Born term accounting for short-range interatomic repulsions (see Equations S10-S16).[66,72] A cutoff of 18 Å is adopted because longer-range dispersion interactions are typically negligible in ionic materials. To enable our classical potential to describe a wide variety of polar materials, each pair-wise term is optimized for the corresponding pair of ions by fitting one free parameter in the Born term. This is performed by imposing that the energy minimum coincides with the equilibrium bond length found in the input bulk structures (see Equation S13). For bonds that only exist at the interface (e.g., $Cs^+$ – $S^{2-}$ in $CsPbBr_3/Pb_4S_3Br_2$), the optimal equilibrium distance is estimated based on the sum of ionic radii extracted from the two parent materials (see Figure S23). Because the fitting procedure relies on the input bulk structures as a reference, we find that the predicted interfacial distances are slightly more accurate when bonds that form between slabs also exist in the bulk materials. However, stable models typically converge to reasonable distances in either case.

We note that there are occasional instances of "*non-bonding*" interfaces, for which the energy *vs* interfacial distance curve is purely repulsive, and no minimum is found within the range considered (see Figure S22). This likely happens because our classical potential does not contain a long-range dispersion term. Because this only occurs when the electrostatic interactions between the two materials are not particularly favorable and the repulsive Born term dominates, we regard such interface configurations as not viable and discard them.

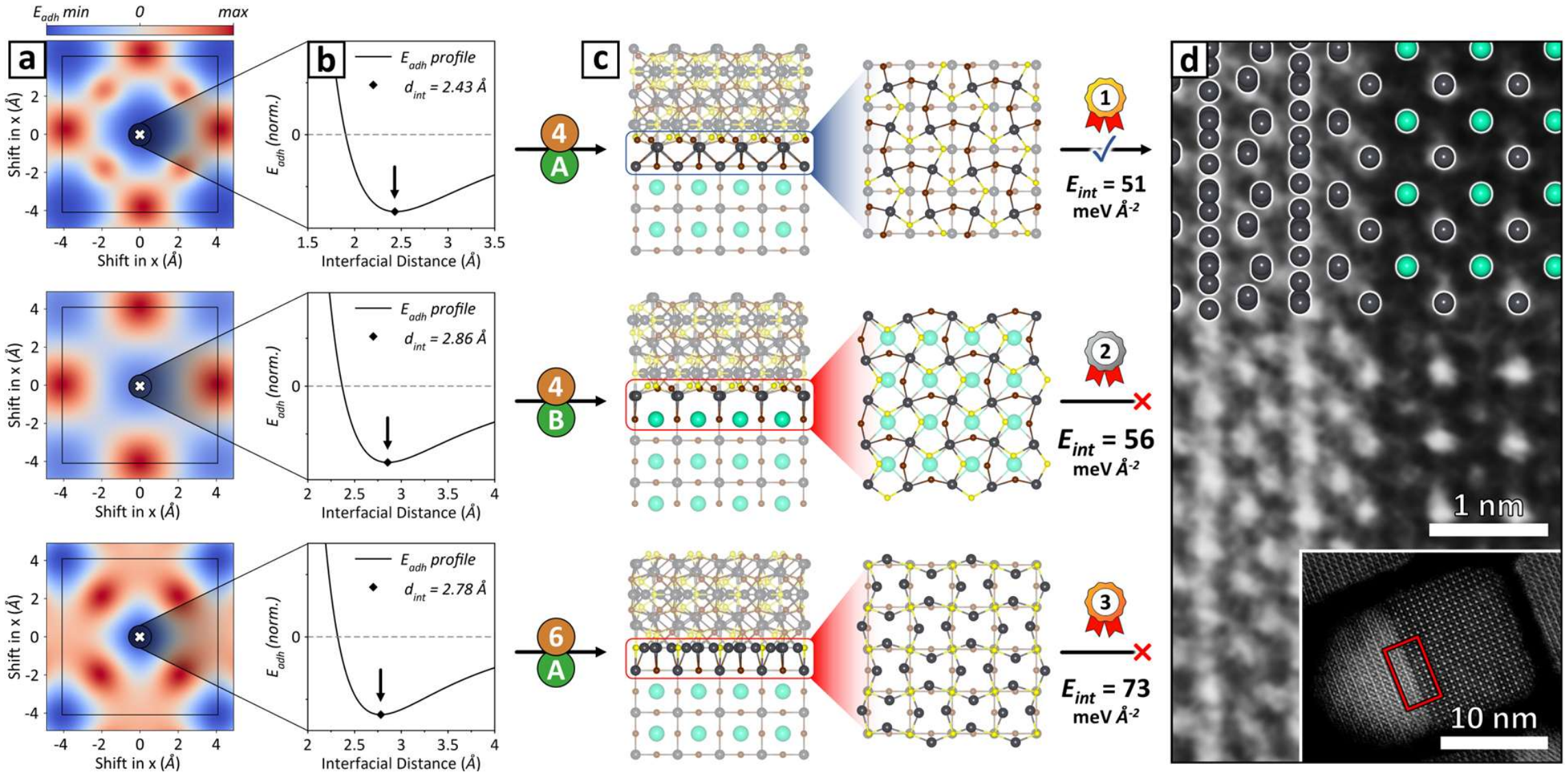

**Figure 4. Surface matching and ranking.** Results are displayed for the 4A, 4B, and 6A models of the (100)⫽(010) – $CsPbBr_3/Pb_4S_3Br_2$ interface, labeled as shown in **Figure 3**. a) Maps of $E_{adh}$ vs *xy*-offset with the optimal epitaxial registry marked by a cross. b) $E_{adh}$ *vs* interfacial distance curves, with the optimal *z*-distance indicated by an arrow. c) Interface models after the *surface matching* step. The side views (left) highlight the formation of bonds between slabs. The top-down views (right) of the interface planes show the atom-to-atom correspondence of anions and cations. The interface energies of the three models are also shown. d) HR-STEM image of a (100)⫽(010) – $CsPbBr_3/Pb_4S_3Br_2$ heterostructure, with superimposed the most stable model predicted by Ogre (see also Figure S4). Only heavy atoms are shown to ease the comparison with electron scattering contrast. Inset: lower magnification view of the same heterostructure. Microscopy data adapted with permission.[73] Copyright 2023, American Chemical Society. Cs atoms are colored in cyan, Pb in gray, S in yellow, and Br in brown.

Due to its purely electrostatic nature, our potential is restricted to compounds in which atoms carry a formal charge. This encompasses the majority of materials found in heterostructures synthesized colloidally with the notable exception of metal domains, that are relevant for plasmonic and catalytic applications.[7,9] However, metals often grow independently of epitaxial constraints due to their high tolerance for defects, to the point that non-substrate-specific growth strategies are available for technologically relevant elements (e.g., Au, Ag).[103,104] Likewise, our potential is not applicable to covalent and van der Waals materials, where non-electrostatic interactions are predominant. We remark that Ogre can predict the structure of interfaces involving non-polar materials by using other methods for surface matching and ranking, including geometric score functions, machine-learned potentials, and DFT, albeit the latter has a significantly higher computational cost.[66–69]

**Figure 4** shows the *surface matching and ranking* results for the three (100)∥(010) – $CsPbBr_3/Pb_4S_3Br_2$ interface models ranked as most stable by our classical potential: **Panels 4a-b** illustrate the search for the optimal epitaxial registry and interface distance, while **Panel 4c** shows the resulting structures (see Figures S3-4 and Table S2 for a full account of the results). In each case, the procedure identifies the optimal bonding configuration, with the two slabs positioned so as to maximize attractive interactions and found within plausible distances. Notably, the two most stable interfaces, labelled 4A and 4B as in **Figure 3**, are of the [0/0] type and are very close in energy (52 vs 57 meV $Å^{-2}$). In contrast, the 6A model corresponds to a charge-unbalanced [+2/0] interface, and is significantly less stable (75 meV $Å^{-2}$). This suggests that a charge accumulation can destabilize the interface despite the good geometric matching between slabs and the presence of favorable $Pb^{2+}$–$S^{2-}$ electrostatic interactions, that are absent in the charge-balanced models 4A and 4B. We note, however, that the classical potential is unable to redistribute charges at the interface, as discussed in **Section 2.1.4**.

As shown in **Figure 4d**, the top-ranked model 4A matches well with experimental observations of the interface.[73] The model produced by Ogre was scaled to match the atomic-resolution image of the heterostructure without any adjustment to the relative position of the two domains, which attests to the accuracy of our prediction (see Figure S4 for a comparison with other models). To provide a practical estimate of the code's performance, the whole workflow illustrated in **Section 2.1** was executed on a mid-tier laptop in about ~100 s.

### 2.1.4. Validation against DFT

To further validate the performance of our classical potential, we compared the *surface matching and ranking* results for the (100)∥(010) – $CsPbBr_3/Pb_4S_3Br_2$ interface to DFT. Because the choice of exchange correlation functional can significantly affect the outcome of DFT simulations, it is important to select one that adequately describes the materials of interest. Here, we have chosen the strongly constrained and appropriately normed (SCAN) functional[105] combined with the revised Vydrov and van Voorhis nonlocal correlation method (rVV10)[106,107] because it has been reported to reliably describe the dynamic tilting of octahedra and the phase transitions in halide perovskites.[108,109] In order to compare the results of the Ogre classical potential to DFT on an equal basis, the same workflow of *surface matching and ranking* was executed using DFT, with no structural relaxation beyond optimizing the relative position of the two rigid slabs.

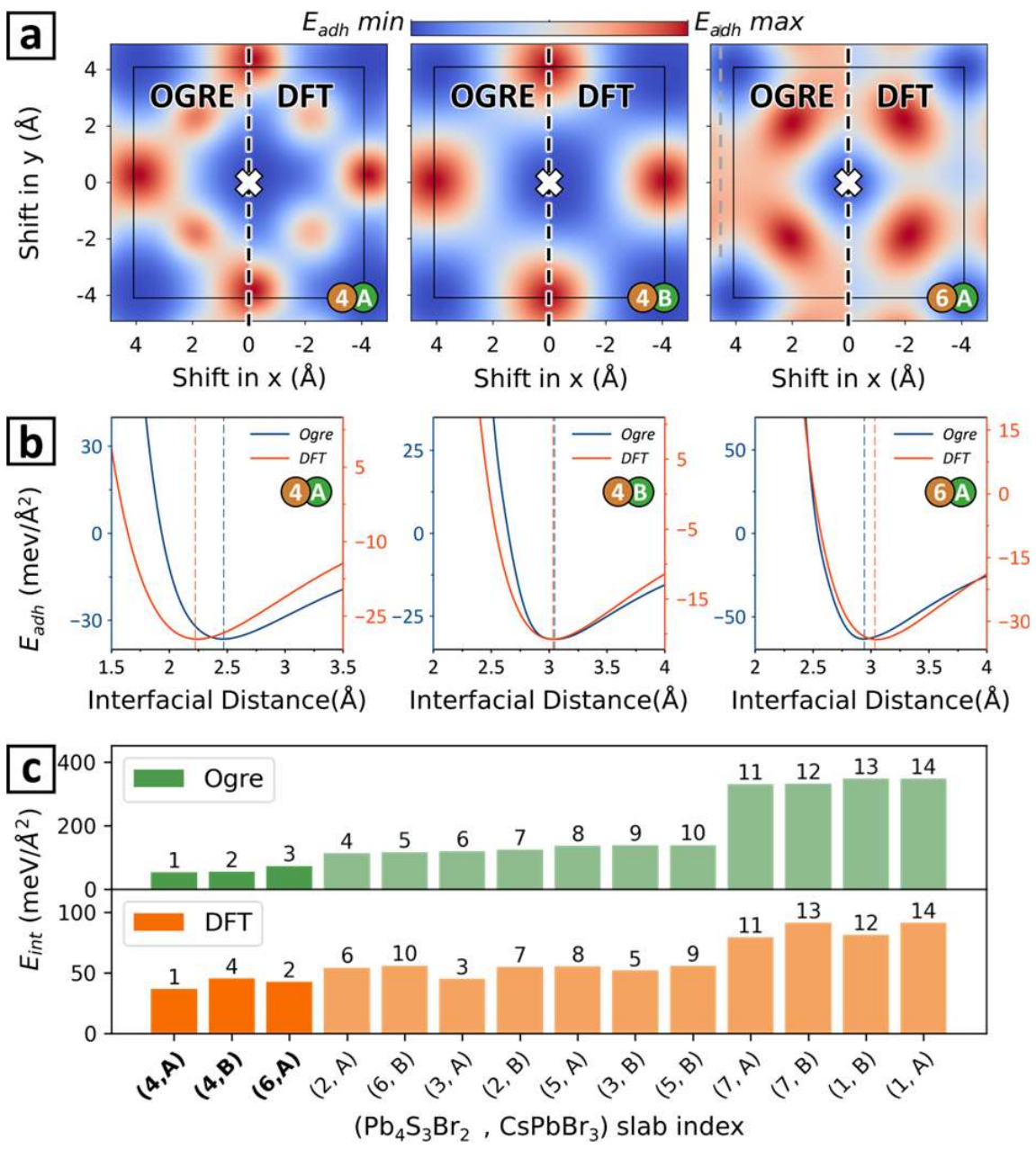

**Figure 5. Validation of *surface matching and ranking* by the Ogre potential against DFT.** a) $E_{adh}$ maps generated with the Ogre classical potential (left) and DFT (right) for the 4A, 4B, and 6A models of the (100)⫽(010) – $CsPbBr_3/Pb_4S_3Br_2$ interface (labeled as in **Figure 4**). Both methods identify the same epitaxial registry, marked by a cross. b) $E_{adh}$ *vs* interfacial distance curves for the same interfaces. c) Interface models ranked by computing $E_{int}$ with the Ogre electrostatic potential (top) and with DFT (bottom) with the 4A, 4B, and 6A models highlighted.

For the 4A, 4B, and 6A interface models shown in **Figure 4**, we compared the 2D energy maps (**Figure 5a**) and the $E_{adh}$ *vs* interfacial distance curves (**Figure 5b**) produced by the Ogre classical potential and by DFT. In all cases, our classical potential is in good agreement with DFT regarding the positions of the energy maxima and minima, meaning that the optimal in-plane epitaxial registry is correctly identified. The interfacial distance is in closer agreement with DFT for configurations 4B and 6A than for 4A, but it is consistently within 0.3 Å from the DFT results. Overall, our potential tends to overestimate the interfacial distances compared to DFT, with increasing inaccuracy as the interface models become less stable. This is likely because the classical potential does not include non-ionic contributions to the binding energy, and does not allow for charge density redistribution (see Table S11 for further discussion).

**Figure 5c** compares the ranking of interface models based on the Ogre classical potential and DFT. Both methods rank the experimental interface 4A as the most stable, and highlight a clear distinction between two groups of interfaces with lower and higher interface energies, separated by the gray dashed line. Within each group the interface energies are similar, leading to some reshuffling in the two rankings. Notably, the high-energy group is formed by models with strong charge imbalance at the interface (7A = 7B = [+4,0]; 1A = 1B = [-4,0]). Conversely, in the lower-energy group the Ogre classical potential favors the two charge-balanced interfaces 4A and 4B, whereas DFT ranks two charge-imbalanced interfaces as second and third (6A = [+2,0] and 3A = [-2/0], respectively). As mentioned above, both the 6A and 3A models feature favorable $Pb^{2+}$–$S^{2-}$ electrostatic interactions. It is possible that with DFT these interfaces are stabilized by charge density redistribution. We also note that local geometry relaxation at the interface may further change the stability ranking. In this respect, the Ogre potential can be used to quickly generate reasonable starting models for DFT-based relaxations.

### 2.2. Using Ogre to support synthetic efforts

In what follows, we demonstrate potential applications of Ogre in the field of colloidal nanomaterials, starting with the interpretation and prediction of possible outcomes for the synthesis of epitaxial heterostructures. Because synthesizing new materials is beyond the scope of this work, we use the results for lead sulfochlorides, reported by some of us, as a case study.[75]

Under typical colloidal synthesis conditions, the Pb-S-Cl system can lead to the growth of two competing materials: the pseudocubic $Pb_3S_2Cl_2$ (a phase with no equivalent for other halogens) and the orthorhombic $Pb_4S_3Cl_2$, which is structurally equivalent to the $Pb_4S_3Br_2$ sulfobromide discussed in **Section 2.1**.[74,75] Indeed, the presence of $CsPbCl_3$ nanocrystals in the reaction medium leads to the growth of $CsPbCl_3$/$Pb_4S_3Cl_2$ heterostructures similar to those in **Figure 4d**. Crucially, their formation coincides with a complete suppression of the pseudocubic competitor $Pb_3S_2Cl_2$. This outcome has been attributed to the growth of $Pb_4S_3Cl_2$ outpacing $Pb_3S_2Cl_2$ thanks to the epitaxial templating action of $CsPbCl_3$ seeds, which enables heterogeneous nucleation.[75] Such an interpretation relies on the assumption that no competing interface can form between the pseudocubic $Pb_3S_2Cl_2$ and $CsPbCl_3$, which we here confirm with Ogre. We note that the concept of promoting the growth of metastable phases by epitaxial templating is widespread in thin film growth,[110–113] but is nascent in colloidal synthesis.

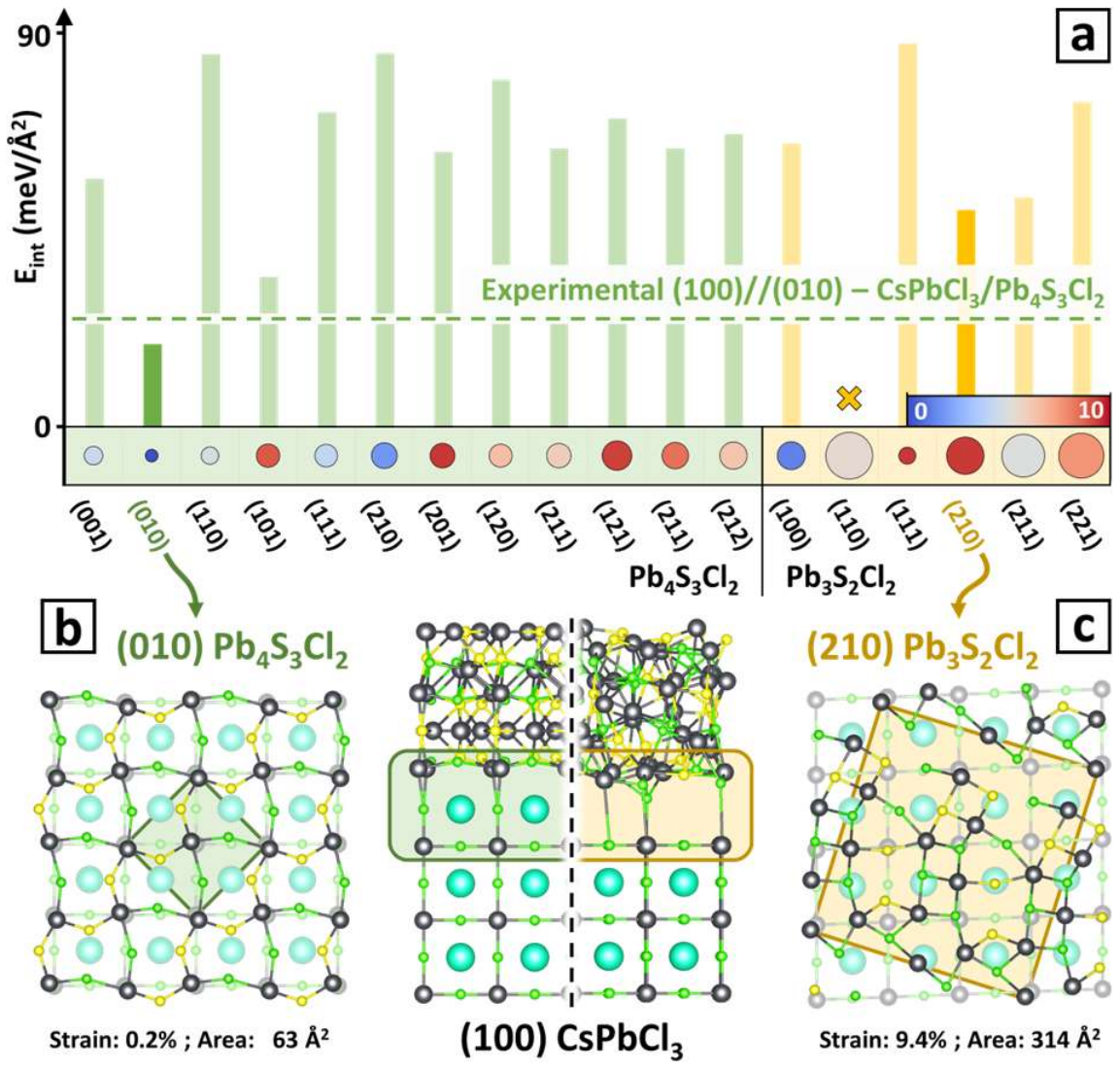

**Figure 6. Stability of $Pb_4S_3Cl_2$ *vs* $Pb_3S_2Cl_2$ on the (100) surface of $CsPbCl_3$.** a) *Lattice matching* and interface ranking results for the growth of $Pb_4S_3Cl_2$ (green) and $Pb_3S_2Cl_2$ (yellow) on the (100) surface of $CsPbCl_3$. The circle size corresponds to the interface area and the color corresponds to the strain. The green dashed line marks the $E_{int}$ value of the experimental interface, which is the 2nd-best model for the (100)⫽(010) – $CsPbCl_3/Pb_4S_3Cl_2$ interface. Non-equivalent epitaxial relations leading to identical *lattice matching* results, like $(h,k,l) = (\bar{h},\bar{k},\bar{l})$ for $Pb_3S_2Cl_2$, have been aggregated for visualization purposes, and only the lowest $E_{int}$ value is shown. All models for the (100)⫽(110) – $CsPbCl_3/Pb_3S_2Cl_2$ interface, marked with an ×, are found to be non-bonding. b) Model of the most stable interface formed between the (100) surface of $CsPbCl_3$ and $Pb_4S_3Cl_2$ (left: interface plane, center: side view). c) Model of the most stable interface formed between the (100) surface of $CsPbCl_3$ and $Pb_3S_2Cl_2$ (center: side view, right: interface plane). The corresponding 2D-supercells are also shown. Cs atoms are colored in cyan, Pb in gray, S in yellow, and Cl in green.

For $Pb_4S_3Cl_2$ no bulk crystal structure was available, and the nanocrystals obtained in Ref. 75 were too small to allow for a proper refinement by X-ray diffraction. Therefore, we extracted the lattice parameters directly from the HR-STEM images of a $CsPbCl_3/Pb_4S_3Cl_2$ heterostructure via Fourier analysis of the lattice periodicity. To ensure maximum accuracy, we calibrated the image using the lattice constant of $CsPbCl_3$ (5.605 Å) as a reference. This resulted in a pseudo-tetragonal unit cell for $Pb_4S_3Cl_2$, with estimated lattice parameters $a = c = 7.94$ Å and $b = 14.94$ Å. The atom positions inside the unit cell were determined starting from the analogous structure of $Pb_4S_3Br_2$ by performing DFT relaxation, using the SCAN+rVV10 functional with the lattice parameters fixed at the estimated values (see Figure S24). Because the $CsPbCl_3$ nanocubes acting as seeds expose the (100) facets, we focused on interfaces of the (100)∥(*hkl*) type with $h,k,l \leq 2$ for both $Pb_4S_3Cl_2$ and $Pb_3S_2Cl_2$ (see Figure S25).

**Figure 6** shows a comparison of Ogre's results for interfaces of $CsPbCl_3$(100) with $Pb_4S_3Cl_2$ and $Pb_3S_2Cl_2$ (see Table S12 for additional information). In the *lattice matching* step, all orientations considered for both $Pb_4S_3Cl_2$ and $Pb_3S_2Cl_2$ resulted in supercells within the specified thresholds for strain and area. Of these, the experimentally observed (100)∥(010) – $CsPbCl_3/Pb_4S_3Cl_2$ epitaxial relation already stands out as the most promising because of its lowest strain of 0.2% and smallest area of 63 $Å^2$. The results of *surface matching and ranking* confirmed that the (100)∥(010) – $CsPbCl_3/Pb_4S_3Cl_2$ orientation is indeed the most stable. The interface structure shown in **Figure 6b** is well-connected, which correlates with its high stability. We note that the lowest $E_{int}$ of 19 meV $Å^{-2}$ is obtained for the CsCl termination of $CsPbCl_3$, rather than the $PbCl_2$ termination observed in experiments. However, the $PbCl_2$-terminated model ranks as the second best for the (100)∥(010) – $CsPbCl_3/Pb_4S_3Cl_2$ interface, with an $E_{int}$ of 27 meV $Å^{-2}$ well below that of all the other interface models (dashed line in **Figure 6a**). This discrepancy in the surface termination is likely due to the tendency of perovskite nanocrystals to express a lead-rich surface,[114–116] which would provide a kinetic advantage to the growth of $PbCl_2$-terminated interfaces. However, other kinetic effects stemming from interactions with the reaction medium, that are not considered in our simulations, might play a role as well.

The most stable interface found for the competing $Pb_3S_2Cl_2$ phase is the (100)∥(210), with $E_{int} = 50$ meV $Å^{-2}$. A visual inspection of the interface connectivity, shown in **Figure 6c**, reveals the presence of several dangling bonds and undercoordinated ions, which explains the lower stability. As the energy of the most stable interface between $CsPbCl_3$ and $Pb_3S_2Cl_2$ is significantly higher than that of the (100)∥(010) – $CsPbCl_3/Pb_4S_3Cl_2$ experimental interface, we conclude that $Pb_3S_2Cl_2$ is unlikely to grow on the surface of the $CsPbCl_3$ seeds because it cannot outcompete the more favorable epitaxial match with $Pb_4S_3Cl_2$. This case demonstrates how Ogre can be used to predict the likely outcome of a synthesis when there are several competing phases, one of which is favored by epitaxial templating.

### 2.3. Using Ogre to interpret experimental interfaces

Another challenge posed by nano-heterostructures is the identification of interfaces obtained experimentally. For example, the determination of the two materials involved can be hindered by the overlap of broad X-ray diffraction signals, and sample-averaged compositional analyses might provide limited information due to the presence of multiple compounds. Such issues can be mitigated with spatially resolved techniques like HR-STEM coupled with energy dispersive X-ray spectroscopy (EDXS). However, these methods have large uncertainties under realistic operational conditions (~ 5-10 %, depending on the elements), and can be misled by cross-element spectral overlap or the residual presence of unreacted precursors, which can increase the concentration measured for some of the elements.[74]

Even when the two domains of a heterostructure can be identified, the question remains of whether or not the interface is epitaxial, and if yes, what structure it adopts. Atomic-resolution images of the interface may provide a definitive answer, but these are challenging to acquire and require advanced instrumentation. For this reason, it is common practice to label interfaces as epitaxial based on matching spots in the Fourier transform of lattice-resolved TEM images,[34,77,117] which however does not provide direct evidence of commensurate matching at the interface and can be easily misinterpreted. In such situations, Ogre can help identify interface models that are in agreement with the available experimental data. To exemplify this, we revisit some published colloidal heterostructures,[34,36,39,76,77] aiming to identify a plausible interface based on information provided in the original publications. We focus on $CsPbBr_3$ because of the wide availability of studies and relevance for the colloidal chemistry community.

The first example is the $CsPbBr_3/Bi_xPb_yS_z$ heterostructures recently reported by some of us (**Figure 7a**).[39] These comprise a $CsPbBr_3$ domain supporting the growth of a lead-bismuth sulfide rod, whose composition could not be identified conclusively. Indeed, the Bi-Pb-S system includes at least five phases with similar stoichiometries and structures: $Bi_2Pb_6S_9$ heyrovskyite, the $Bi_2Pb_3S_6$ polymorphs lillianite and xilingolite, $Bi_2Pb_2S_5$ cosalite, and $Bi_2PbS_4$ galenobismuthite. Moreover, $Bi_2S_3$-PbS solid solutions with non-stoichiometric compositions can also form.[118] Adding to the challenge, the presence of lead in both domains of the heterostructure and its spectral overlap with sulfur in EDXS made the compositional analysis unreliable, yielding a non-charge-balanced composition of 19.6% Pb, 29.2% Bi, and 51.2% S.[39]

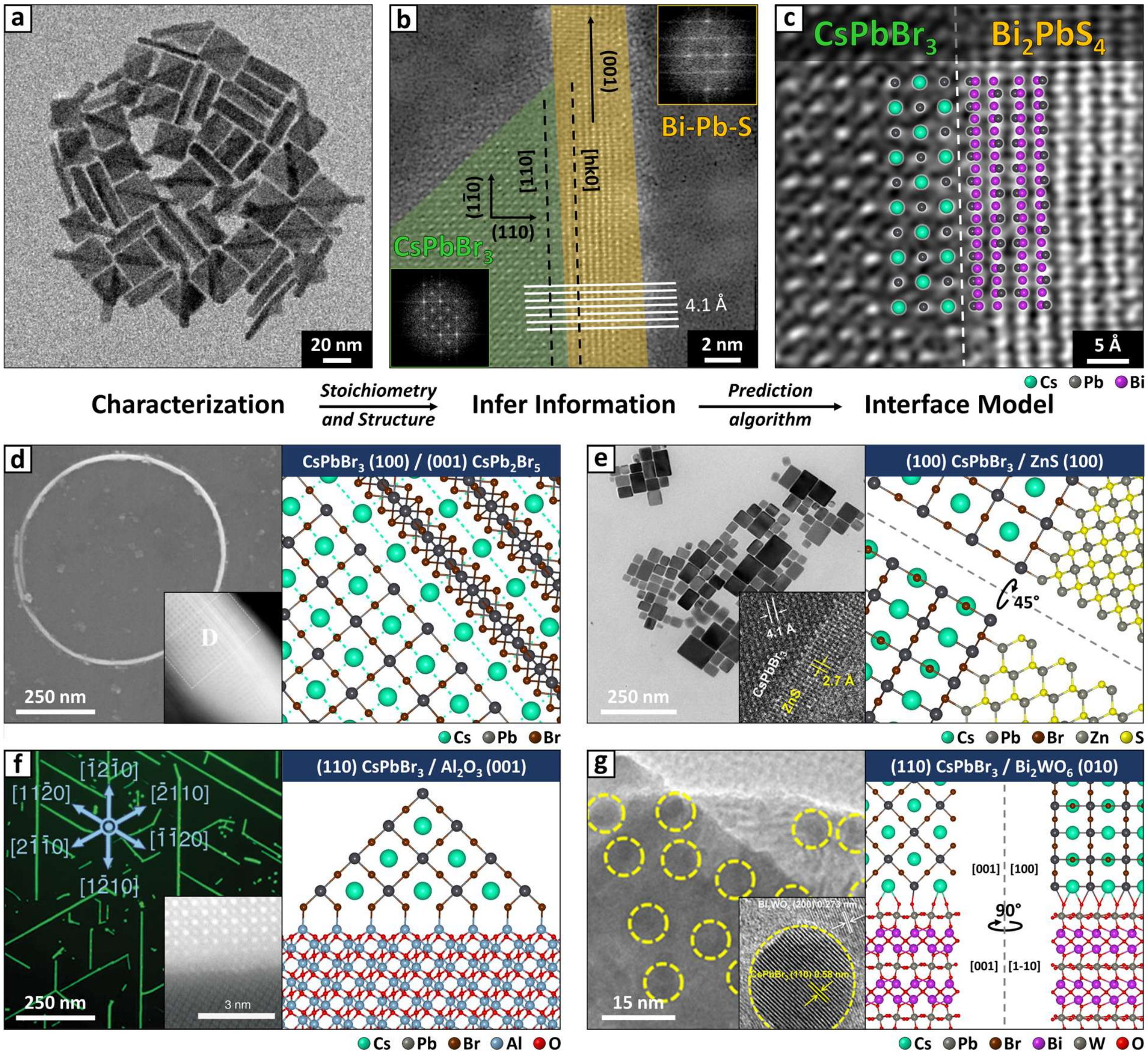

**Figure 7. Heterostructures involving $CsPbBr_3$.** a) Low-resolution TEM image of $CsPbBr_3/Bi_xPb_yS_z$ heterostructures. b) Lattice-resolved TEM image of a $CsPbBr_3/Bi_xPb_yS_z$ heterostructure, showing the flat epitaxial interface. The relative orientation of $CsPbBr_3$ and $Bi_xPb_yS_z$ domains is narrowed down via the Fourier transform analysis of their lattices (insets). Model of the (110)∥(100) – $CsPbBr_3/Bi_2PbS_4$ galenobismuthite interface proposed by Ogre superimposed to a magnified HR-STEM image of the interface (adapted with permission.[39] Copyright 2023, the Authors). d-g) Images of other heterostructures involving $CsPbBr_3$ (left and inset), compared with models of the interface produced by Ogre (right). Adapted with permission.[34,36,76,77] From left to right: (100)∥(001) – $CsPbBr_3/CsPb_2Br_5$ (d, Copyright 2020, Springer Nature);[36] (100)∥(100) – $CsPbBr_3$/ZnS (e, Copyright 2020, American Chemical Society),[34] (110)∥(001) – $CsPbBr_3/Al_2O_3$ (f, Copyright 2020, the Authors),[76] (110)∥(001) – $CsPbBr_3/Bi_2WO_6$ (g, Copyright 2020, American Chemical Society).[77]

Lattice-resolved TEM images of the heterostructures pointed to the formation of a sharp epitaxial interface. Their Fourier transform was exploited to identify the orientation of the $CsPbBr_3$ domain and determine that it exposes the (110) facet at the interface, as shown in **Figure 7b**. In this orientation the $(2\bar{2}0)$ planes of $CsPbBr_3$ are perpendicular to the interface, and their ~4.1 Å periodicity matches with that of the (001) planes of all the $Bi_xPb_yS_z$ phases considered, which share similar columnar structures with *c* as the preferred growth axis. It was also established that the *c*-axis of $Bi_xPb_yS_z$ is parallel to the interface, meaning that the sulfide rod matches the perovskite with a (*hk*0) plane. Based on this evidence, we constrained the search to a $(110)/\!/(hk0)$ – $CsPbBr_3/Bi_xPb_yS_z$ interface, with the additional requirement that the $[1\bar{1}0]$ lattice vector of $CsPbBr_3$ and the [001] vector of $Bi_xPb_yS_z$ must be parallel (i.e., $[1\bar{1}0]\Uparrow[001]$). This allowed us to lower the number of epitaxial relations between $CsPbBr_3$ and the five $Bi_xPb_yS_z$ phases considered from more than 100 (area based on **Equation 1**, strain ≤ 10%, *h,k,l* ≤ 2) to just 10 supercells that meet all the requirements (see Figure S26 and related discussion).

After performing *surface matching and ranking*, the $(110)/\!/(100)$ – $CsPbBr_3/Bi_2PbS_4$ interface with galenobismuthite emerges as the most stable, with $E_{int}$ = 46 meV Å$^{-2}$ (see Figure S27-28 and Table S13 for a full account of the results). This interface appears well-connected (see Figure S27), and the model accurately captures the positions of heavy atoms in the $Bi_xPb_yS_z$ rod observed by atomic-resolution TEM, supporting the identification of galenobismuthite (**Figure 7c**). Two additional candidate interfaces between $CsPbBr_3$ and the (100) and (110) planes of $Bi_2Pb_2S_5$ cosalite were found to be only slightly less stable. However, we consider them unlikely owing to the significantly higher strain and interface area, and because the atom positions do not match the experiment as closely as the galenobismuthite (see Figure S27). Notably, Patra et al. have later independently reported similar heterostructures, which they also identified as $CsPbBr_3/Bi_2PbS_4$ galenobismuthite, thus reinforcing our findings.[119] We remark that this conclusion would have been challenging to extract from TEM images alone because all $Bi_xPb_yS_z$ phases feature similar columns of heavy elements, whose apparent spacing under TEM depends on the orientation of the crystal.

A similar approach of exploiting morphological data to inform simulations was adopted to reassess the interfaces shown in **Figure 7d-g**, reported in Refs. 34,36,76,77. In all these cases, the epitaxial relations proposed in the original studies lead to small commensurate domains with reasonable strain and supercell area in Ogre's lattice matching step (see Figures S29-31).

Therefore, we proceed with the proposed interface orientation and focus on providing a plausible model of the interface. For instance, the lattice-resolved image of a nanoring in **Figure 7d** pointed to a (100)⫽(001) – $CsPbBr_3/CsPb_2Br_5$ interface,[36] for which the most stable model produced by Ogre (see Figure S29 and Table S14), is consistent with the empirical principle of $Cs^+$-sublattice continuity observed for cesium lead halide heterointerfaces (marked by dashed cyan lines).[120] For the $CsPbBr_3$/ZnS particles in **Figure 7e**,[34] our findings (see Figure S30 and Table S15) confirm that ZnS can passivate the facets of perovskite nanocubes by forming a (100)⫽(100) epitaxial shell, despite their different crystal structures.[13] Finally, **Figure 7f** provides an example of epitaxy on a single-crystalline substrate, where the $CsPbBr_3$ microwires grow along the lattice vectors of $Al_2O_3$.[76] The resulting (110)⫽(001) – $CsPbBr_3/Al_2O_3$ relation (see Figure S31 and Table S16) features a remarkably high strain and supercell area (see Table S5), which is likely possible thanks to the softness of halide perovskites.[121,122]

The case of $Bi_2WO_6/CsPbBr_3$ heterostructures in **Figure 7g** is different,[77] because the TEM images do not indicate a clear epitaxial relation between the tungstate and the perovskite. Indeed, Ref. 77 does not advance claims on the nature of the interface. However, the morphology of $Bi_2WO_6$ offers some clues, as this material tends to grow in wide nanosheets exposing the (010) surface due to its layered crystal structure.[123] Moreover, the lattice fringes indexed by the authors of Ref. 77 (**Figure 7g**, inset) suggest a (010)⫽(110) – $Bi_2WO_6/CsPbBr_3$ interface. Indeed, this orientation produces a promising match with a 5.0% strain and an area of 90 Å$^2$ in the Ogre's *lattice matching* step (see Figure S32). Following *surface matching and ranking*, the proposed (010)⫽(110) – $Bi_2WO_6/CsPbBr_3$ orientation produces a set of well-connected interface models. **Figure 7g** shows the third most stable ($E_{int}$ = 293 meV Å$^{-2}$) because it presents an oxide-rich termination for $Bi_2WO_6$, which we consider more likely based on kinetic considerations. Although the two top candidates are energetically more favorable ($E_{int}$ = 248 and 257 meV Å$^{-2}$, see Table S17), their formation would require a Bi-termination for the tungstate, which we deem experimentally unlikely because the perovskite was grown on pre-formed $Bi_2WO_6$ nanosheets.[77] Unfortunately, in the absence of atom-resolved TEM images it is not possible to validate this hypothesis.

## 2.4. Validation for oxide interfaces

To showcase Ogre's applicability beyond metal halides, we assess its performance for a set of known oxide-oxide interfaces, which are the most widely studied class of polar materials for heteroepitaxy.[78–81] We focus on thin films because colloidal heterostructures are not as common for oxides, which limits the availability of high quality atomic-resolution images. For all the examples presented in **Figure 8** we adopt the epitaxial orientation reported in the original publication, for which Ogre indeed finds low strain and a small supercell area (see 2D-supercells in Figures S33-36). Subsequently, the interface models produced by the *surface matching and ranking* procedure (see Tables S18-21) were compared to the experimental TEM images. All models were scaled isotropically without adjusting the epitaxial registry and interfacial distance. The small differences in the atomic positions may be attributed to strain, image aberrations, and the approximations used in the classical potential for predicting interface distances.

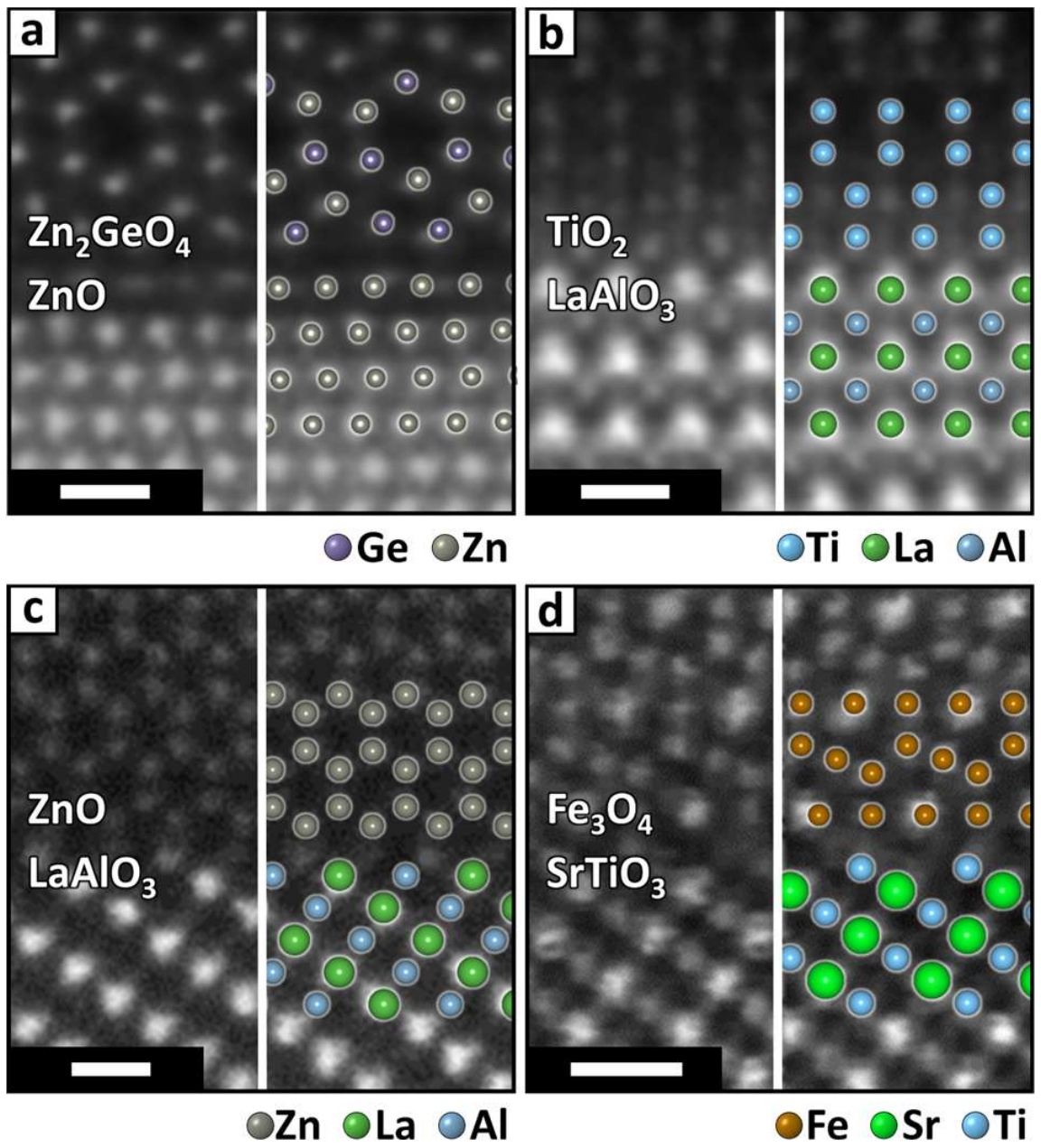

**Figure 8. Validation for oxide interfaces.** Atomic resolution images of epitaxial interfaces from the literature with the structure predicted by Ogre superimposed. TEM images adapted with permission.[78–81] Left to right, top to bottom: $(001)/\!/(\bar{1}10)$ – $ZnO/Zn_2GeO_4$ [a, Copyright 2019, IOP Publishing];[78] $(100)/\!/(001)$ – $LaAlO_3/TiO_2$ [b, Copyright 2022, the Authors];[79] $(112)/\!/(100)$ – $LaAlO_3/ZnO$ [c, Copyright 2013, Elsevier B.V.];[80] $(\bar{1}11)/\!/(111)$ – $Fe_3O_4/SrTiO_3$ [d, Copyright 2016, the Authors].[81] All models were scaled to match the TEM images without adjusting epitaxial registry and interfacial distance. Oxygen atoms are omitted to ease the comparison with electron scattering contrast. The contrast and sharpness of TEM images have been adjusted to enhance the visibility of atoms. All scalebars are 5 Å.

In three out of four cases (**Figure 8a-c**, see also Figures S33-35 and Tables S18-20), the model ranked as the most stable by Ogre yields the best match to the experimental TEM images. The sole exception is the $SrTiO_3/Fe_3O_4$ interface (**Figure 8d**), for which the model ranked second, featuring an oxygen-rich surface for the $Fe_3O_4$ domain, aligns better with the experiment than the top-ranking model, where $Fe_3O_4$ is iron-terminated (see Figure S36 and Table S21). As noted above for the (100)∥(010) – $CsPbCl_3/Pb_4S_3Cl_2$ interface, Ogre does not consider growth kinetics, reaction conditions (e.g., temperature, pressure), and interactions with the environment (e.g., the reaction atmosphere or solvent). Therefore, Ogre may predict interface configurations that are thermodynamically stable, but not kinetically favorable under the specific experimental conditions. This means that Ogre may fail to predict the outcome of epitaxial growth when the structure adopted by an interface is strongly influenced by factors that are beyond the reach of simulations, such as the native surface passivation of the substrate or the reactivity of precursors. Conversely, Ogre may offer valuable insights for optimizing the conditions of an experiment to promote the growth of a particular interface. For example, learning that the $SrTiO_3/Fe_3O_4$ interface can adopt two competing structures, differentiated by the presence of oxygen at the surface of $Fe_3O_4$, might prompt one to condition the substrate with oxygen or, conversely, apply a reducing pre-treatment to select between the two structures.

### 2.5. User best practices

Finally, for the benefit of prospective users, we provide a series of guidelines for assessing whether Ogre is the correct tool for their needs and help them make the best use of the code:

1. *Scope of simulations.* Ogre is designed to identify epitaxial relations between two given materials and propose plausible interface models. The results can inform the synthesis of new systems, help rationalize the result of experiments, and serve as a starting point for more advanced simulations based on DFT. Be mindful that Ogre does not account for variables like temperature, pressure, interactions with the reaction medium and kinetics.

2. *Selection of input CIFs.* Ogre does not check the input structures provided by the user. Hence, unreliable CIFs may lead to unreliable lattice matching results and classical potential parameters. The user is advised to check the temperature and pressure at which the structure was refined, as they influence the cell parameters and can lead to polymorphs. CIFs with partial occupancies are not supported.

3. *Lattice matching.* Select the (*hkl*) indices based on the surfaces exposed by the seed/substrate, and allow higher strain for nanomaterials. Interfaces with small supercells are more likely to succeed, even with (reasonably) higher strain. The *lattice matching* step works also for non-polar materials, as interactions are not considered.

4. *Surface matching and ranking, assessment of results.* The classical potential implemented here is only applicable to polar materials, in which formal charges can be assigned to atoms. The $E_{int}$ values it produces are valid only for relative comparisons, and DFT calculations are recommended for further refinement. We also advise the user to inspect the final models and ensure that they are chemically sound, as Ogre only verifies numerical convergence. We recommend watching out for undercoordinated ions, and for ions of the same sign that are found too close to each other at the interface.

## 4. Conclusion

Given two materials, Ogre identifies domain-matched epitaxial relations and produces structural models of the corresponding interfaces. The prediction workflow consists of three steps: *lattice matching, interface generation,* and *surface matching and ranking*. In the *lattice matching* step, combinations of lattice planes from the two materials are scanned to find commensurate epitaxial relations with low strain and supercell area. For the orientations that provide promising epitaxial matches, the *interface generation* step produces candidate models by combining all the possible surface terminations for the two materials. Finally, the *surface matching and ranking* step determines the optimal epitaxial registry and interfacial distance for each interface model, and ranks their stability by evaluating the interface energy.

A significant advancement is the implementation of two developments specifically intended for the fast prediction of interfaces between ionic or polar materials. The first is a preliminary screening at the *interface generation* stage based on charge balancing, which reduces the number of models by eliminating candidates that would be unstable due to repulsive electrostatic interactions at the interface. The second is a classical electrostatic potential parameterized based on the bulk structures of the two materials, which allows to optimize and rank the interface models at a fraction of the computational cost of DFT. We note that our classical potential is designed for ionic and polar materials, and is not intended for interfaces involving covalent, metallic, and van der Waals materials (alternative methods are available in Ogre for these cases). Thanks to these advancements, the full prediction workflow for polar interfaces can now be executed in just a few minutes on a simple laptop. To further lower the accessibility barrier for non-specialized users, the Ogre algorithm can also be executed using the desktop application *OgreInterface*, available for Windows, Linux and Mac (Figures S37-39, see Data Availability for the installation wizards).

To demonstrate its utility, we applied Ogre to a wide variety of polar interfaces, with a focus on colloidal nano-heterostructures formed by lead halide perovskites and oxide thin films grown by physical deposition methods. In the case of known interfaces, Ogre's predictions aligned well with experimental evidence, confirming its reliability. In cases where the nature of the interfaces was not yet fully established, Ogre provided useful insights to elucidate experimental observations. For example, we used Ogre to explain the formation of a specific epitaxial interface in the presence of many potential competitors, and we were able to propose plausible interface models for colloidal heterostructures that were reported but not fully characterized hitherto. Finally, the interface structures produced by Ogre can be used as starting models to pursue further geometry relaxation, stability evaluation, and prediction of electronic, magnetic, and topological properties by DFT.[124–128]

Based on the results presented here, we propose Ogre as a quick and user-friendly tool for designing experiments and assigning structures to the resulting interfaces. Example applications include assessing if two materials are likely to form a heterostructure, selecting the most promising substrate for growing a target material, and providing atomistic models to support the interpretation of high-resolution TEM images when new interfaces are obtained. Finally, we envision Ogre as a component of high-throughput workflows for the exploration of prospective epitaxial interfaces, where it may be used for the screening of material pairs from databases of inorganic structures.

We thus conclude that Ogre is a powerful, versatile, and accessible tool for the structure prediction of epitaxial interfaces between polar materials, with great potential for the discovery of new interfaces and the interpretation of experimental results. In the hands of experimentalists, Ogre could significantly boost advancements in all fields that rely on epitaxial interfaces. This includes the rapidly expanding field of colloidal nano-heterostructures, as well as epitaxial interfaces grown by thin film deposition methods for applications in electronic devices and photocatalysis.

## 5. Methods

**DFT calculations.** DFT calculations were performed using the Vienna ab Initio Simulation Package (VASP) version 6.4.2[129–133] with the projector-augmented wave (PAW) method.[129,134] The strongly constrained and appropriately normed (SCAN) meta-generalized gradient approximated (meta-GGA) was employed for the description of the exchange-correlation interaction between electrons, and the revised Vydrov-van Voorhis (rVV10) nonlocal correlation functional was used to describe van der Waals interactions. The relevant VASP INCAR tags for SCAN+rVV10 are METAGAA = SCAN, LUSE_VDW = True, BPARAM = 15.7, CPARAM = 0.0093, and LASPH = True.[107] A plane-wave cutoff of 350 eV was used. A Monkhorst *k*-point grid with a density of 3.5 points per $Å^{-1}$ was used to sample the Brillouin zone. All calculations were converged to a total energy change of less than $1 \times 10^{-5}$ eV (EDIFF = 1E-5). In all calculations involving slab structures (i.e. surfaces and interfaces), dipole corrections were applied along the *c*-axis (IDIPOL = 3, and LDIPOLE = True),[135] and a vacuum region of 60 Å was added to avoid interactions between periodic images along the surface normal. All $CsPbBr_3$ slabs were 31.8 Å thick, and all $Pb_4S_3Br_2$ slabs were 41.7 Å thick for both interface and stand-alone surface models. No surface passivation was applied to any slab or interface model in this study.

### Data Availability

For reproducibility, the version of Ogre used for this work (1.2.16) is available on GitHub at the link https://github.com/DerekDardzinski/OgreInterface/tree/ionic_heterostructures_paper. The raw output of all Ogre simulations in this work is available on Zenodo at the link https://zenodo.org/records/13472738, together with the Python scripts required to reproduce such simulations and the installation wizards for the *OgreInterface* desktop application (available for Windows, Linux, and Mac). The most updated version of the *OgreInteface* library can be installed using the Python Package Index (PyPI). Other versions of Ogre are available for download from www.nomarom.com. Raw outputs and VESTA[101] atomistic models for all interfaces discussed here are provided as Supporting Materials. Reference CIFs for all the materials studied here are provided as Supporting Material.

**Supporting Material**

The Supporting Information document is available free of charge. It contains a full report of all simulations discussed in this work, as well as more in-depth discussion on the technical aspects behind each step of the Ogre prediction workflow. The Supporting Material is available at the link https://zenodo.org/records/13472738, and contains:

- Ogre simulations for all interfaces in this work, provided as raw output.
- Installation wizards for the *OgreInterface* application (Windows, Linux, and Mac).
- Jupyter Notebook interface to run the Ogre library with more flexibility.
- Jupyter Notebook interface to reproduce all simulations discussed in this work.
- Reference CIFs for all the materials discussed in this work.
- VESTA atomistic models of all the interfaces discussed in the Main Text.

**Acknowledgements**

Stefano Toso and Derek Dardzinski contributed equally to this work. Stefano Toso and Liberato Manna acknowledge for funding the Project IEMAP (Italian Energy Materials Acceleration Platform) within the Italian Research Program ENEA-MASE (Ministero dell'Ambiente e della Sicurezza Energetica) 2021-2024 "Mission Innovation" (agreement 21A033302 GU n. 133/5-6-2021). Liberato Manna acknowledges funding from the European Research Council through the ERC Advanced Grant NEHA (grant agreement no. 101095974)Derek Dardzinski and Noa Marom acknowledge funding provided by the U.S. Department of Energy through Grant DE-SC-0019274. Computational resources were provided by the National Energy Research Scientific Computing Center (NERSC), a U.S. Department of Energy Office of Science User Facility, operated under Contract No. DE-AC02-05CH11231. We thank Dr. Gabriele Saleh, Dr. Mesoun Jabrane, and Dr. Carlo Giansante for their useful discussion and constructive feedback.

Supporting Information for

# Structure Prediction of Ionic Epitaxial Interfaces with Ogre Demonstrated for Colloidal Heterostructures of Lead Halide Perovskites

*Stefano Toso[1,†], Derek Dardzinski[2,†], Liberato Manna[1]*, Noa Marom,[2,3,4]**

[1] Nanochemistry Department, Istituto Italiano di Tecnologia, Genova, 16163, Italy

[2] Department of Materials Science and Engineering, Carnegie Mellon University, Pittsburgh, PA, 15213, United States

[3] Department of Physics, Carnegie Mellon University, Pittsburgh, PA, 15213, United States

[4] Department of Chemistry, Carnegie Mellon University, Pittsburgh, PA, 15213, United States

[†] These authors contributed equally

* E-mail: liberato.manna@iit.it, nmarom@andrew.cmu.edu

## S1. Cubic vs orthorhombic $CsPbBr_3$

### S1.1. Cubic vs orthorhombic description for CsPbX3 perovskites.

All $CsPbX_3$ (X = Cl, Br, I) perovskites are orthorhombic at room temperature, both in bulk form[1–3] and as nanocrystals.[4] They crystallize in the space group #62, for which both the standard setting *P*nma and the non-standard setting *P*bnm are adopted in the literature. In this work, the reference structure for $CsPbBr_3$ is the *P*nma ICSD-143617.[1] However, this structure differs from the ideal cubic prototype only for mild distortions. Hence, crystallographic features (*i.e.*, lattice directions and planes) that in the orthorhombic setting are technically not equivalent are nevertheless extremely similar. Therefore, we chose to adopt the pseudocubic structure for all the predictions presented in the Main Text. **Table S1** allows to convert relevant Miller indices from the pseudocubic *P*m-3m to the orthorhombic *P*nma and *P*bnm settings.

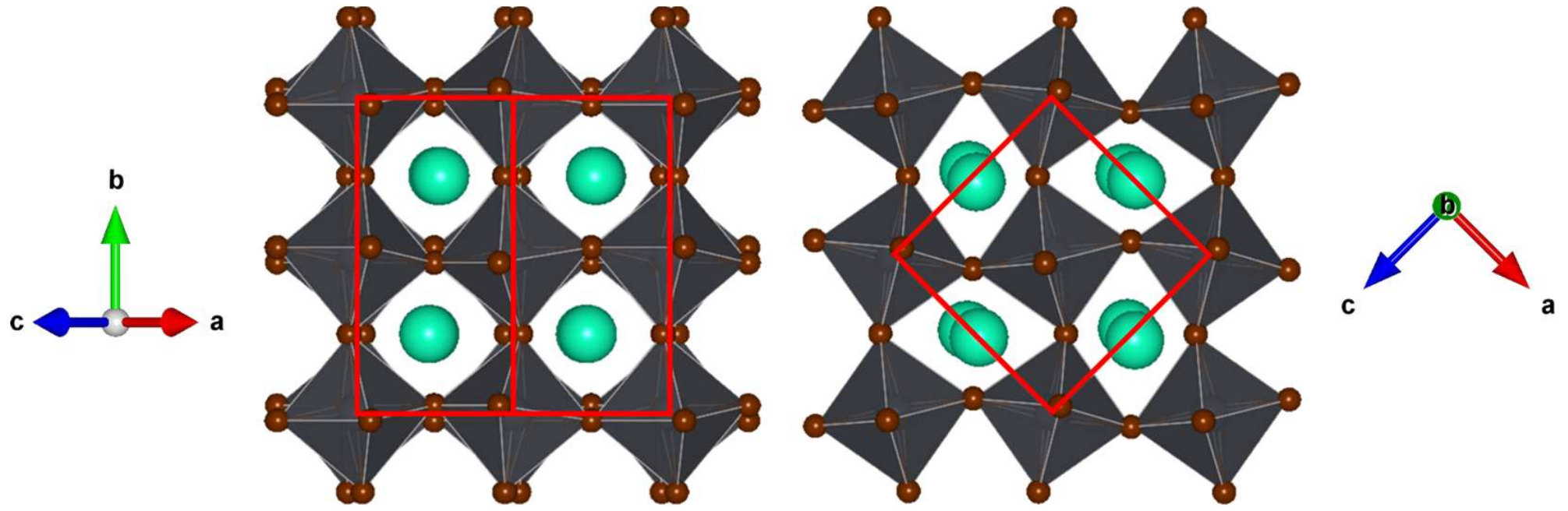

**Figure S1. $CsPbBr_3$ structure.** The orthorhombic structure of $CsPbBr_3$ (*P*nma) as seen from two non-equivalent lattice directions: [101] and [010]. Their similarities are effectively captured by the pseudocubic notation [100], where they become equivalent.

**Table S1. $CsPbBr_3$ orthorhombic to pseudocubic conversion.**

| Pseudocubic (*P*m-3m) | Orthorhombic - standard (*P*nma) | Orthorhombic – non-standard (*P*bnm) |
|---|---|---|
| (100) | (101) ; (020) | (110) ; (002) |
| (110) | (121) ; (200) ; (002) | (112) ; (200) ; (020) |
| (111) | (220) ; (022) | (202) ; (022) |

Adopting the cubic setting simplifies the discussion, as planes that are different in the orthorhombic setting become equivalent in the cubic setting (**Figure S1**). For consistency, the pseudocubic lattice parameter was calculated from the *P*nma reference through **Equation S1**:

$$a_{PC} = \sqrt[3]{\frac{Z_{Pm\bar{3}m} \cdot V_{Pnma}}{Z_{Pnma}}} = \sqrt[3]{\frac{1 \cdot 796.24\ \text{Å}^3}{4}} = 5.839\ \text{Å} \quad \textbf{(S1)}$$

## S1.2. Control simulations with Pnma CsPbBr3

To assess the impact of choosing the pseudocubic *P*m-3m over the *P*nma structure for $CsPbBr_3$, we repeated the simulations for our $CsPbBr_3/Pb_4S_3Br_2$ test interface in both settings. Since the orthorhombic structure offers two similar planes, we eventually compared three different interface models, denoted as follows:

- (100)⫽(010) – $CsPbBr_3/Pb_4S_3Br_2$ [*P*m-3m cubic setting]
- (010)⫽(010) – $CsPbBr_3/Pb_4S_3Br_2$ [*P*nma orthorhombic setting, orientation 1]
- (101)⫽(010) – $CsPbBr_3/Pb_4S_3Br_2$ [*P*nma orthorhombic setting, orientation 2]

### S1.2.1 Cubic vs Orthorhombic lattice matching results

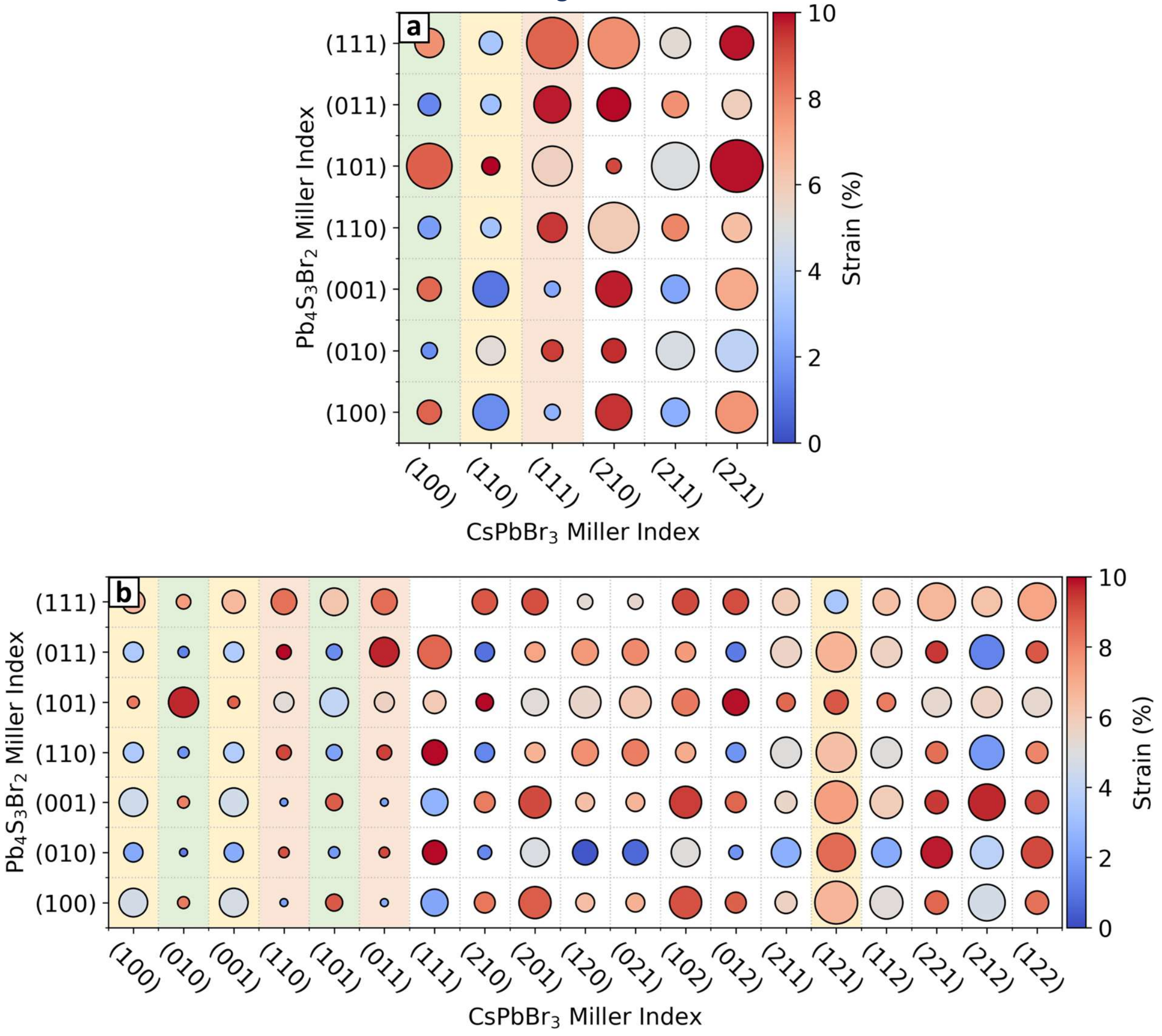

**Figure S2. *P*m-3m vs *P*nma lattice matching.** Lattice matching results for $CsPbBr_3/Pb_4S_3Br_2$ with the perovskite domain described as cubic (a) and as orthorhombic (b). Colored columns help visualize how one Miller index for the *P*m-3m cubic setting splits into similar, but not identical sets of matches for the *P*nma orthorhombic setting.

### S1.2.2 Cubic: (100)⫽(010) – $CsPbBr_3/Pb_4S_3Br_2$

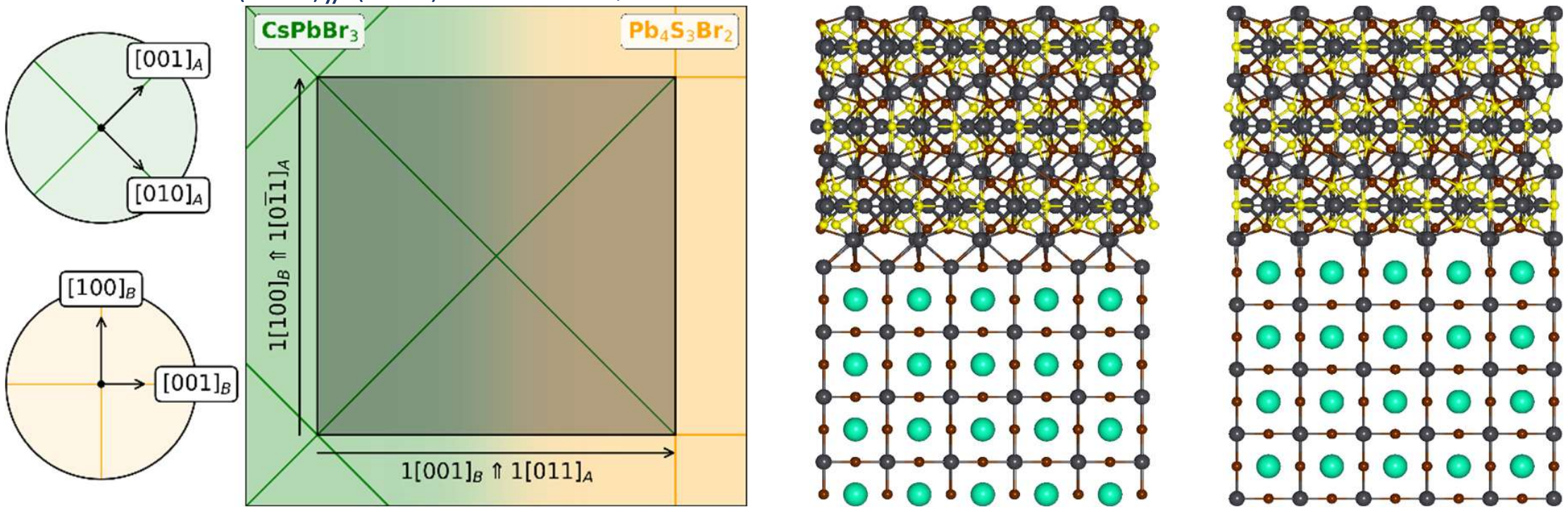

**Figure S3. (100)⫽(010) – $CsPbBr_3/Pb_4S_3Br_2$ interface (*P*m-3m).** 2D-supercell (left), $PbBr_2$-terminated model (center) and CsBr-terminated model (right) of the interface. Corresponding data are highlighted in **Table S2** (green = $PbBr_2$ termination, blue = CsBr termination). The letters in the supercell identify the two materials: *A* = first material of the pair (here $CsPbBr_3$), *B* = second material of the pair (here $Pb_4S_3Br_2$). This notation is used consistently throughout the work. The $[hkl]_A⇑[h'k'l']_B$ notation at the edges of the 2D-supercells identifies lattice vectors that lie in the plane of the interface and are parallel to each other. It is adopted in conjunction with the $(hkl)_A⫽(h'k'l')_B$ notation for planes parallel at the interface to unambiguously identify the relative orientation of the two materials. This notation is also adopted in **Table S5**. Reference structure: CCDC-2181721 ($Pb_4S_3Br_2$).

**Table S2. *Interface ranking* results.** Green = $PbBr_2$-terminated interface model. Blue = CsBr-terminated interface model. The indexes in the first two columns match the labels in **Figure 2**.

| $Pb_4S_3Br_2$ slab index | $CsPbBr_3$ slab index | Interfacial dist. [Å] | $Pb_4S_3Br_2$ charge | $CsPbBr_3$ charge | $E_{int}$ [meV Å$^{-2}$] |
|---|---|---|---|---|---|
| 4 | A | 2.47 | 0 | 0 | 52 |
| 4 | B | 3.07 | 0 | 0 | 57 |
| 6 | A | 2.97 | +2 | 0 | 75 |
| 2 | A | 3.24 | -2 | 0 | 115 |
| 6 | B | 3.74 | +2 | 0 | 118 |
| 3 | A | 3.00 | -2 | 0 | 121 |
| 2 | B | 3.80 | -2 | 0 | 126 |
| 5 | A | 3.35 | +2 | 0 | 137 |
| 3 | B | 3.97 | -2 | 0 | 138 |
| 5 | B | 3.64 | +2 | 0 | 139 |
| 7 | A | 2.55 | +4 | 0 | 332 |
| 7 | B | 3.14 | +4 | 0 | 333 |
| 1 | B | 3.91 | -4 | 0 | 347 |
| 1 | A | 4.01 | -4 | 0 | 348 |

**Notes:** this is the interface discussed in the Main Text, **Figures 3-5**.

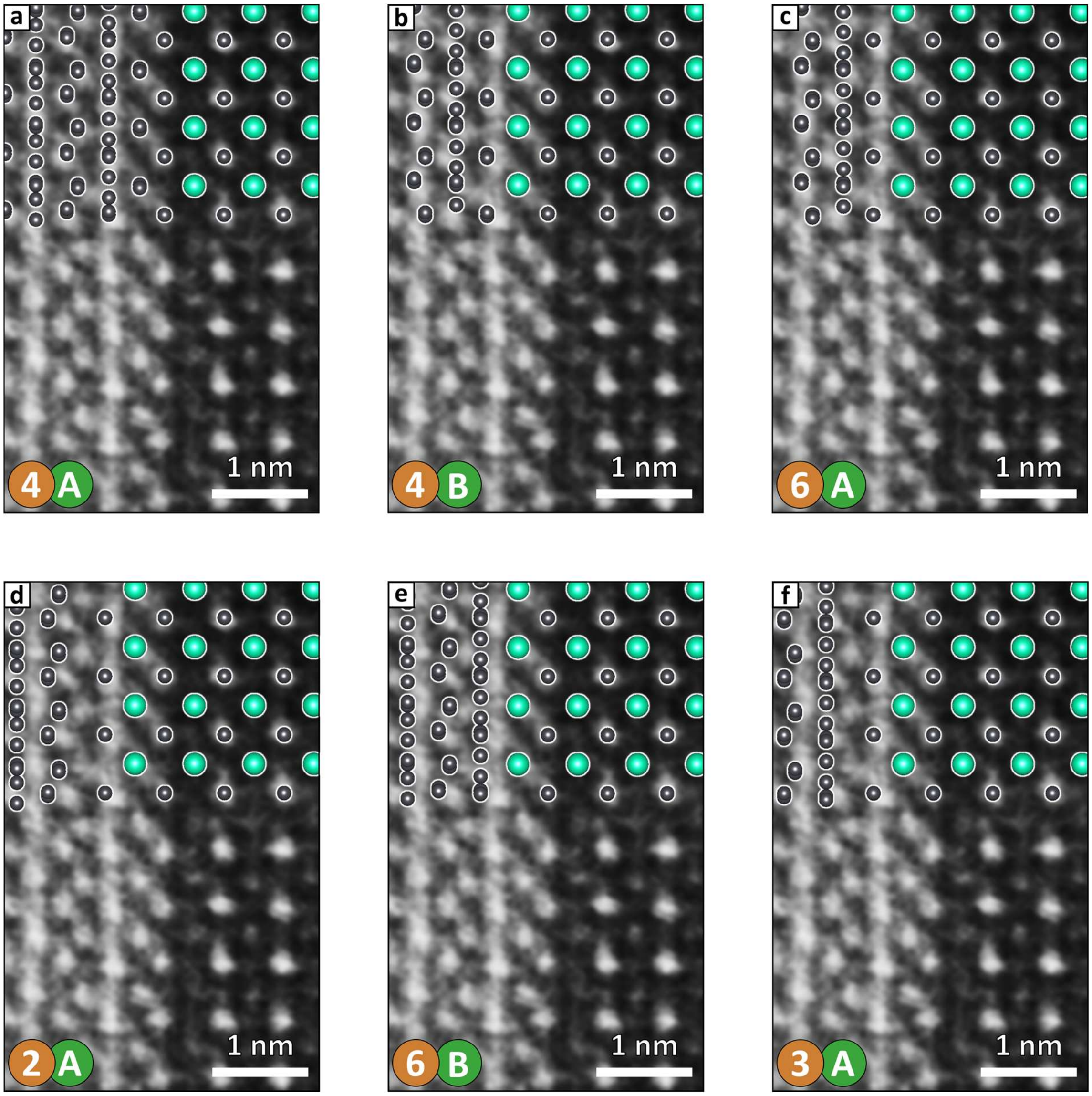

**Figure S4. Top six ranked models for the (100)⫽(010) – $CsPbBr_3/Pb_4S_3Br_2$ interface.** The models were ranked by stability and labeled as shown in **Figure 4** of the Main Text, and are here superimposed to an atomic-resolution TEM image of a heterostructure. Only heavy atoms are shown to ease the comparison with electron scattering contrast (Pb = gray, Cs = cyan). Microscopy data adapted with permission.[5] Copyright 2023, American Chemical Society. Atoms color legend: Cs = cyan; Pb = gray; S = yellow; Br = brown.

### S1.2.3 Orthorhombic 1: (010)⫽(010) – $CsPbBr_3/Pb_4S_3Br_2$

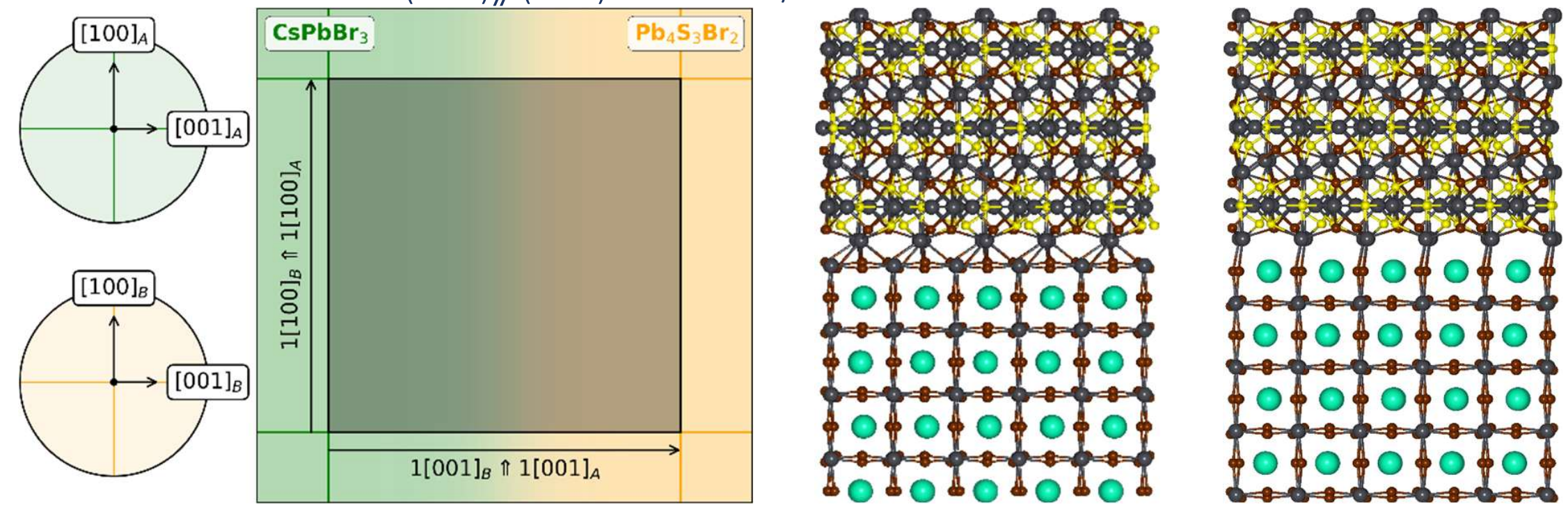

**Figure S5. (010)⫽(010) – $CsPbBr_3/Pb_4S_3Br_2$ interface (*P*nma).** 2D-supercell (left), $PbBr_2$-terminated model (center) and CsBr-terminated model (right) of the interface. Corresponding data are highlighted in **Table S3** (green = $PbBr_2$ termination, blue = CsBr termination).

**Table S3. *Interface ranking* results.** Green = $PbBr_2$-terminated interface model. Blue = CsBr-terminated interface model.

| $Pb_4S_3Br_2$ slab index | $CsPbBr_3$ slab index | Interfacial dist. [Å] | $Pb_4S_3Br_2$ charge | $CsPbBr_3$ charge | $E_{int}$ [meV Å$^{-2}$] |
|---|---|---|---|---|---|
| 0 | 0 | 2.06 | 0 | 0 | 43 |
| 0 | 1 | 3.08 | 0 | 0 | 58 |
| 3 | 0 | 2.70 | +2 | 0 | 79 |
| 2 | 0 | 2.60 | -2 | 0 | 109 |
| 3 | 1 | 3.42 | +2 | 0 | 111 |
| 4 | 0 | 3.23 | -2 | 0 | 121 |
| 4 | 1 | 3.37 | -2 | 0 | 121 |
| 5 | 0 | 2.71 | +2 | 0 | 126 |
| 2 | 1 | 3.54 | -2 | 0 | 133 |
| 5 | 1 | 3.18 | +2 | 0 | 134 |
| 6 | 1 | 2.78 | +4 | 0 | 334 |
| 6 | 0 | 3.19 | +4 | 0 | 334 |
| 1 | 1 | 3.51 | -4 | 0 | 348 |
| 1 | 0 | 3.30 | -4 | 0 | 349 |

**Notes:** the most stable model here is structurally equivalent to the $PbBr_2$-terminated model for the cubic setting (see **Paragraph S1.2.2**). The two interface models are also comparable by $E_{int}$.

### S1.2.4 Orthorhombic 2: (101)⫽(010) – $CsPbBr_3/Pb_4S_3Br_2$

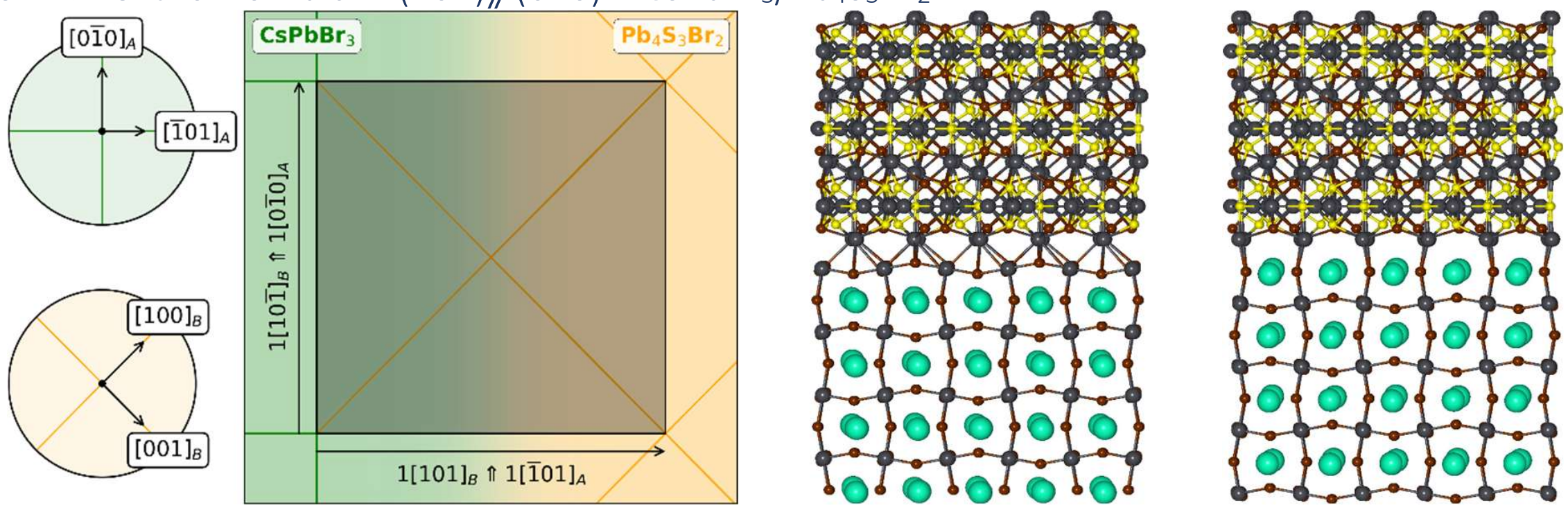

**Figure S6. (101)⫽(010) – $CsPbBr_3/Pb_4S_3Br_2$ interface (*P*nma).** 2D-supercell (left), $PbBr_2$-terminated model (center) and CsBr-terminated model (right) of the interface. Corresponding data are highlighted in **Table S4** (green = $PbBr_2$ termination, blue = CsBr termination).

**Table S4. *Interface ranking* results.** Green = $PbBr_2$-terminated interface model. Blue = CsBr-terminated interface model.

| $Pb_4S_3Br_2$ slab index | $CsPbBr_3$ slab index | Interfacial dist. [Å] | $Pb_4S_3Br_2$ charge | $CsPbBr_3$ charge | $E_{int}$ [meV Å$^{-2}$] |
|---|---|---|---|---|---|
| 0 | 1 | 2.83 | 0 | 0 | 59 |
| 0 | 0 | 2.25 | 0 | 0 | 60 |
| 3 | 0 | 2.62 | +2 | 0 | 88 |
| 3 | 1 | 3.27 | +2 | 0 | 113 |
| 2 | 0 | 2.93 | -2 | 0 | 123 |
| 4 | 1 | 2.52 | -2 | 0 | 123 |
| 4 | 0 | 3.27 | -2 | 0 | 124 |
| 5 | 0 | 2.83 | +2 | 0 | 135 |
| 2 | 1 | 3.10 | -2 | 0 | 137 |
| 5 | 1 | 3.45 | +2 | 0 | 138 |
| 6 | 0 | 2.25 | +4 | 0 | 333 |
| 6 | 1 | 2.93 | +4 | 0 | 334 |
| 1 | 1 | 3.41 | -4 | 0 | 349 |
| 1 | 0 | 3.93 | -4 | 0 | 353 |

**Notes:** different from the two cases above, the most stable model here corresponds to a CsBr-termination for $CsPbBr_3$. However, the $PbBr_2$-terminated model is ranked second and with a negligible interface energy difference (0.3 meV Å$^{-2}$), indicating that in this setting the two interface models are equivalent in terms of stability.

### S1.2.4 Cubic and Orthorhombic results compared

As seen in **Figure S7**, both the cubic and orthorhombic settings for the perovskite produced comparable models for interfaces formed by $PbBr_2$-terminated (top row) and CsBr-terminated (bottom row) slabs of $CsPbBr_3$. All models were comparable by $E_{int}$, with a mild preference for $PbBr_2$-terminated models in two cases out of three, suggesting that the two terminations are comparable in terms of stability. The fact that only the $PbBr_2$-terminated interface was reported experimentally is likely due to the influence of experimental factors,[5] such as the tendency of $CsPbBr_3$ nanocrystals to adopt a Pb-rich surface layer.[6,7] Overall, these results confirm that selecting the pseudocubic *P*m-3m setting instead of the more accurate but complex *P*nma description for $CsPbBr_3$ does not impact substantially the results of simulations.

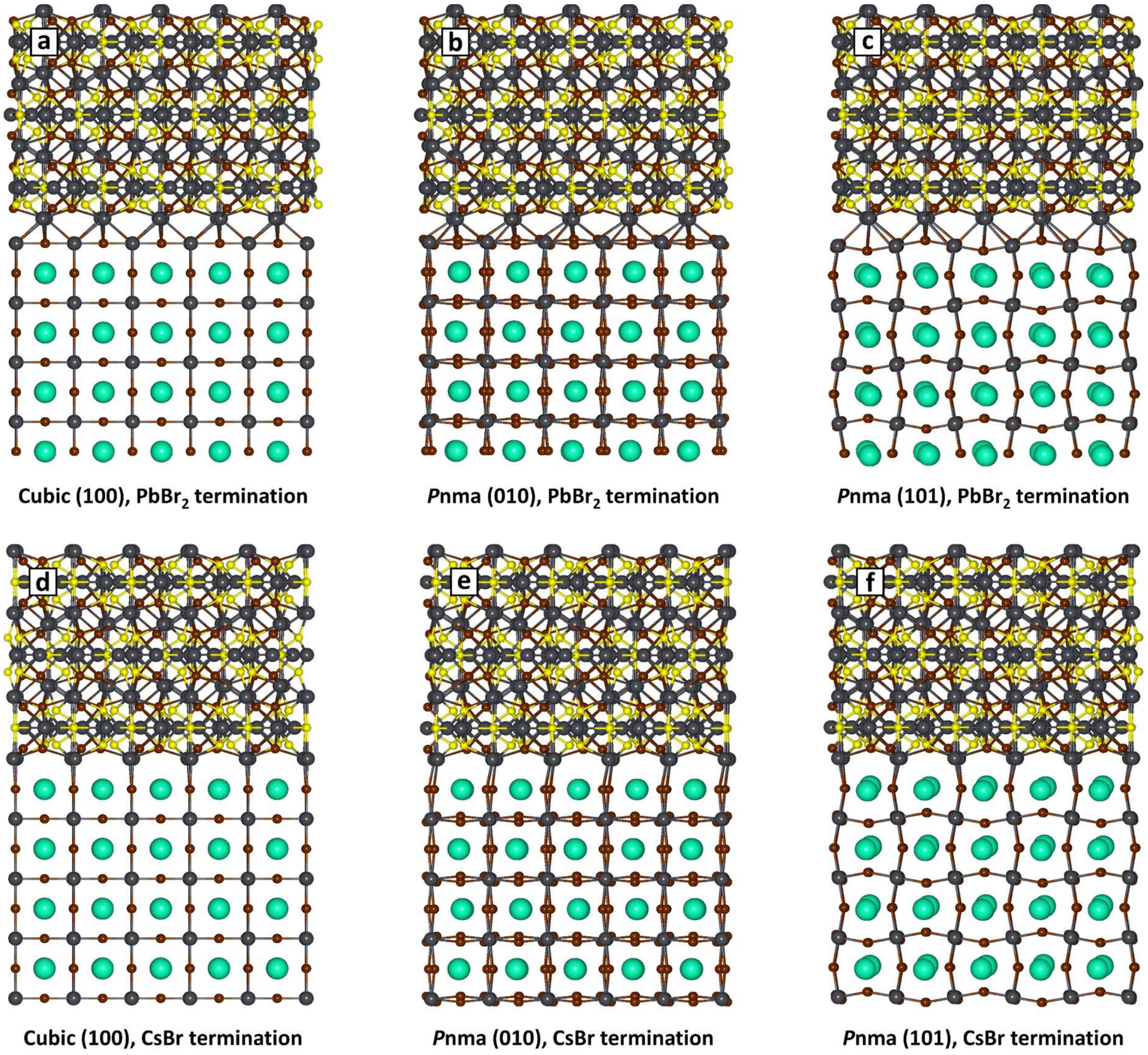

**Figure S7. Interfaces formed by $Pb_4S_3Br_2$ with $PbBr_2$- and CsBr-terminated slabs of $CsPbBr_3$ in the cubic and orthorhombic settings examined.** a-c) Interfaces formed by $PbBr_2$-terminated $CsPbBr_3$ slabs. d-f) Interfaces formed by CsBr-terminated $CsPbBr_3$ slabs.

## S2. Interfaces discussed in the Main Text

**Table S5. Experimental interfaces discussed in the Main Text.** The $[hkl]_A \Uparrow [h'k'l']_B$ notation indicates lattice vectors parallel to each other in the plane of the interface, and is included to avoid ambiguity in the choice of different supercells for the same $(hkl)_A /\!/ (h'k'l')_B$ pair.

| Interface | $(hkl)_A /\!/ (h'k'l')_B$<br>$[hkl]_A \Uparrow [h'k'l']_B$ | Strain (%) | Area<br>(Å$^2$) |
|---|---|---|---|
| $CsPbBr_3$ / $Pb_4S_3Br_2$<br>[Figs. 2-4, Ref. 5] | $(100)/\!/(010)$<br>$[011]\Uparrow[001]$ | 1.6 | 68 |
| $CsPbCl_3$ / $Pb_4S_3Cl_2$<br>[Fig. 6, Ref. 8] | $(100)/\!/(010)$<br>$[011]\Uparrow[001]$ | 0.2 | 63 |
| $CsPbBr_3$ / $Pb_3S_2Cl_2$<br>[Fig. 6, hypothetical] | $(200)/\!/(201)$<br>$[013]\Uparrow[010]$ | 9.4 | 314 |
| $CsPbBr_3$ / $Bi_2PbS_4$<br>[Fig. 7a-c, Ref. 9] | $(110)/\!/(100)$<br>$[1\bar{1}0]\Uparrow[001]$ | 0.9 | 96 |
| $CsPbBr_3$ / $CsPb_2Br_5$<br>[Fig. 7d, Ref. 10] | $(100)/\!/(001)$<br>$[011]\Uparrow[\bar{1}00]$ | 2.7 | 68 |
| $CsPbBr_3$ / ZnS<br>[Fig. 7e, Ref. 11] | $(100)/\!/(100)$<br>$[010]\Uparrow[001]$ | 8.0 | 34 |
| $CsPbBr_3$ / $Al_2O_3$<br>[Fig. 7f, Ref. 12] | $(110)/\!/(001)$<br>$[1\bar{1}0]\Uparrow[\bar{1}\bar{2}0]$ | 6.0 | 145 |
| $CsPbBr_3$ / $Bi_2WO_6$<br>[Fig. 7g, Ref. 13] | $(110)/\!/(010)$<br>$[001]\Uparrow[00\bar{1}]$ | 5.0 | 96 |
| ZnO / $Zn_2GeO_4$<br>[Fig. 8a, Ref. 14] | $(001)/\!/(\bar{1}10)$<br>$[100]\Uparrow[1\bar{1}\bar{1}]$ | 11.9 | 82 |
| $LaAlO_3$ / $TiO_2$<br>[Fig. 8b, Ref. 15] | $(100)/\!/(001)$<br>$[010]\Uparrow[\bar{1}00]$ | 0.2 | 14 |
| $LaAlO_3$ / ZnO<br>[Fig. 8d, Ref. 16] | $(112)/\!/(100)$<br>$[1\bar{1}0]\Uparrow[001]$ | 2.2 | 35 |
| $Fe_3O_4$ / $SrTiO_3$<br>[Fig. 8e, Ref. 17] | $(-111)/\!/(111)$<br>$[0\bar{1}1]\Uparrow[0\bar{1}1]$ | 7.5 | 31 |

## S3. Lattice matching

### S3.1. Quantification of strain

A common expression for strain at an epitaxial interface is shown in **Equation S2**:

$$S = \frac{|a_{epilayer} - a_{substrate}|}{a_{substrate}} \cdot 100\% \qquad \textbf{(S2)}$$

where $a$ indicates the lattice parameter. This is well-suited for epitaxial films involving cubic bulk structures, where the predominant role of the substrate makes it reasonable to select its lattice parameter as a reference. However, in colloidal heterostructures the distinction between substrate and epilayer is not always clear, and the strain can equally affect both materials at the interface. Hence, it is better to take as a reference the average of lattice parameters instead:

$$S = 2 \cdot \frac{|a_{epilayer} - a_{substrate}|}{a_{epilayer} + a_{substrate}} \cdot 100\% \qquad \textbf{(S3)}$$

Nevertheless, **Equations S3-4** only work for the very simple case of two cubic materials being matched along the same lattice direction, that is $(hkl)_A = (h'k'l')_B$. This is a common situation for simple interfaces like sphalerite CdS/CdSe,[18] but cannot be applied to pairs of structurally diverse materials growing along arbitrary orientations. Therefore, we introduced a new strain parameter inspired by **Equation S3**, but based on the concept of 2D-supercells, that can be applied to any materials and crystallographic orientations.

In short, given a $(hkl)_A /\!/ (h'k'l')_B$ interface we apply the Zur-McGill algorithm[19] to construct a common 2D-supercell starting from the single-material 2D-cells of the $(hkl)_A$ and $(h'k'l')_B$ planes. As the 2D-supercell must describe the periodicity of both materials at the interface, it can be constructed from the lattice vectors of either. This results in two almost equivalent descriptions for the same supercell, that have as base vectors $X_A$ and $X_B$ respectively:

$$X_A = \begin{bmatrix} a_x^A & a_y^A \\ b_x^A & b_y^A \end{bmatrix} = \begin{bmatrix} \vec{a}_A \\ \vec{b}_A \end{bmatrix} \qquad \textbf{(S4a)}$$

$$X_B = \begin{bmatrix} a_x^B & a_y^B \\ b_x^B & b_y^B \end{bmatrix} = \begin{bmatrix} \vec{a}_B \\ \vec{b}_B \end{bmatrix} \qquad \textbf{(S4b)}$$

If the interface had no strain, both sets of base vectors would describe exactly the same supercell, and $X_A = X_B$. If strain is present, instead, and therefore when $X_A \approx X_B$, there must exist a (2×2) strain matrix $S$ that transforms one set into the other:

$$X_A \cdot S = X_A \cdot \begin{bmatrix} s_0 & s_1 \\ s_2 & s_3 \end{bmatrix} = X_B \;\rightarrow\; S = X_A^{-1} \cdot X_B \tag{S5}$$

Knowing $S$, we can compute the strain $\varepsilon$ as the euclidean norm of the difference between the identity matrix $I$ and the strain matrix $S$:

$$\varepsilon = |I - S|_2 \tag{S6}$$

One advantage of **Equation S6** is that it is equivalent to **Equation S3** when applied to simple isostructural interfaces like the CdS/CdSe mentioned above, and is therefore comparable with a classical description of strain. We warn the reader that this approach was used in the Main Text to reassess the strain for heterostructures reported in the literature, which might cause small differences between our values and those reported in the publications (if present). For example, Ref. 20 reports 10.7% strain for the InAs/ZnS interface, based on **Equation S2** and with the InAs substrate as reference. We instead indicate 12.0%, based on **Equations S4-6**.

In the *interface generation* and *interface ranking* steps of the algorithm, the user is given a choice on how to distribute the effects of strain between substrate and epilayer. This is described by the *strain_distribution* parameter, which affects the 2D-supercell:

- *strain_distribution = 0* → the film is unchanged, while the epilayer absorbs all strain.
- *strain_distribution = 1* → the epilayer is unchanged, while the film absorbs all strain.
- *strain_distribution = 0.x* → the 2D-supercell is a weighted average of the two above.

This parameter is meant to better represent different situations of epitaxial growth: for example, a thin film grown on a bulk material will likely see the substrate little affected by strain, while the epilayer would be more heavily deformed (*strain_distribution* = 0). Conversely, two chemically similar materials involved in a colloidal interface like that shown in **Figure 4d** of the Main Text will likely share the deformation equally at the interface (*strain_distribution* = 0.5). The user might want to set this parameter depending on their knowledge of the system, for example when interfacing materials with remarkably high or low Young modulus (that is, deformability). In this work, the strain distribution is assumed to be 0.5 for all interfaces.

### S3.2. Supercell area threshold

Unlike a high strain, a large supercell area does not necessarily imply an unfavorable interface. Nevertheless, smaller supercells are more likely to result in good atom-to-atom correspondence between the two slabs, and therefore to yield stable interfaces. Indeed, an interface is stable when it maximizes the attractive interactions between the two slabs that are being matched, minimizes the repulsive ones, and leaves no dangling bonds. This usually happens when the two materials being matched feature terminations with ions positioned in similar or complementary patterns, as this will allow to fulfill all the three conditions above. Hence, well-matched materials will likely have similar 2D-periodicities, and the resulting supercell will be small. Conversely, a large 2D-supercell indicates that the two materials feature dissimilar periodicities and possibly terminations, and therefore any bond formed at the interface will repeat identical to itself only after many lattice steps of both materials. Such an interface could still be stable if enough favorable but not equivalent interactions are formed in the space of such repetition, but this circumstance becomes less likely as the supercell area increases.

It is worth clarifying that we are referring here to the "*primitive*" 2D-supercell of an interface, that is the smallest supercell required to fully describe its periodicity. Indeed, an interface can always be described by larger, non-primitive 2D-supercells with no upper limit to their extension (see for example cells *a*, *b*, and *c* in **Figure S8**), but this would not affect the actual atomic structure of the interface, nor its stability. For this reason, the 2D-supercells proposed by our lattice matching algorithm are always primitive.

That said, the choice of a threshold value for the supercell area is somehow arbitrary. By default, the algorithm uses the following cutoff value (**Equation S7** = **Equation 1** in the Main Text):

$$S_T = 2 \cdot \max[S_A(hkl);\ S_B(h'k'l')] \qquad \textbf{(S7)}$$

where $S_A(hkl)$ and $S_B(h'k'l')$ are the surfaces of the 2D-cells describing the lattice planes being matched at the interface, namely $(hkl)_A$ and $(h'k'l')_B$. The logic is to include 2D-supercells that can be formed by two single-material 2D-cells positioned side-by-side, thus allowing for a more extended repeating unit. An example is seen in **Figure 2c** of the Main Text, where the 2D-supercell of the interface is exactly twice as large as the 2D-cell of the (010) – $Pb_4S_3Br_2$ plane. However, the user is free to tune such threshold depending on their needs.

### S3.3. One $(hkl)_A$//$(h'k'l')_B$ pair, many 2D-supercells

Just defining a pair of lattice planes $(hkl)_A /\!/ (h'k'l')_B$ is insufficient to fully describe the relative orientation of two materials. Indeed, the two structures can still revolve around an axis perpendicular to the interface, potentially resulting in infinite non-equivalent relative orientations. **Figure S8** shows some examples of 2D-supercells all consistent with the (100)⫽(010) – $CsPbBr_3/Pb_4S_3Br_2$ label. Of these, panels *a*, *b*, and *c* describe the same interface (*a* = primitive, *b,c* = non-primitive), as seen by the identical strain and relative orientation of the lattice vectors. The interface described in panel *d* is instead different.

The presence of multiple possible interfaces per $(hkl)_A /\!/ (h'k'l')_B$ pair imposes a choice on which interface to optimize. Based on the considerations outlined in **Paragraphs S2.1-S2.2**, we opted for prioritizing the smallest supercell among those with strain below the search threshold. We found that this strategy better reflects the direct correlation *excessive strain = instability*, while assigning a ranking role to the supercell area. In **Table S5** we reported the extended notation $(hkl)_A /\!/ (h'k'l')_B + [hkl]_A \Uparrow [h'k'l']_B$, where the first indices identify the two planes that form the interface, while the second indicates a pair of lattice vectors that are parallel to each other in the plane of the interface, thus removing any rotational ambiguity.

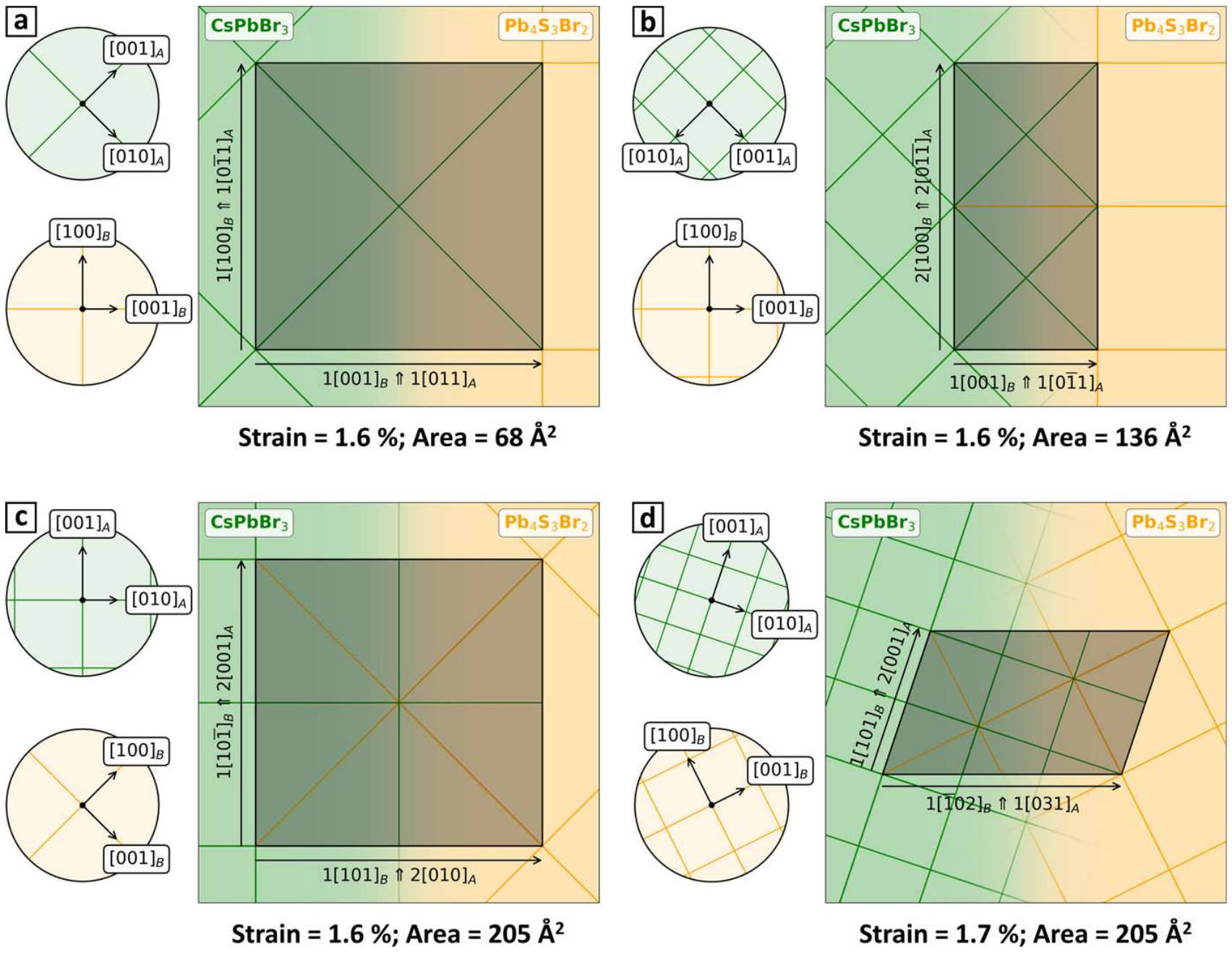

**Figure S8. Possible (100)∥(010) – $CsPbBr_3/Pb_4S_3Br_2$ supercells.** Supercells in panels (b) and (c) are "*non-primitive"* representations of interface (a), as seen by the identical strain and relative orientation of the two materials (see circular dials). The supercell (d) represents instead a different interface, characterized by a different strain and relative orientation of the two materials. Legend: $CsPbBr_3$ = green lattice and "*A*" label in captions; $Pb_4S_3Br_2$ = orange lattice and "*B*" label in captions. *Lattice matching* settings: strain < 10%; supercell area < 250 Å$^2$.

## S4. Other $CsPbBr_3/Pb_4S_3Br_2$ reported interfaces

### S4.1. Extended $CsPbBr_3/Pb_4S_3Br_2$ *lattice matching*

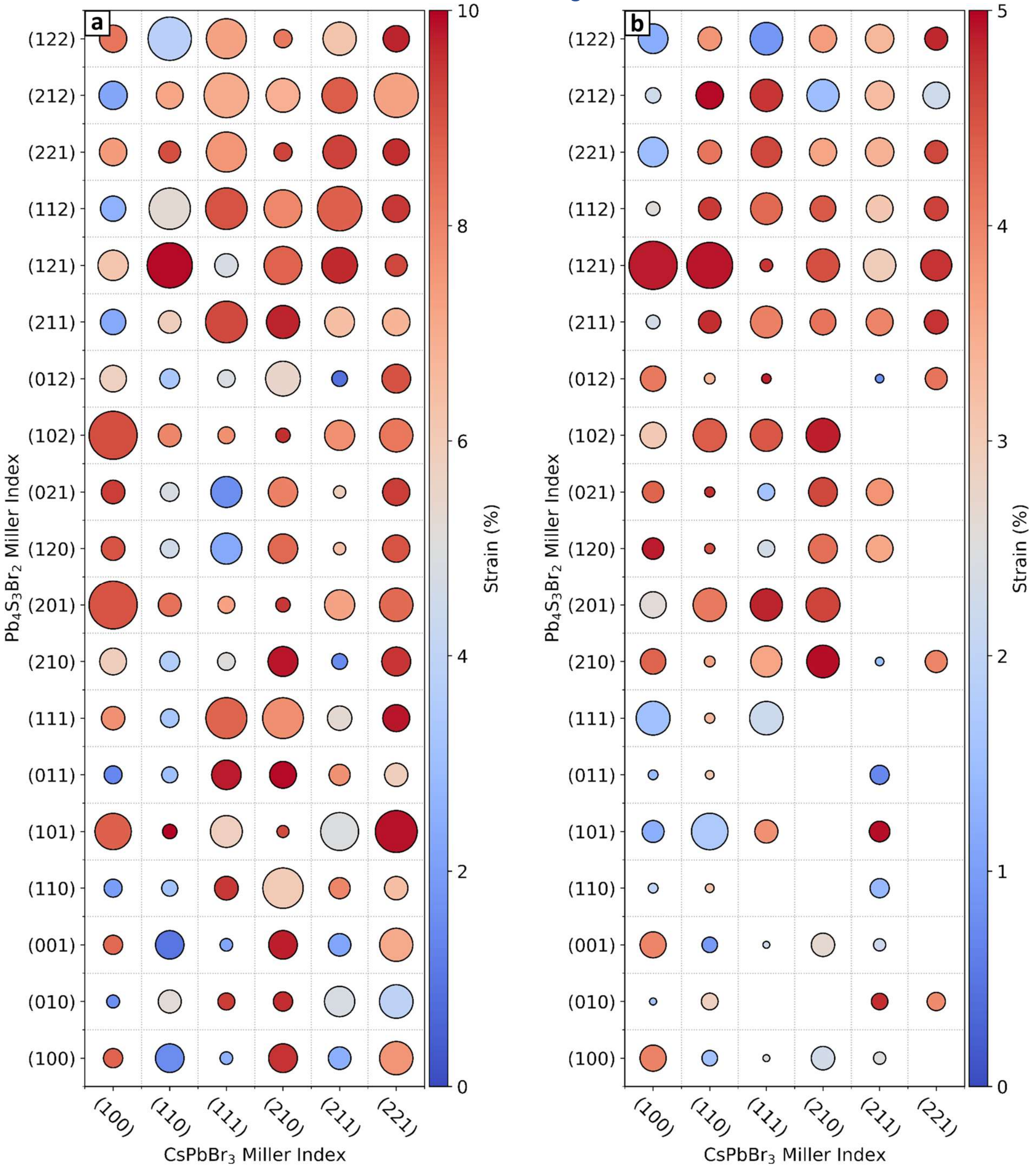

**Figure S9. Extended $CsPbBr_3$/$Pb_4S_3Br_2$ *lattice matching*.** a) A more extended version of the *lattice matching* results presented in **Figure 2** of the Main Text, here including Miller indices up to *h,k,l* = *2* for both materials. Constraints: strain < 10 %, area → **Equation S7 = Equation 1** in the Main Text. b) *Lattice matching* results obtained by setting a more limiting strain < 5% threshold. As the number of epitaxial matches increases rapidly with the Miller indices, it is advisable to consider only those lattice planes that can realistically be exposed by the substrate. In colloidal heterostructures this means limiting the planes to the facets of nanocrystal seeds, while in thin films one should focus on the lattice plane exposed by the substrate, if known.

### S4.2. (100)∥(001) – $CsPbBr_3$/$Pb_4S_3Br_2$

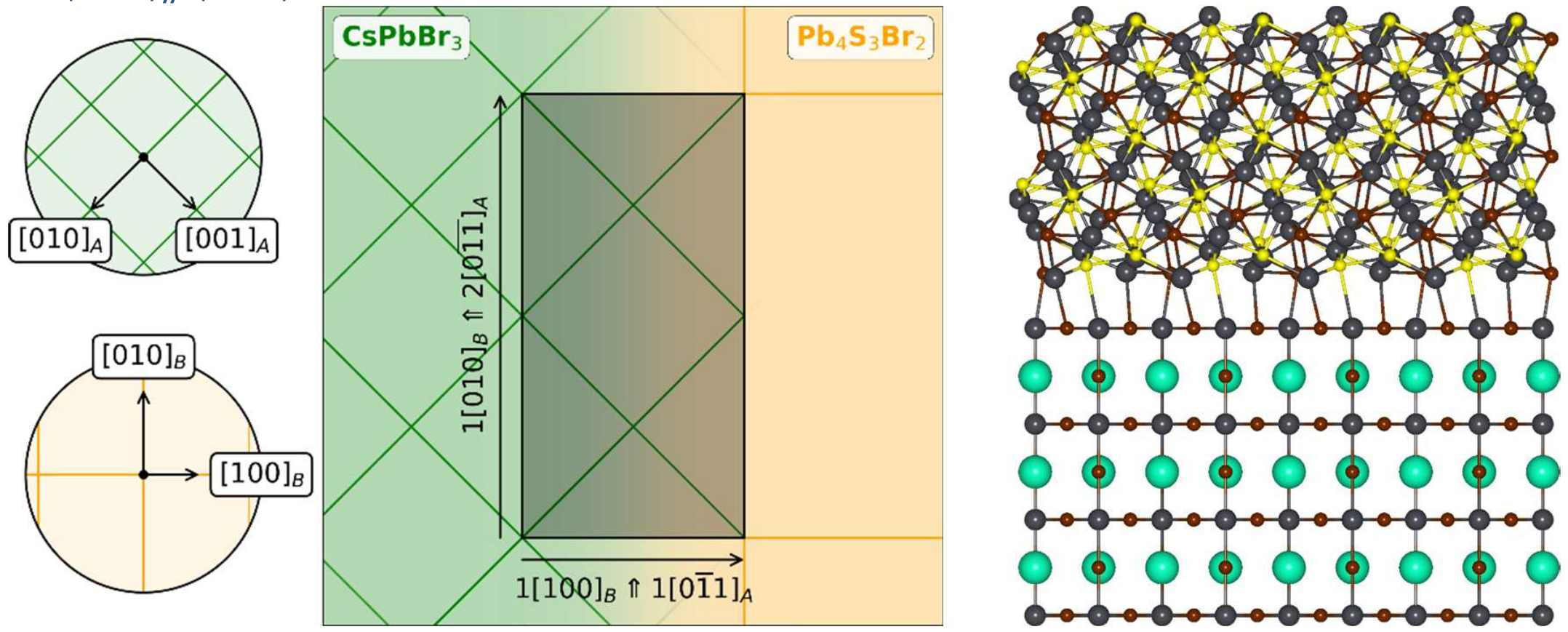

**Figure S10. (100)∥(001) – $CsPbBr_3$/$Pb_4S_3Br_2$ interface.** Supercell (left) and most stable interface model (right). The corresponding data are indicated in green in **Table S6**.

**Table S6. *Interface ranking* results**. Green indicates the most stable interface.

| $Pb_4S_3Br_2$ slab index | $CsPbBr_3$ slab index | Interfacial dist. [Å] | $Pb_4S_3Br_2$ charge | $CsPbBr_3$ charge | $E_{int}$ [meV Å$^{-2}$] |
|---|---|---|---|---|---|
| 6 | 0 | 3.28 | 0 | 0 | 77 |
| 6 | 1 | 3.57 | 0 | 0 | 77 |
| 0 | 0 | 2.97 | 0 | 0 | 105 |
| 3 | 0 | 3.41 | +2 | 0 | 111 |
| 9 | 1 | 3.60 | -2 | 0 | 111 |
| 0 | 1 | 3.62 | 0 | 0 | 111 |
| 3 | 1 | 3.67 | +2 | 0 | 112 |
| 9 | 0 | 3.55 | -2 | 0 | 112 |
| 10 | 0 | 3.17 | +2 | 0 | 113 |
| 10 | 1 | 3.56 | +2 | 0 | 115 |
| 8 | 0 | 3.28 | 0 | 0 | 116 |
| 2 | 0 | 3.24 | -2 | 0 | 116 |
| 5 | 1 | 3.30 | +2 | 0 | 117 |
| 4 | 0 | 3.51 | 0 | 0 | 117 |
| 8 | 1 | 3.71 | 0 | 0 | 117 |
| 4 | 1 | 3.91 | 0 | 0 | 117 |
| 5 | 0 | 3.21 | +2 | 0 | 117 |
| 7 | 0 | 3.61 | -2 | 0 | 118 |
| 7 | 1 | 4.16 | -2 | 0 | 119 |
| 2 | 1 | 4.22 | -2 | 0 | 119 |
| 1 | 1 | 3.24 | +2 | 0 | 131 |
| 11 | 0 | 3.19 | -2 | 0 | 133 |
| 1 | 0 | 3.14 | +2 | 0 | 133 |
| 11 | 1 | 3.77 | -2 | 0 | 135 |

## S4.3. (100)⫽(100) – $CsPbBr_3/Pb_4S_3Br_2$

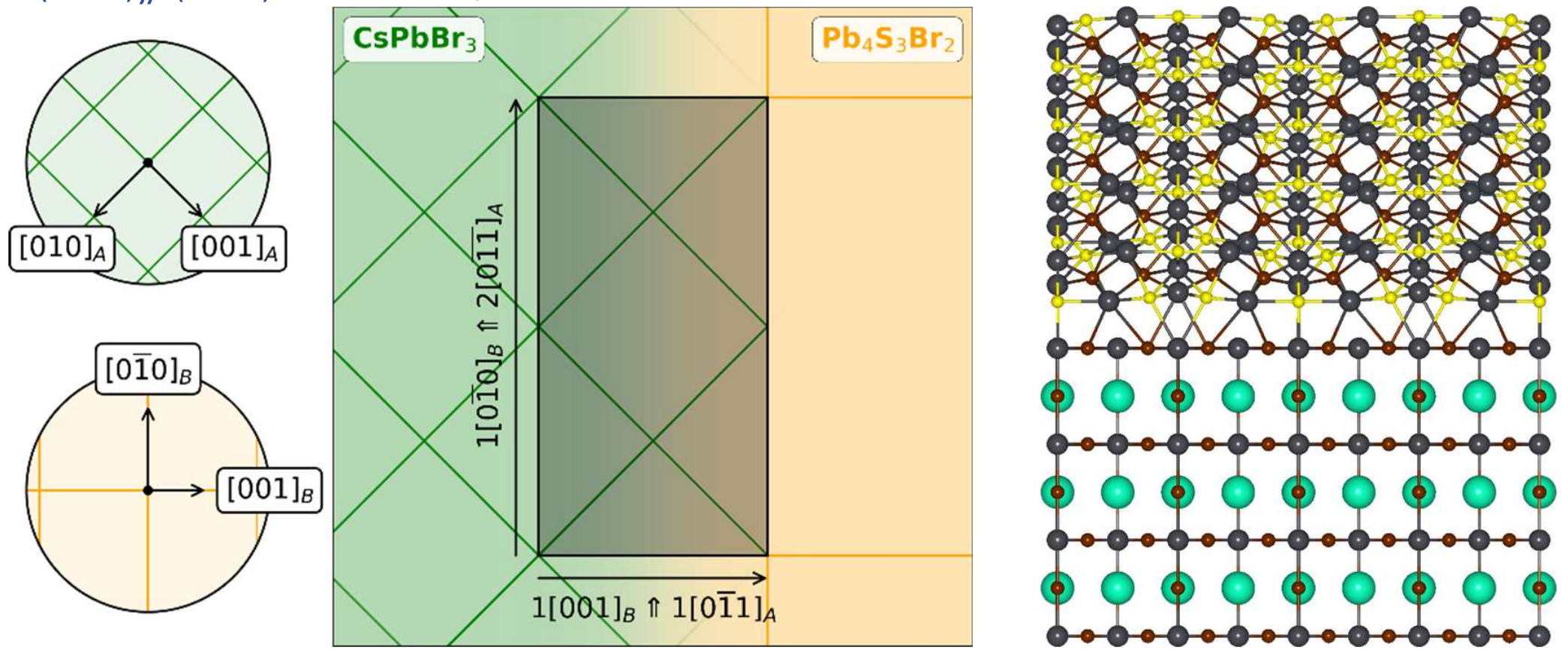

**Figure S11. (100)⫽(100) – $CsPbBr_3/Pb_4S_3Br_2$ interface.** Supercell (left) and most stable interface model (right). The corresponding data are indicated in green in **Table S7**.

**Table S7. *Interface ranking* results**. Green indicates the most stable interface.

| $Pb_4S_3Br_2$ slab index | $CsPbBr_3$ slab index | Interfacial dist. [Å] | $Pb_4S_3Br_2$ charge | $CsPbBr_3$ charge | $E_{int}$ [meV $Å^{-2}$] |
|---|---|---|---|---|---|
| 5 | 0 | 3.02 | 0 | 0 | 50 |
| 5 | 1 | 3.63 | 0 | 0 | 57 |
| 2 | 0 | 3.26 | 0 | 0 | 65 |
| 8 | 1 | 3.36 | 0 | 0 | 66 |
| 2 | 1 | 3.72 | 0 | 0 | 67 |
| 8 | 0 | 3.62 | 0 | 0 | 68 |
| 0 | 1 | 3.40 | 0 | 0 | 68 |
| 0 | 0 | 4.10 | 0 | 0 | 69 |
| 1 | 0 | 2.59 | 2 | 0 | 71 |
| 1 | 1 | 3.07 | 2 | 0 | 72 |
| 9 | 1 | 4.00 | -2 | 0 | 79 |
| 9 | 0 | 3.91 | -2 | 0 | 79 |
| 7 | 1 | 3.07 | 2 | 0 | 102 |
| 7 | 0 | 2.88 | 2 | 0 | 104 |
| 3 | 0 | 3.50 | -2 | 0 | 106 |
| 3 | 1 | 3.83 | -2 | 0 | 106 |
| 4 | 0 | 3.30 | 2 | 0 | 114 |
| 4 | 1 | 3.78 | 2 | 0 | 115 |
| 6 | 0 | 3.21 | -2 | 0 | 115 |
| 6 | 1 | 3.46 | -2 | 0 | 116 |

**Notes:** this interface and that in **Paragraph S3.2** have very similar supercells, and would be challenging to tell apart based on the Fourier Transform analysis of HRTEM images.[21] Nevertheless, they lead to different structures due to the diverse orientation of $Pb_4S_3Br_2$. Both are reasonably connected, but the (100)⫽(100) interface shown here appears more favorable by

$E_{int}$, suggesting that this might be the actual orientation of the $Pb_4S_3Br_2$ domain. Unfortunately, no atomic-resolution images of the interface are available to date to confirm this prediction.

### S4.4. (110)∥(001) – $CsPbBr_3$/$Pb_4S_3Br_2$ (low strain)

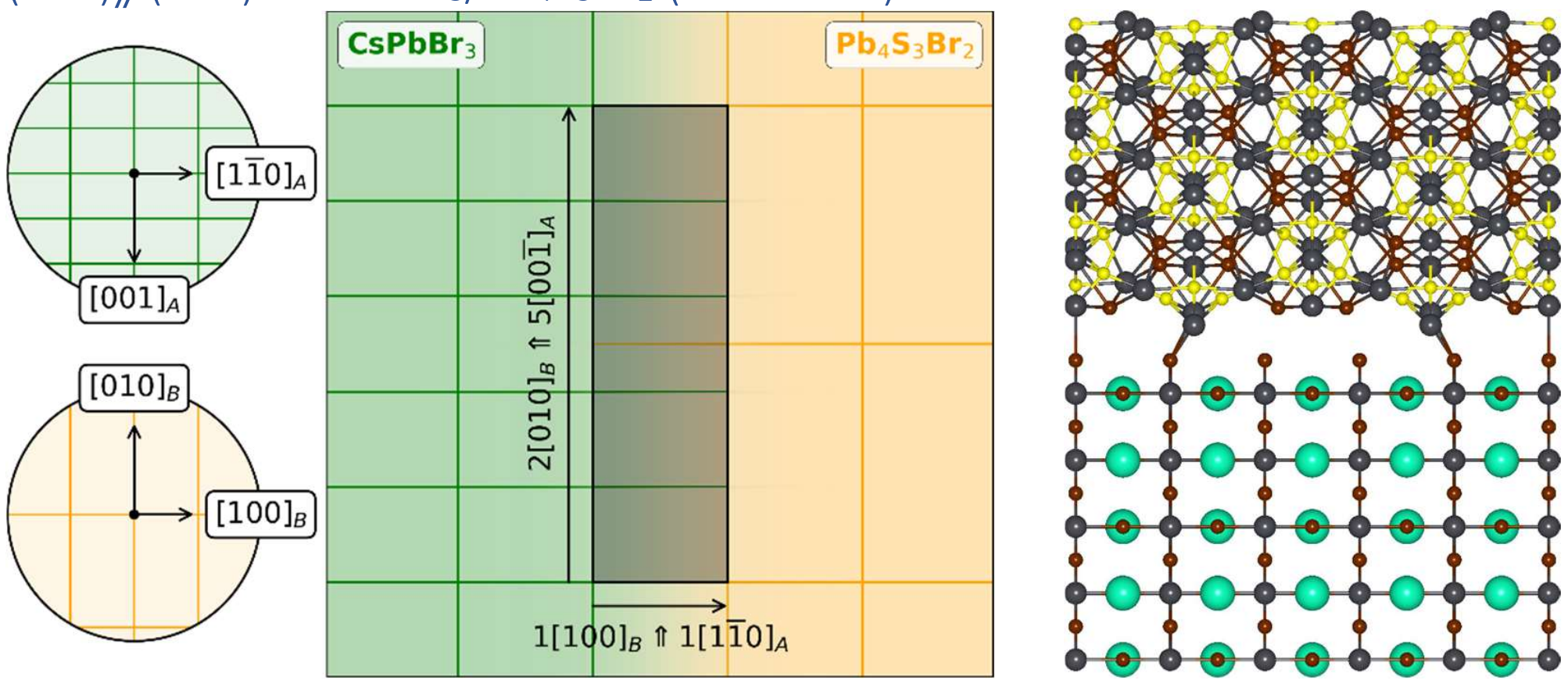

**Figure S12. (110)∥(001) – $CsPbBr_3$/$Pb_4S_3Br_2$ interface.** Supercell (left) and most stable interface model (right). The corresponding data are indicated in green in **Table S8**. The strain parameter for this interface model is 0.94%.

**Table S8. *Interface ranking* results**. Green indicates the most stable interface.

| $Pb_4S_3Br_2$ slab index | $CsPbBr_3$ slab index | Interfacial dist. [Å] | $Pb_4S_3Br_2$ charge | $CsPbBr_3$ charge | $E_{int}$ [meV Å$^{-2}$] |
|---|---|---|---|---|---|
| 5 | 1 | 2.20 | +2 | -1 | 102 |
| 6 | 1 | 3.54 | 0 | -1 | 107 |
| 9 | 0 | 2.97 | -2 | +1 | 109 |
| 3 | 1 | 2.74 | +2 | -1 | 110 |
| 6 | 0 | 4.22 | 0 | +1 | 115 |
| 10 | 1 | 2.78 | +2 | -1 | 116 |
| 2 | 0 | 3.07 | -2 | +1 | 117 |
| 7 | 0 | 3.17 | -2 | +1 | 125 |
| 1 | 1 | 2.68 | +2 | -1 | 131 |
| 11 | 0 | 2.97 | -2 | +1 | 140 |
| 8 | 1 | 3.87 | 0 | -1 | 149 |
| 4 | 0 | 4.22 | 0 | +1 | 149 |
| 0 | 0 | 4.22 | 0 | +1 | 150 |
| 0 | 1 | 4.22 | 0 | -1 | 150 |
| 8 | 0 | 4.22 | 0 | +1 | 150 |
| 4 | 1 | 4.22 | 0 | -1 | 150 |

**Notes:** the large number of dangling bonds visible in **Figure S12** make this interface unlikely to form experimentally. However, there appears to be a pattern of alternated areas of good and poor atom-to-atom matching, which suggests that a more strained but better connected model

might exist. Indeed, by increasing the strain threshold to < 15% we could identify a smaller 2D-supercell which resulted in a more stable interface model (see **Paragraphs S3.5-6**).

## S4.5. (110)∥(001) – $CsPbBr_3/Pb_4S_3Br_2$ (high strain)

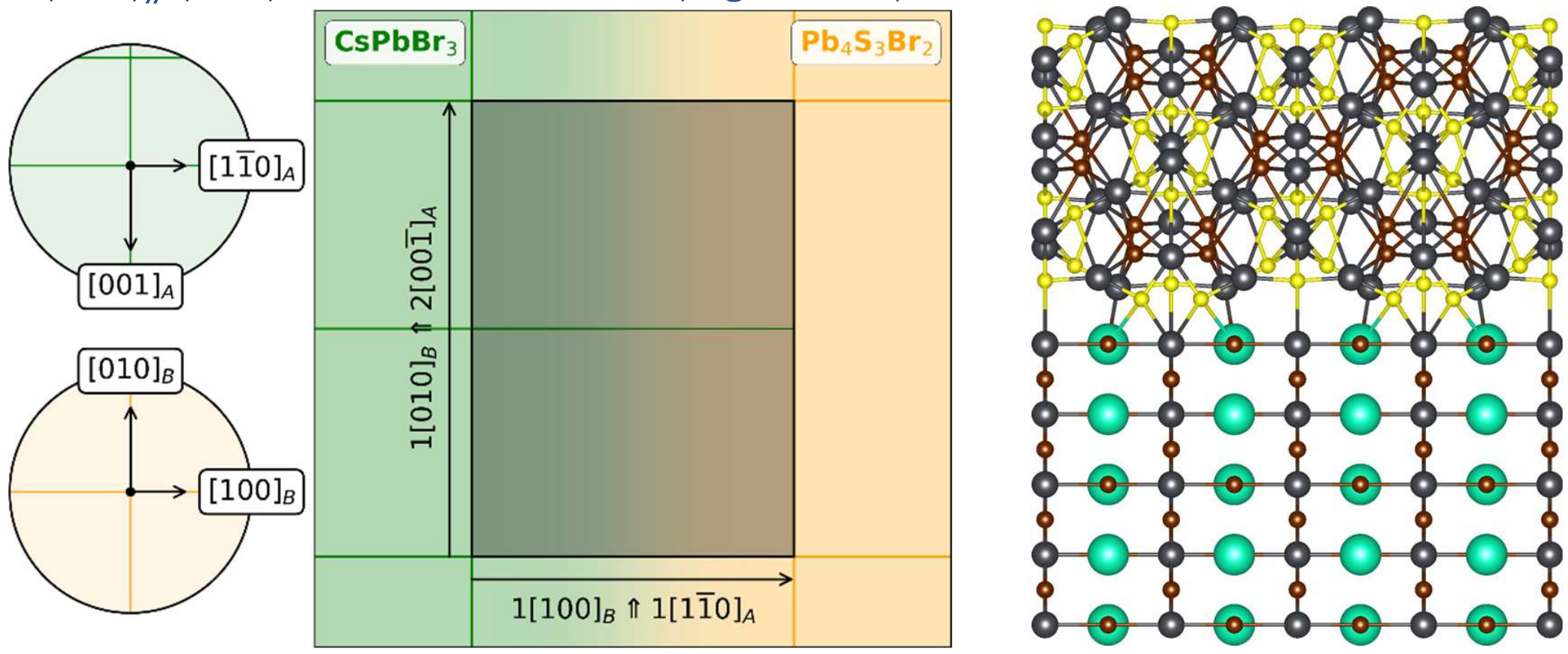

**Figure S13. (110)∥(001) – $CsPbBr_3/Pb_4S_3Br_2$ interface.** Supercell (left) and most stable interface model (right). The corresponding data are indicated in green in **Table S9**. The strain parameter for this interface model is 14.7%.

**Table S9. *Surface ranking* results**. Green indicates the most stable interface.

| $Pb_4S_3Br_2$ slab index | $CsPbBr_3$ slab index | Interfacial dist. [Å] | $Pb_4S_3Br_2$ charge | $CsPbBr_3$ charge | $E_{int}$ [meV Å$^{-2}$] |
|---|---|---|---|---|---|
| 9 | 0 | 2.36 | -2 | +1 | 68 |
| 10 | 1 | 2.32 | +2 | -1 | 68 |
| 5 | 1 | 2.40 | +2 | -1 | 84 |
| 11 | 0 | 2.34 | -2 | +1 | 90 |
| 3 | 1 | 2.56 | +2 | -1 | 91 |
| 6 | 1 | 3.28 | 0 | -1 | 100 |
| 2 | 0 | 2.89 | -2 | +1 | 103 |
| 6 | 0 | 3.28 | 0 | +1 | 110 |
| 7 | 0 | 2.86 | -2 | +1 | 119 |
| 0 | 0 | 2.92 | 0 | +1 | 119 |
| 1 | 1 | 2.59 | +2 | -1 | 124 |
| 0 | 1 | 2.64 | 0 | -1 | 136 |
| 4 | 0 | 3.10 | 0 | +1 | 140 |
| 4 | 1 | 2.91 | 0 | -1 | 143 |
| 8 | 1 | 2.62 | 0 | -1 | 143 |
| 8 | 0 | 2.89 | 0 | +1 | 144 |

**Notes:** despite being better connected that the model shown in **Paragraph S3.4**, this interface model requires a much higher strain. As this interface was reported in nano-heterostructures[21] where the contact surface between the two domains was just ~20×20 nm$^2$, it is possible that

such level of strain might be tolerable, especially considering the intrinsic softness of metal halides.[22,23] This considered, we deem the reported interface plausible, albeit perhaps not particularly favorable. However, we cannot exclude a case of misidentification, as the orientation of the two domains was assigned based on non-atomic-resolution TEM images.

### S4.6. (110)∥(100) – $CsPbBr_3$/$Pb_4S_3Br_2$ (high strain)

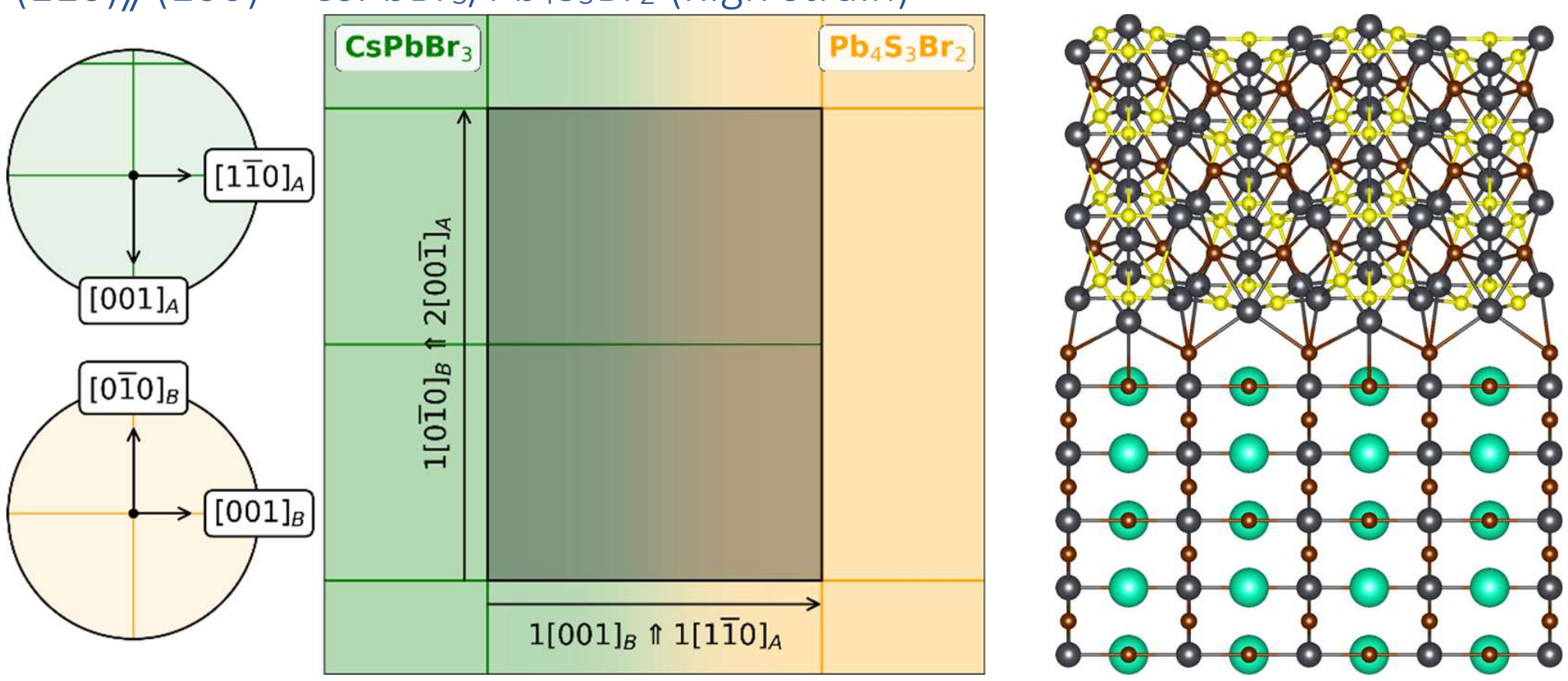

**Figure S14. (110)∥(100) – $CsPbBr_3$/$Pb_4S_3Br_2$ interface.** Supercell (left) and most stable interface model (right). The corresponding data are indicated in green in **Table S10**. The strain parameter for this interface model is 14.7%.

**Table S10. *Surface ranking* results**. Green indicates the most stable interface.

| $Pb_4S_3Br_2$ slab index | $CsPbBr_3$ slab index | Interfacial dist. [Å] | $Pb_4S_3Br_2$ charge | $CsPbBr_3$ charge | $E_{int}$ [meV Å$^{-2}$] |
|---|---|---|---|---|---|
| 1 | 1 | 1.73 | +2 | -1 | 48 |
| 9 | 0 | 2.80 | -2 | +1 | 64 |
| 5 | 0 | 2.98 | 0 | +1 | 66 |
| 3 | 0 | 2.71 | -2 | +1 | 69 |
| 7 | 1 | 2.29 | +2 | -1 | 72 |
| 4 | 1 | 2.32 | +2 | -1 | 72 |
| 5 | 1 | 2.97 | 0 | -1 | 85 |
| 2 | 1 | 2.52 | 0 | -1 | 86 |
| 8 | 1 | 2.80 | 0 | -1 | 89 |
| 2 | 0 | 2.72 | 0 | +1 | 91 |
| 0 | 0 | 2.97 | 0 | +1 | 91 |
| 8 | 0 | 2.77 | 0 | +1 | 91 |
| 6 | 0 | 2.75 | -2 | +1 | 100 |
| 0 | 1 | 4.22 | 0 | -1 | 104 |

**Notes:** this interface and that in **Paragraph S3.5** have very similar supercells, and would be challenging to tell apart based on the Fourier Transform analysis of HRTEM images.[21] Nevertheless, they lead to different structures due to the diverse orientation of $Pb_4S_3Br_2$. Both

are reasonably connected, but the (110)∥(100) interface shown here appears more favorable by $E_{int}$, suggesting that this might be the actual orientation of the $Pb_4S_3Br_2$ domain. Unfortunately, no atomic-resolution images of the interface are available to date to confirm this prediction.

**S5. Interface generation**

## S5.1. Generation of non-equivalent slabs

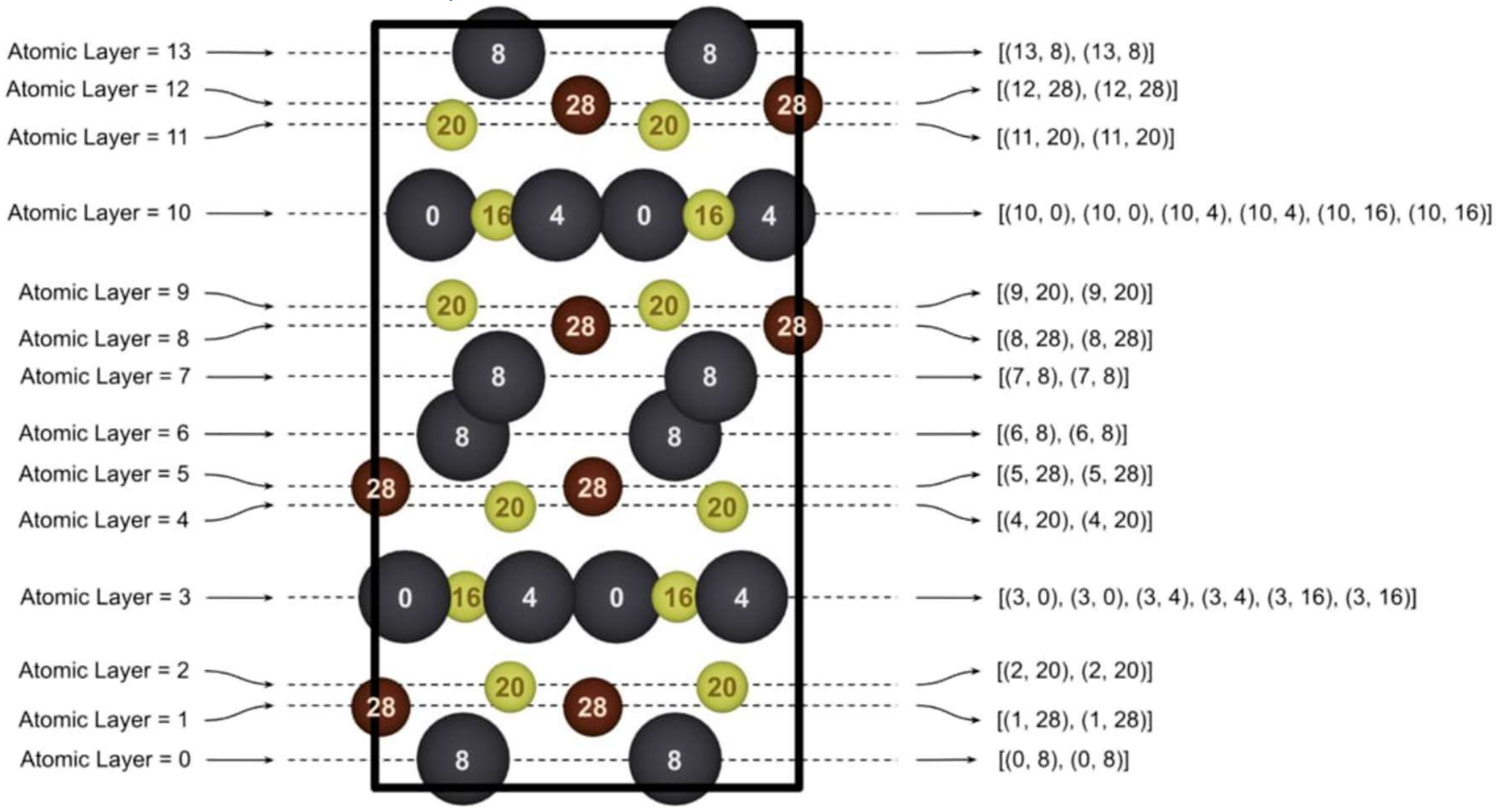

**Figure S15. Generating non-equivalent (010) slabs for $Pb_4S_3Br_2$.** Slabs are generated by cleaving the bulk structure of a material parallel to the interface contact plane. Unless clustering is applied (see **Paragraph S4.2**), the bulk is sliced at every atomic layer (left). To check if two of the resulting slabs are equivalent by symmetry, the algorithm compares the content of all atomic layers at and above the interface (i.e., moving away from the slab surface). If these are composed of atoms that are equivalent by symmetry and the layers alternate in the same exact sequence (right), then the two slabs are considered equivalent.

Given a material and a [*hkl*] lattice direction, a comprehensive set of slabs is generated by slicing the bulk structure along (*hkl*) planes positioned to cut at each atomic layer. Layers are identified as slabs of atoms with identical or similar vertical coordinates: for example, layer 3 in **Figure S15** contains 4 $Pb^{2+}$ ions (grey) and 2 $S^{2-}$ ions (yellow). When atoms are not lying exactly flat on the same plane, a tolerance threshold is applied: the motivations and clustering method are discussed in **Paragraph 4.2**, with the *P*nma structure of $CsPbBr_3$ as an example.

Once all slabs have been generated, the algorithm checks if any of them are equivalent by symmetry, meaning that they would produce identical interfaces if matched with the same slab of another material. To do so, the algorithm labels each atom in the slab based on the position they occupy in the bulk structure of the parent material using the Python package *spglib*:[24] atoms equivalent by symmetry (i.e., that are mapped one onto another by the space group symmetry operations), are given the same label. This allows to express the composition of each atomic layer in terms of equivalent atoms.

However, due to the directionality of the epitaxial growth, comparing only the composition of the surface layer is not enough to check if two slabs are equivalent. For example, layers 6 and 7 in **Figure S15** are not, despite being both terminated by equivalent $Pb^{2+}$ ions, because choosing one or the other slab would result in a different number of $Pb^{2+}$ ions at the interface. Instead, the algorithm checks all the atomic layers at and above the slab surface: if they are identical by composition and sequence (e.g., layers 3 and 10), then the two slabs are equivalent. For instance, the $Pb_4S_3Br_2$ structure shown in **Figure S15** features a $2_1$ screw axis along the [010] direction, which makes its 14 possible terminations equivalent two-by-two. Hence, we can consider just 7 non-equivalent slabs, obtained by cleaving the structure at layers 0-6.

## S5.2. Atom clustering algorithm

Unless high-symmetry structures are considered, atoms seldom lie perfectly flat on the same plane. Therefore, if we were to slice a bulk material in correspondence of every ion, the number of possible terminations would grow rapidly with its complexity. For example, slicing the *P*nma orthorhombic structure of $CsPbBr_3$ would result in either *8* (**Figure S16b**) or *10* different slabs (**Figure S16c**), depending on the direction of the cut. However, a comparison with the *P*m-3m cubic structure (**Figure S16a**) quickly reveals that most of these slabs would be meaningless for the purpose of generating epitaxial interfaces, as they would feature highly defective surface terminations with systematically missing atoms (see for example **Figure S16g-i**, where the missing $Br^-$ ions are highlighted in red).

Therefore, we introduced a clustering algorithm that regroups ions with similar distance into the same atomic layer. By default, atoms are part of the same layer if they are found within 0.15 × [max vertical distance between consecutive atoms]. For example, atoms in **Figure S16b** have a maximum vertical distance of 2.65 Å along the [010] direction (i.e., Br-to-Cs vertical distance, magenta arrow), and hence the grouping tolerance is 0.15 × 2.65 Å = 0.40 Å. As a result, the clustering algorithm could ensure that the $PbBr_2$ and CsBr atomic slabs were preserved as solid entities upon slicing, reducing the number of slabs to just 4 (two $PbBr_2$-terminated and two

CsBr-terminated). Subsequently, the symmetry equivalency check described in **Paragraph 4.1** further reduced this number to two, effectively identifying the same slabs found for the cubic *P*m-3m $CsPbBr_3$ structure (**Figure S7d-f**). See **Section S1** for the results of *lattice matching* and *interface optimization* when considering the orthorhombic *P*nma structure for $CsPbBr_3$.

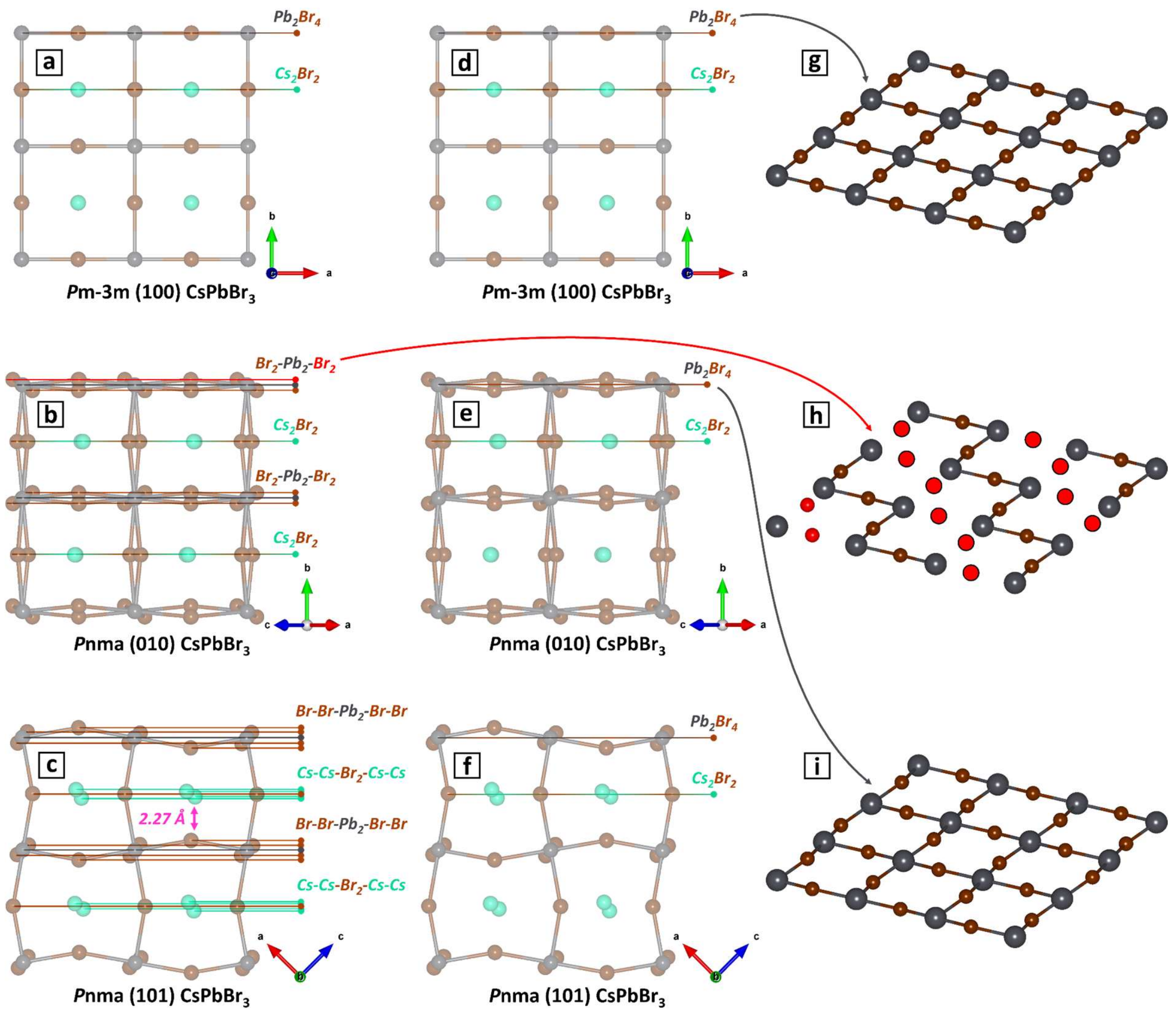

**Figure S16. Effects of atom clustering on *P*nma $CsPbBr_3$.** a-c) Surface terminations identified in the absence of atom clustering and equivalency check for a) *P*m-3m $CsPbBr_3$ along the [100] direction, b) *P*nma $CsPbBr_3$ in the [010] orientation, and c) *P*nma $CsPbBr_3$ in the [101] orientation. d-e) Surface terminations identified for the same structures when both atom clustering and equivalency check are implemented. g-i) Surface terminations of slabs resulting from the cuts highlighted. Noticeably, the surface shown in panel (h) has systematically missing $Br^-$ ions, and would therefore result in a highly defective model if used to assemble an epitaxial interface. For this reason, it is excluded by our atom clustering algorithm.

### S5.3. On the slabs surface charge $Q$

To assign a surface charge to each of the two slabs forming an interface, we must consider that:

1. Each slab should emulate the behavior of a semi-infinite solid, as real-world interfaces can extend for very long distances away from the interface plane (>> 1 unit cell).

2. Therefore, all slabs must be overall neutral to avoid unintended electrostatic interactions due to a non-zero total charge. This is ensured by constructing slabs whose thickness is a multiple of the (*hkl*) plane periodicity, which forces them to contain an integer multiple of the material's formula unit (e.g., 16 × $Pb_4S_3Br_2$). This also allows to compare the relative energy of different slabs, as the total composition is constant.

3. The surface charge has meaning and effect only locally, close to the interface, and not on average on the entire slab structure.

Given these premises, the surface charge attributed to a slab should consider both the charges of ions and their position. Ions closer to the slab surface should exert a stronger influence, while those farther away should gradually diminish their impact, as their contribution is compensated by the infinite bulk located behind them. This can be captured by constructing an *"effective surface dipole" P*, where the charges of ions are weighted over their distance from the interface:

$$P = \sum_i q_i \cdot (D - d_i) \qquad \textbf{(S8)}$$

where $i$ iterates over all ions in the slab, $q_i$ is the $i^{th}$ ion's charge as automatically assigned by PyMatGen,[25] $d_i$ is the ion's distance from the slab's surface, and $D$ is the slab thickness. By construction, $D$ is a multiple of $d_{(hkl)}$, that is the periodicity of the material perpendicular to the (*hkl*) interface. Note that the base quantity of $P$ is that of an *electric dipole = charge · length*. Therefore, we can retrieve an "*effective surface charge" Q* by dividing $P$ by its length $D$:

$$Q = \frac{P}{D} = \frac{\sum_i q_i \cdot (D - d_i)}{D} = \sum_i q_i - \frac{\sum_i q_i \cdot d_i}{D} = -\frac{\sum_i q_i \cdot d_i}{D} \qquad \textbf{(S9)}$$

which yields **Equation S9** = **Equation 2** in the Main Text. Note that $\sum_i q_i = 0$ because each slab is neutral by construction. This is also why the slab's thickness must be an integer multiple of $d_{(hkl)}$, as this ensures that the slab contains a multiple of the material's formula unit.

Conceptually, two slabs with surface charges $Q$ and $Q'$ will, in the far-field approximation (i.e., where slabs are so distant that only electrostatic interactions matter), repel or attract each other as if they were two point charges $Q$ and $Q'$ located at the slab's surfaces. This is why comparing the signs of $Q$ and $Q'$ is a convenient screening tool to exclude electrostatically unfavorable interfaces. Likewise, it intuitively explains why [0/0], [0/+], and [0/-] interfaces cannot be excluded: their far-field electrostatic interaction is null, but when put in close proximity the formation of bonds or the effect of other attractive interactions (e.g., van der Waals) might stabilize them. Finally, it also accounts for [+/-] interfaces that are not charge-balanced: the two slabs will indeed attract each other, and if the system has a way to accommodate that local charge accumulation (e.g., by introducing defects or locally altering the oxidation state of species), a stable interface might form.

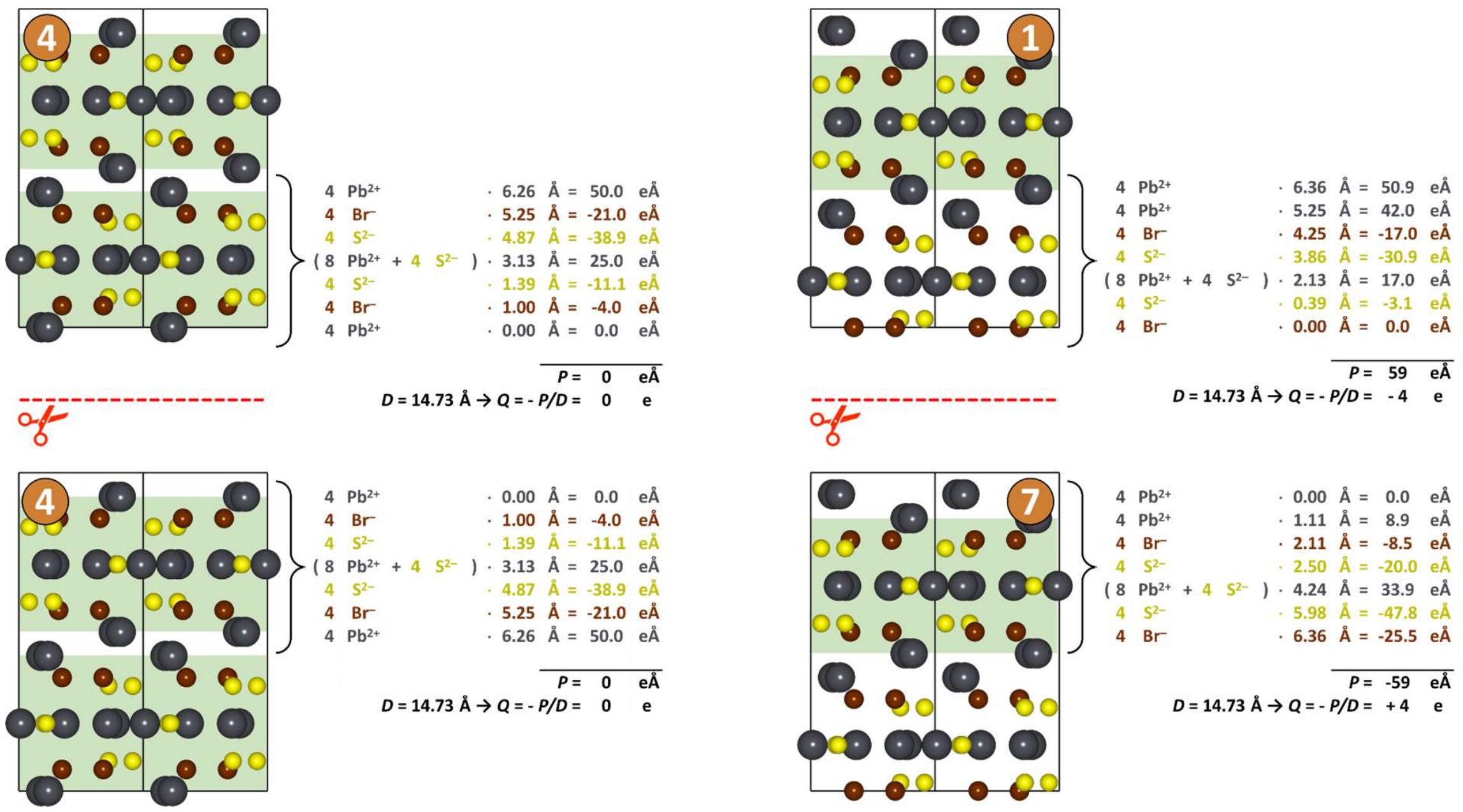

**Figure S17. Calculation of surface charges demonstrated on (010) slabs of $Pb_4S_3Br_2$.** Two (010)∥(010) – $Pb_4S_3Br_2/Pb_4S_3Br_2$ homo-interfaces are here obtained by cleaving the bulk at different planes (dashed red lines). The left interface is of the [0/0] type, while the right one is of the [+/-] type. Note that in the latter case the charges of the two slabs have opposite sign and equal magnitude, as expected when slicing a neutral bulk structure (see tables on the side, which represent the application of **Equation S9**). The green shading identifies slabs that are overall neutral. In the case of #7, this helps visually identify the source of the residual + charge in the extra layer of $Pb^{2+}$ cations on the top. All labels are consistent with **Figure 3** of the Main Text.

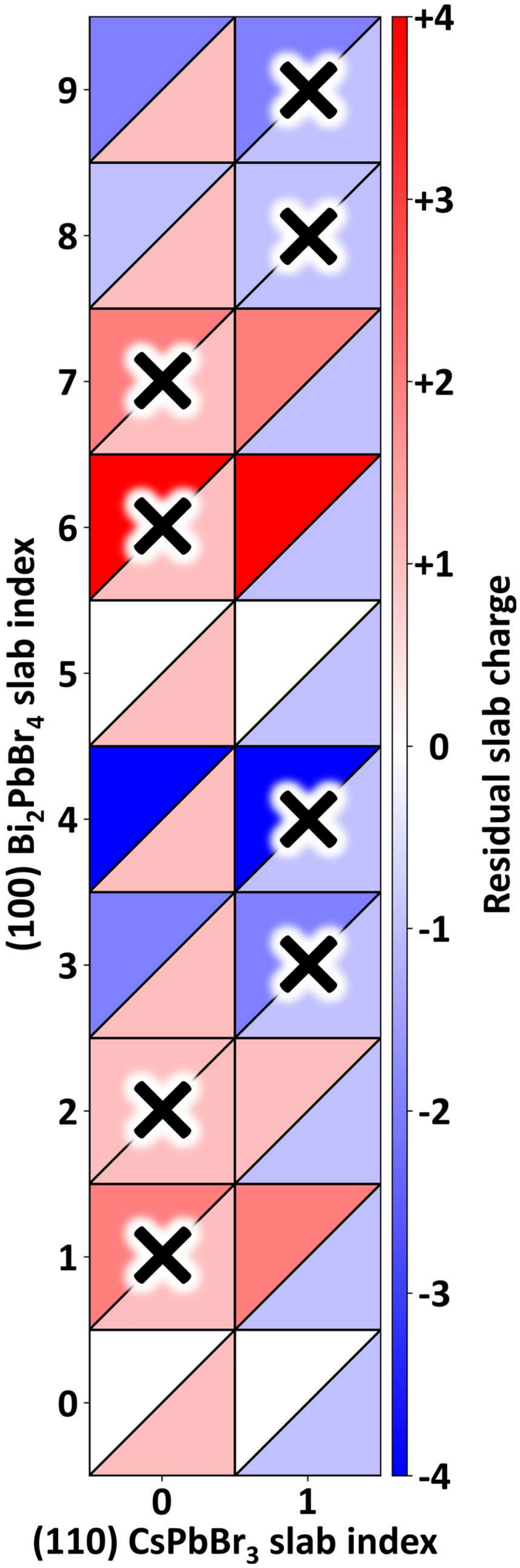

**Figure S18. Example of charge balance screening.** Results of charge balance screening for the (110)//(100) – $CsPbBr_3/Bi_2PbS_4$ interface discussed in **Section 2.3** of the Main Text and in **Section S7** of the SI. Each tile represents a pair of slabs, identified by a numerical index. The pairs marked with an × are excluded from the *surface matching and ranking* step due to their repulsive [+/+] or [–/–] nature, thus reducing the number of interface models from 20 to 12.

**S6. Surface Matching and Ranking**

## S6.1 Visual representation of $E_{int}$ and $E_{adh}$

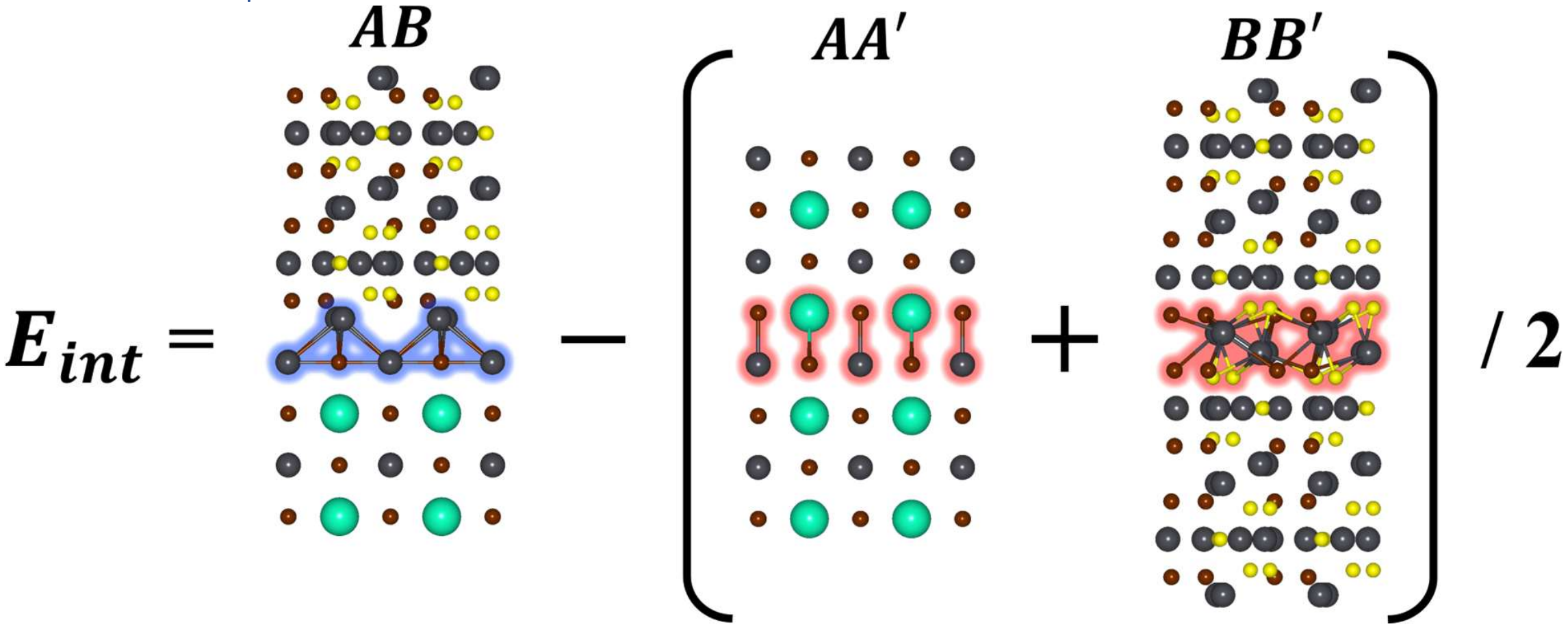

**Figure S19. Interface energy visualized for the (010)⫽(010) – $CsPbBr_3/Pb_4S_3Br_2$ interface.** $E_{int}$ compares the energy of the interface with that of the two bulk materials. In essence, it indicates how (un)favorable it is to cleave the bonds of the bulk (highlighted in red) and in exchange form those at the interface (highlighted in blue). The bulk terms are divided by two to ensure that the total energy of bonding interactions is compared over the same interface area.

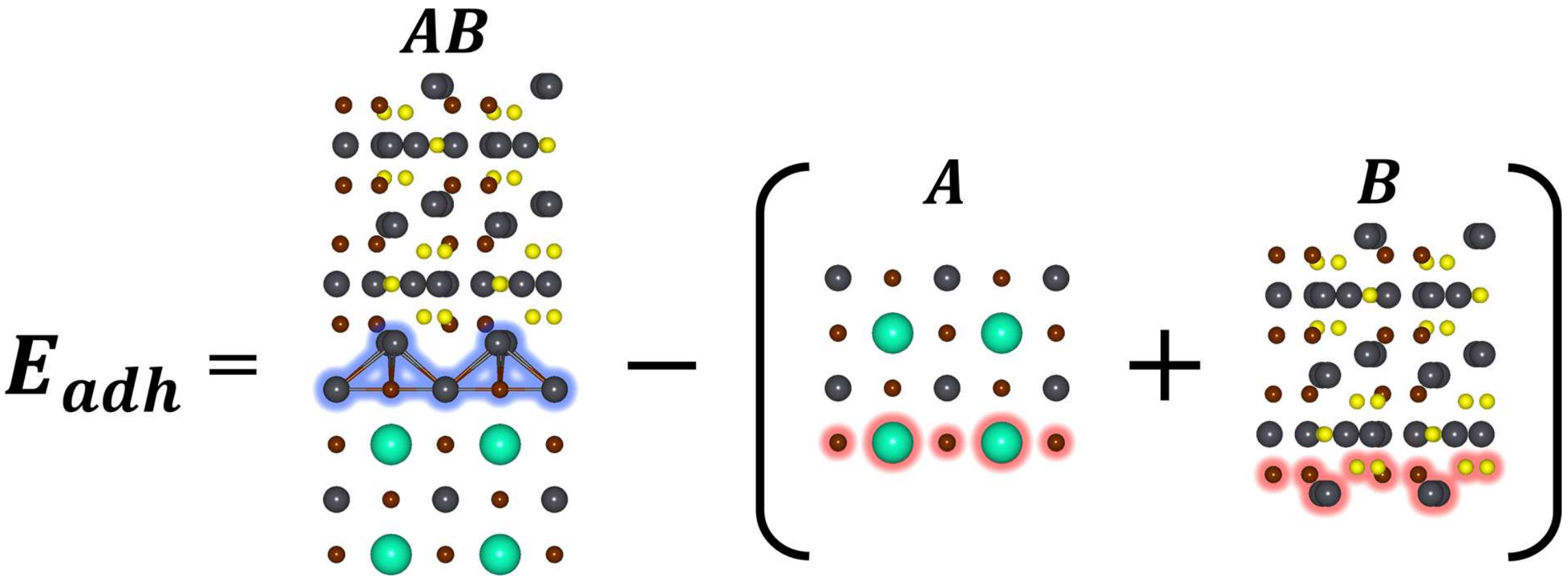

**Figure S20. Adhesion energy visualized for the (010)⫽(010) – $CsPbBr_3/Pb_4S_3Br_2$ interface.** $E_{adh}$ compares the energy of the interface with that of the two isolated slabs. In essence, it indicates how favorable it is for the two unterminated slab surfaces (highlighted in red) to bond together and form the interface (highlighted in blue). $E_{adh}$ is positive when the interactions between slabs are repulsive, and vice versa.

## S6.2 Adhesion energy maps

### S6.2.1. Constructing $E_{adh}$ maps

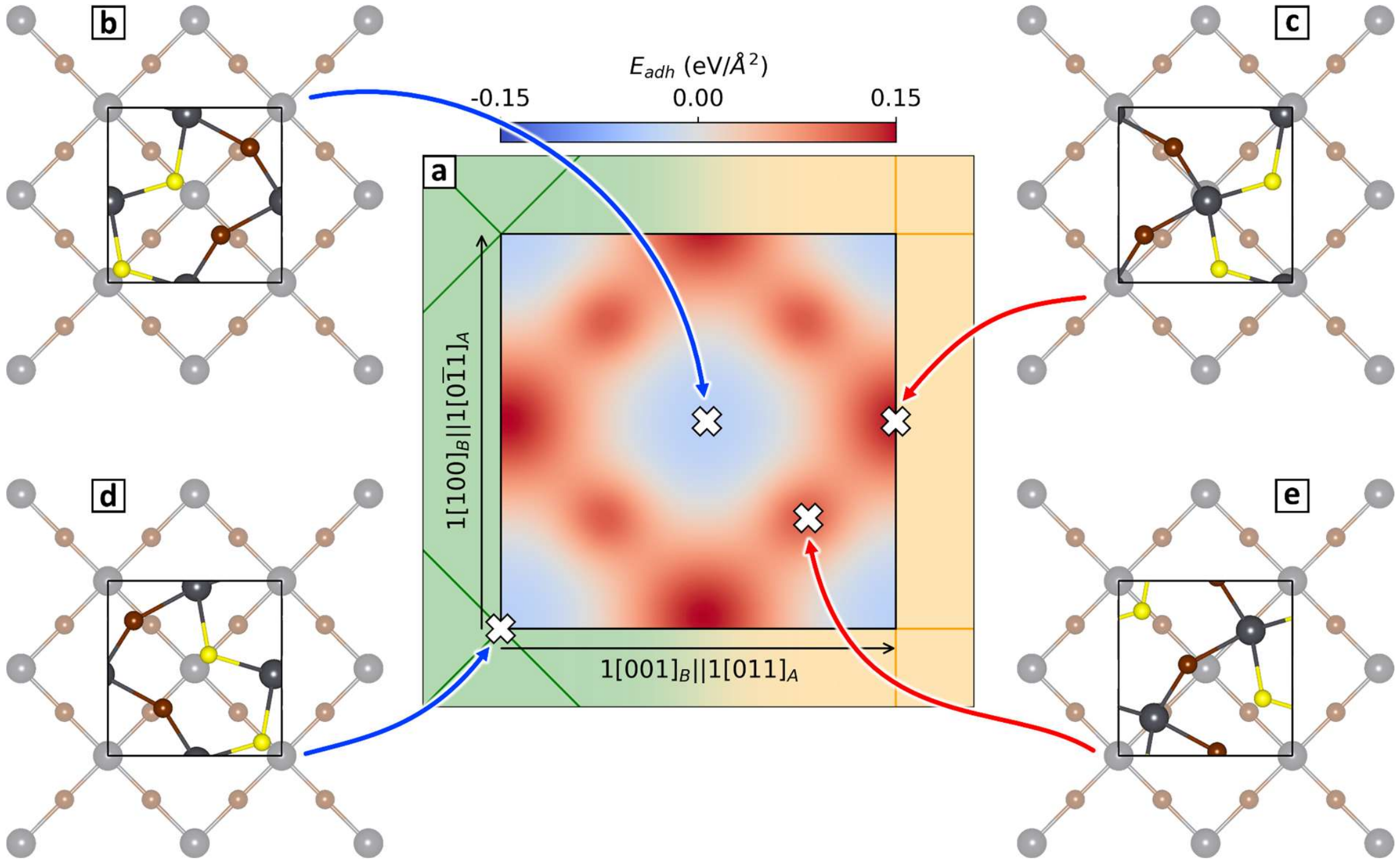

**Figure S21. Adhesion energy map for (010)⫽(010) – $Pb_4S_3Br_2$/$Pb_4S_3Br_2$.** a) The $E_{adh}$ map is shown overlaid to the 2D-supercell of the corresponding interface (see **Figure 2b** of the Main Text). The × symbols mark high-symmetry positions in the supercell, for which a top-down view of the interface is shown in panels (b-e).

To ensure an efficient optimization of the interface, Ogre employs a particle swarm algorithm[26,27] to find the energy minimum in the 3D-space formed by the *xy*-lateral shift and *z*-distance of the two slabs. Hence, both the epitaxial registry and the interfacial distance are optimized at the same time, without the need to explicitly construct $E_{adh}$ maps.

However, $E_{adh}$ maps can be useful to visually assess the quality of interface models (see **Section S5.2.2**). These are constructed by pinning one slab (material *A*), and shifting the other laterally (material *B*) to explore all the possible epitaxial registries (i.e., lateral shifts), while keeping the interfacial distance fixed at the value identified by the particle swarm optimization. **Figure S21** shows an example of $E_{adh}$ map and the corresponding epitaxial registries. We note that constructing energy maps based on $E_{adh}$ and $E_{int}$ would be equivalent, as these energies differ only by a rigid shift.

### S5.2.2. Energy *vs* interfacial distance curves and non-bonding interfaces

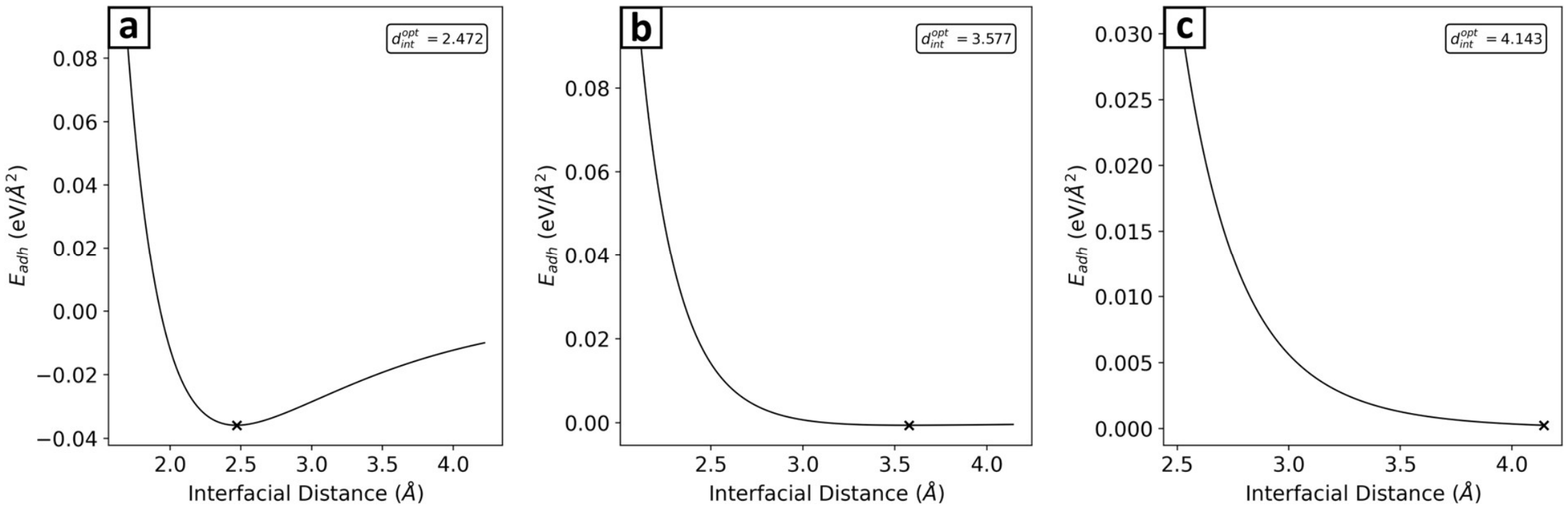

**Figure S22. Examples of $E_{adh}$ *vs* interfacial distance curves.** Three typical examples of energy *vs* interfacial distance curve profiles. From left to right: a) well-defined energy minimum (b) shallow energy minimum; c) no energy minimum, indicating a "*non-bonding*" interface.

The $E_{adh}$ *vs* interfacial distance curves can be a useful tool to interpret the results of the simulation. By default, Ogre computes such curves up to 1.2 × the maximum sum of ionic radii across all ions in materials *A* and *B*. This approach ensures that the longest possible bonding distance between the slabs always falls within the calculation range.

Normally, stable and highly favorable interface models exhibit a curve like the one shown in **Figure S22a** (from the (100)⫽(010) – $CsPbBr_3/Pb_4S_3Br_2$ interface, see **Figure 4** in the Main Text), which displays a well-defined energy minimum. In other cases, the energy minimum might appear shallow like in **Figure S22b** (from the (100)⫽(210) – $CsPbCl_3/Pb_3S_2Cl_2$ interface, see **Figure 6** in the Main Text). This might be an indication of a problematic interface, which should prompt a further, visual scrutiny of the atomistic model, and in general suggests that the calculated interfacial distance might become less accurate. Finaly, for some interface models the electrostatic term of our classical potential might be simply insufficient to counter the Born repulsive term within the given interfacial distance range, as shown in **Figure S22c** (from the (100)⫽(110) – $CsPbCl_3/Pb_3S_2Cl_2$ interface, see **Figure 6** in the Main Text). Those interfaces are considered "*non-bonding*", and should be discarded. In the Ogre output files, models of this kind are flagged as "*converged = False*".

## S6.3 Parametrized classical potential

To achieve a fast prediction of interfaces between polar materials, we implemented in Ogre a pair-wise interatomic potential consisting of the sum of an electrostatic term, which can be repulsive or attractive, and a Born term, that is always repulsive. Traditionally, the electrostatic potential of fully periodic systems is calculated through the Ewald summation, of which many optimized open-source Python implementations are available for 3D-periodic structures (e.g., PyMatGen,[25] SchNetPack,[28,29] and DScribe[30,31]). However, many of the computational strategies employed to speed up the 3D Ewald summation are not valid for 2D-periodic systems like epitaxial interfaces, and there are currently no Python-based Ewald sum implementations optimized specifically for 2D-periodic models.[32] Therefore, we implemented instead a non-Ewald algorithm known as *"damped shifted force"* (DSF), which was introduced by Fennell and Gezelter.[33] This potential has the advantage of producing results comparable to the Ewald summation without imposing any requirements on the periodicity of the system, and most importantly its complexity scales linearly as $O(N)$, while the most efficient Ewald summation implementations scale as $O(N \cdot \log(N))$.

The DSF potential is defined in **Equation S10**, where $q_i$ and $q_j$ are the charges of ions $i$ and $j$, and $d_{ij}$ is the distance between them. Here, $R_c$ is the cutoff radius of the potential, while $\alpha$ is a dampening parameter used to accelerate the convergence of the sum.

$$V_{Coulomb}^{DSF} = q_i q_j \left( \frac{erfc(\alpha d_{ij})}{d_{ij}} - \frac{erfc(\alpha R_c)}{R_c} + \left( \frac{erfc(\alpha R_c)}{R_c^2} + \frac{2\alpha}{\sqrt{\pi}} \frac{e^{-\alpha^2 R_c^2}}{R_c} \right) (d_{ij} - R_c) \right) \quad \forall d_{ij} \leq R_c \quad \textbf{(S10)}$$

Since the Coulomb potential is monotonic in nature, the optimal bonding distance between ions of opposite charge would be zero in the absence of repulsive contributions. Hence, to ensure a non-zero optimal bonding distance we included a purely repulsive Born term into the potential, here described by **Equation S11**:

$$V_{Born} = \frac{B_{ij}(d_{0,ij})}{d_{ij}^n} \quad \textbf{(S11)}$$

Here, $B_{ij}(d_{0,ij})$ is a structure-dependent constant chosen to reproduce the optimal bonding distance $d_{0,ij}$ between a given pair of ions (see **Equations S13-16** below), while $n$ controls how sharp the energy minimum will be. In this work $n$ is set to 12, which we found to produce good results in homo-interfaces (i.e., one material cleaved and reassembled) simulated for testing. **Equation S12** now defines the total electrostatic + Born interatomic pairwise potential, where $B_{ij}(d_{0,ij})$ is the only system-dependent parameter to be determined:

$$V_{Total} = V_{Coulomb}^{DSF} + V_{Born} \tag{S12}$$

To optimize $B_{ij}(d_{0,ij})$ for a given interface, we can find the $B_{ij}(d_{0,ij})$ value for which the derivative of the total potential with respect to the ion-ion distance $d_{ij}$ is zero at the optimal bonding distance (i.e., $d_{0,ij}$), assuming that the two ions have opposite signs:

$$0 = \frac{d}{d(d_{ij})}(V_{Total})\big|_{d_{ij}=d_{0,ij}} = \frac{d}{d(d_{ij})}(V_{Coulomb}^{DSF} + V_{Born})\big|_{d_{ij}=d_{0,ij}} \tag{S13}$$

$$0 = \frac{d}{d(d_{ij})}\left[-|q_i|\,|q_j|\left(\frac{erfc(\alpha d_{ij})}{d_{ij}} - \frac{erfc(\alpha R_c)}{R_c} + \left(\frac{erfc(\alpha R_c)}{R_c^2} + \frac{2\alpha}{\sqrt{\pi}}\frac{e^{-\alpha^2 R_c^2}}{R_c}\right)(d_{ij} - R_c)\right) + \frac{B_{ij}(d_{0,ij})}{d_{ij}^n}\right]_{d_{ij}=d_{0,ij}} \tag{S14}$$

$$0 = \left[-|q_i||q_j|\left(-\frac{erfc(\alpha d_{ij})}{d_{ij}^2} - \frac{2\alpha}{\sqrt{\pi}}\frac{e^{-\alpha^2 d_{ij}^2}}{d_{ij}} + \frac{erfc(\alpha R_c)}{R_c^2} + \frac{2\alpha}{\sqrt{\pi}}\frac{e^{-\alpha^2 R_c^2}}{R_c}\right) - n\frac{B_{ij}(d_{0,ij})}{d_{ij}^{n+1}}\right]_{d_{ij}=d_{0,ij}} \tag{S15}$$

$$B_{ij}(d_{0,ij}) = -|q_i||q_j|\frac{d_{ij}^{n+1}}{n}\left[-\frac{erfc(\alpha d_{0,ij})}{d_{ij}^2} - \frac{2\alpha}{\sqrt{\pi}}\frac{e^{-\alpha^2 d_{0,ij}^2}}{R_c} + \frac{erfc(\alpha R_c)}{R_c^2} + \frac{2\alpha}{\sqrt{\pi}}\frac{e^{-\alpha^2 R_c^2}}{R_c}\right] \tag{S16}$$

To determine the optimal bonding distance between each pair of ions we assume that the two input bulk structures are stable, and have equilibrium lattice constants and ion coordinates. Then, we use the CrystalNN nearest-neighbor algorithm implemented in PyMatGen[25,34] to find the coordination number of each site and the minimum bonding distance (i.e., $d_{0,ij}$) between each pair of symmetrically unique sites of both bulk structures. Once $d_{0,ij}$ and the coordination numbers of the $i$ and $j$ ions have been determined, $d_{0,ij}$ can be decomposed into the individual ionic radii of the anion ($r^-$) and cation ($r^+$) using the ratio between the ionic radii tabulated by Shannon and Baloch et. al.[35,36]

Once $r^+$ or $r^-$ are assigned to each ion, the optimal bonding distance for ion pairs across the interface can be predicted simply by summing the two individual ionic radii (i.e., $d_{0,ij} = r^+ + r^-$). Hence, $B_{ij}(d_{0,ij})$ can be determined through **Equation S15**, and the pairwise potential between $i$ and $j$ is finally described as per **Equations S10-11**. Crucially, this allows to reliably predict the potential and optimal bond length also for pairs of ions that are not found in the two individual bulk structures (e.g., Cs-S for the $CsPbBr_3/Pb_4S_3Br_2$ interface models shown in **Figure 4c** of the Main Text). The workflow for decomposing the equilibrium distance $d_{0,ij}$ into individual radii is visually described in **Figure S23**.

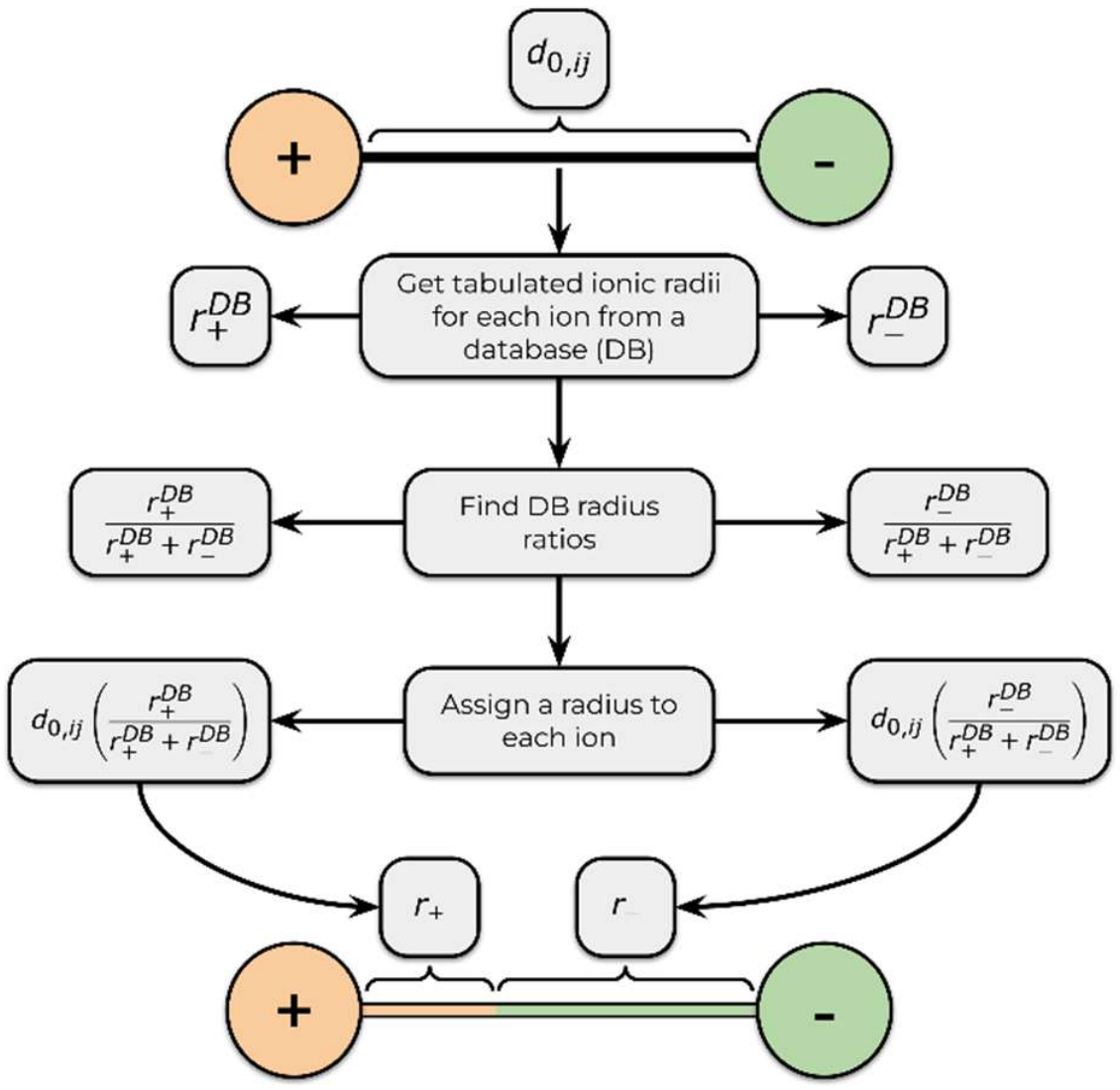

**Figure S23. Decomposing $d_{0,ij}$ into $r^+ + r^-$ for a given pair of ions.** The equilibrium distance between two ions $i$ and $j$ (i.e., $d_{0,ij}$) is extracted from the input bulk structures, together with the coordination environments of the same ions. Then, the ionic radii for all the elements and coordination environments are extracted from databases, and used to calculate the fractional contribution of each ion to the bond length $d_{0,ij}$. Based on such fractional contributions, $d_{0,ij}$ is finally decomposed as $d_{0,ij} = r^+ + r^-$, which are then used to estimate the optimal bond lengths for ion pairs that are absent in the two input bulk structures (i.e., at the interface).

## S7. Electrostatic potential vs DFT

To compare the performances of our electrostatic potential with DFT, we repeated the *surface matching and ranking* procedure for the (100)∥(010) – $CsPbBr_3/Pb_4S_3Br_2$ interface with the SCAN+rVV10 functional, which is known to produce reliable results for metal halide perovskites.[37,38] Given the large number of atoms in the model, a full 3D-optimization of the interface geometry would have exceeding computational costs. Therefore, we trusted the epitaxial registry identified by the electrostatic potential, as the *xy*-position of the energy minimum is generally easy to identify thanks to the geometric arrangement of atoms at the interface. Conversely, the interfacial distance was re-optimized independently by DFT, and we also computed the $E_{adh}$ maps for the three top-ranking interfaces to visually compare how Ogre and DFT perform far from the energy minimum (see **Figure 5b-c** of the main text). **Table S11** provides detailed results, including the $E_{int}$ values plotted in **Figure 5a** of the Main Text.

**Table S11. (100)∥(010) – $CsPbBr_3/Pb_4S_3Br_2$, electrostatic potential *vs* DFT**. Interface models are listed as ranked by the Ogre electrostatic potential. Green indicates the most stable interface. Data referring to the Ogre potential are identical to those in **Table S2**.

| $Pb_4S_3Br_2$ frag. | $CsPbBr_3$ frag. | Charge [epi/sub] | Ogre Int. Dist.[Å] | DFT Int. Dist.[Å] | Ogre $E_{int}$ [meV Å$^{-2}$] | DFT $E_{int}$ [meV Å$^{-2}$] | DFT Ranking |
|---|---|---|---|---|---|---|---|
| 4 | A | [0/0] | 2.47 | 2.25 | 52 | 37 | 1 |
| 4 | B | [0/0] | 3.07 | 3.03 | 57 | 46 | 4 |
| 6 | A | [+2/0] | 2.97 | 3.06 | 75 | 42 | 2 |
| 2 | A | [-2/0] | 3.24 | 2.89 | 115 | 54 | 6 |
| 6 | B | [+2/0] | 3.74 | 3.34 | 118 | 56 | 10 |
| 3 | A | [-2/0] | 3.00 | 2.74 | 121 | 45 | 3 |
| 2 | B | [-2/0] | 3.80 | 3.19 | 126 | 55 | 7 |
| 5 | A | [+2/0] | 3.35 | 3.01 | 137 | 55 | 8 |
| 3 | B | [-2/0] | 3.97 | 3.34 | 138 | 52 | 5 |
| 5 | B | [+2/0] | 3.64 | 2.97 | 139 | 56 | 9 |
| 7 | A | [+4/0] | 2.55 | 2.35 | 332 | 79 | 11 |
| 7 | B | [+4/0] | 3.14 | 3.22 | 333 | 91 | 13 |
| 1 | B | [-4/0] | 3.91 | 2.95 | 347 | 82 | 12 |
| 1 | A | [-4/0] | 4.01 | 3.30 | 348 | 91 | 14 |

On average, we note that our electrostatic potential tends to overestimate the interfacial distances compared to DFT, with increasing inaccuracy as the interface models become less favorable. This is likely due to the lack of non-ionic contributions to the binding energy and to interface phenomena like charge density redistribution, which can be handled by DFT but not by a classical potential where the charge of ions is defined a-priori.

## S8. Lead sulfochloride/$CsPbBr_3$ interfaces

### S8.1 On the $Pb_4S_3Cl_2$ crystal structure

We know from HAADF-STEM images of our previous works[5,8] that the two sulfohalides $Pb_4S_3Br_2$ and $Pb_4S_3Cl_2$ share similar crystal structures, and can both match epitaxially the corresponding Cs-Pb halide perovskite. However, the bulk structure of $Pb_4S_3Cl_2$ was never reported in the literature, and the nanocrystals obtained in our previous work were too small to allow for a proper refinement of the structure by X-ray diffraction.[8]

To construct a CIF file for the simulation of interfaces involving $Pb_4S_3Cl_2$, we therefore adapted the published structure of $Pb_4S_3Br_2$ by extracting the lattice parameters directly from the HAADF-STEM images of a $CsPbCl_3/Pb_4S_3Cl_2$ heterostructure, via Fourier analysis of the lattice periodicity. To ensure maximum accuracy, we calibrated the image using the lattice constant of $CsPbCl_3$ as a reference (5.605 Å). This resulted in a pseudo-tetragonal unit cell for $Pb_4S_3Cl_2$, with estimated lattice parameters $a = c = 7.941$ Å and $b = 14.943$ Å. Finally, the coordinates of atoms inside the unit cell were optimized by DFT, performed using the SCAN+rVV10 functional and while keeping the lattice parameters fixed to the estimated values. **Figure S24** summarizes the process.

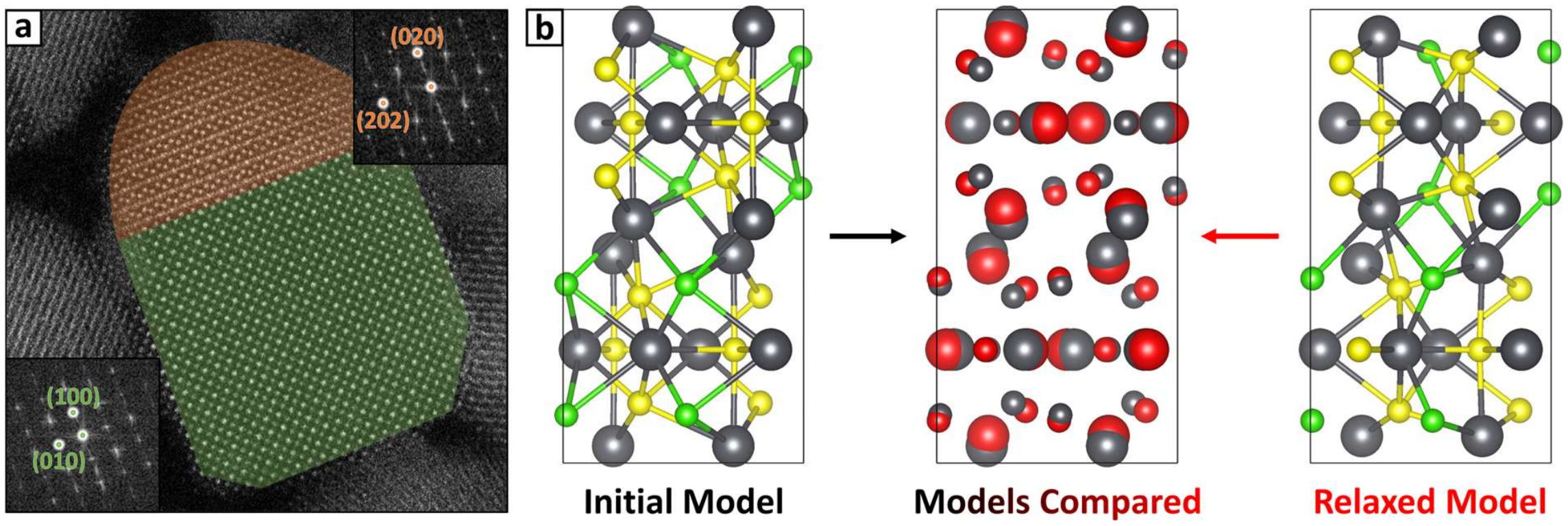

**Figure S24. Obtaining a structural model for $Pb_4S_3Cl_2$.** a)The unit cell parameters were extracted via the Fourier transform of and HAADF-STEM image of a $Pb_4S_3Cl_2/CsPbCl_3$ heterostructure. The $CsPbCl_3$ domain is highlighted in red, that of in $Pb_4S_3Cl_2$ orange. b) The unit cell was then populated according to the structure reported for $Pb_4S_3Br_2$ (left), followed by a DFT relaxation of the atomic coordinates (center). The process resulted in a mild shift of ions from the starting position (middle). HAADF-STEM image reproduced with permission, Copyright 2022, the Authors.[8]

## S8.1 Lead sulfochloride/$CsPbBr_3$ lattice matching

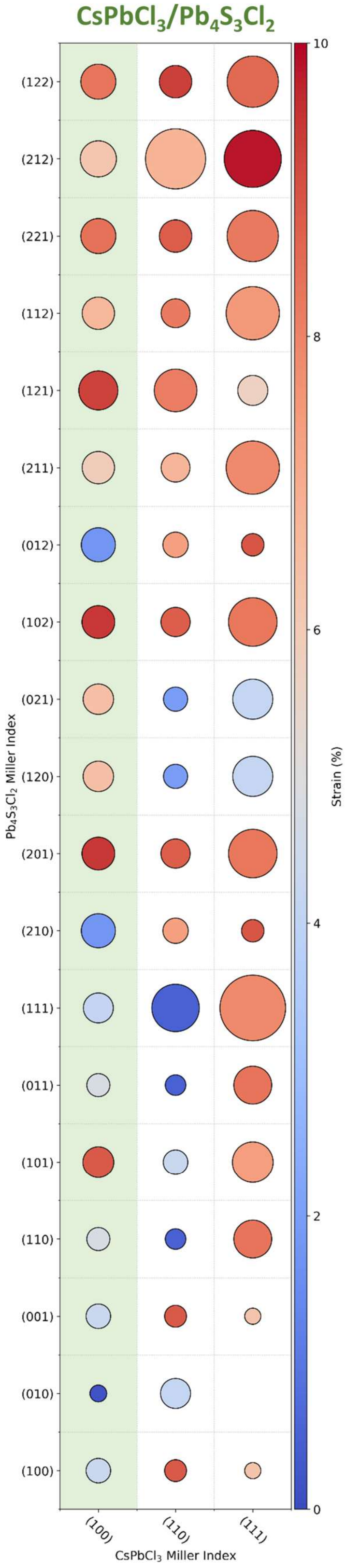

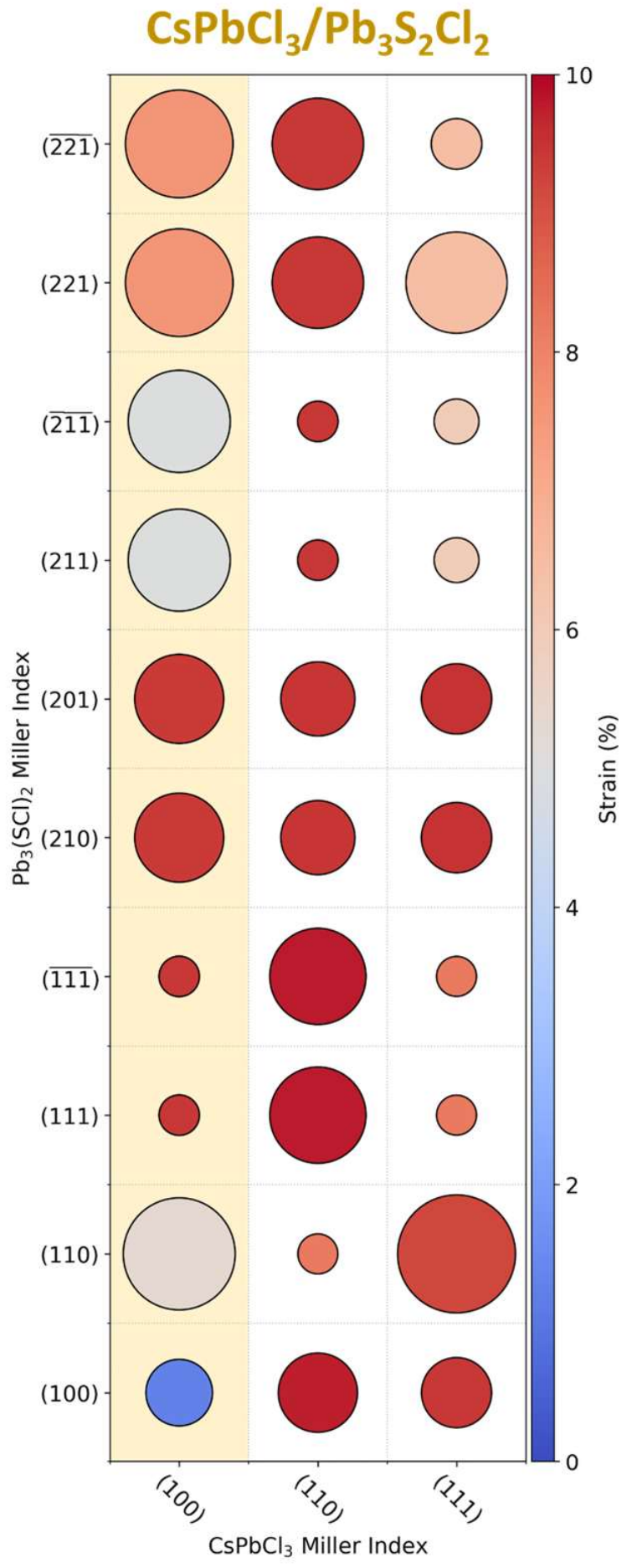

**Figure S25. Lattice matching results for $CsPbCl_3/Pb_4S_3Cl_2$ and $CsPbCl_3/Pb_3S_2Cl_2$**. The two colored columns contain the interfaces that have been optimized and ranked (see **Table S12**).

**Table S12. Summary of all interfaces between (100) $CsPbCl_3$ and lead sulfohalides.** Green indicates interfaces with $Pb_4S_3Cl_2$, yellow those with $Pb_3S_2Cl_2$. Darker colors identify the most stable model for both structures. The $E_{int}$ value in parentheses refer to the experimental interface, which is the second most stable model for the (100)(010) – $CsPbCl_3$∥$Pb_4S_3Cl_2$ interface. Reference structures: CSD-2181723 ($Pb_3S_2Cl_2$, transformed to pseudocubic setting) ; ICSD-201250 ($CsPbCl_3$). For $Pb_4S_3Cl_2$ see **Section S7.1**.

| $Pb_xS_yCl_z$ sulfohalide | $Pb_xS_yCl_z$ (hkl) | $Pb_xS_yCl_z$ slab charge | $CsPbCl_3$ slab charge | Strain [%] | Area [Å$^2$] | Interfacial dist. [Å] | $E_{int}$ [meV Å$^{-2}$] |
|---|---|---|---|---|---|---|---|
| $Pb_4S_3Cl_2$ | 010 | 0 | 0 | 0.2 | 63 | 3.08 | 19 (27) |
| $Pb_4S_3Cl_2$ | 011 | 0 | 0 | 4.8 | 126 | 2.88 | 35 |
| $Pb_4S_3Cl_2$ | 100 | 0 | 0 | 4.3 | 126 | 3.09 | 57 |
| $Pb_4S_3Cl_2$ | 101 | 0 | 0 | 8.9 | 188 | 3.38 | 60 |
| $Pb_4S_3Cl_2$ | 201 | 0 | 0 | 9.4 | 251 | 2.88 | 63 |
| $Pb_4S_3Cl_2$ | 211 | -2 | 0 | 5.9 | 251 | 3.06 | 64 |
| $Pb_4S_3Cl_2$ | 212 | +1 | 0 | 6.2 | 314 | 2.97 | 67 |
| $Pb_4S_3Cl_2$ | 112 | +4 | 0 | 6.7 | 251 | 2.75 | 68 |
| $Pb_4S_3Cl_2$ | 121 | -4 | 0 | 9.3 | 377 | 2.94 | 71 |
| $Pb_4S_3Cl_2$ | 111 | +2 | 0 | 4.1 | 188 | 3.07 | 72 |
| $Pb_4S_3Cl_2$ | 221 | -2 | 0 | 8.4 | 283 | 3.06 | 74 |
| $Pb_4S_3Cl_2$ | 001 | 0 | 0 | 4.3 | 126 | 3.51 | 78 |
| $Pb_4S_3Cl_2$ | 021 | 0 | 0 | 6.4 | 188 | 3.04 | 80 |
| $Pb_4S_3Cl_2$ | 122 | 0 | 0 | 8.3 | 283 | 3.26 | 85 |
| $Pb_4S_3Cl_2$ | 110 | +2 | 0 | 4.8 | 126 | 2.69 | 86 |
| $Pb_4S_3Cl_2$ | 012 | +2 | 0 | 1.8 | 251 | 2.96 | 86 |
| $Pb_4S_3Cl_2$ | 120 | 0 | 0 | 6.4 | 188 | 3.69 | 92 |
| $Pb_4S_3Cl_2$ | 102 | +2 | 0 | 9.4 | 251 | 3.57 | 97 |
| $Pb_4S_3Cl_2$ | 210 | +4 | 0 | 1.8 | 251 | 3.24 | 99 |
| $Pb_3S_2Cl_2$ | 201 | -1 | 0 | 9.4 | 314 | 3.31 | 50 |
| $Pb_3S_2Cl_2$ | -2-1-1 | +2 | 0 | 5.0 | 377 | 3.96 | 53 |
| $Pb_3S_2Cl_2$ | 210 | -1 | 0 | 9.4 | 314 | 3.58 | 61 |
| $Pb_3S_2Cl_2$ | 100 | +1 | 0 | 1.3 | 157 | 3.29 | 65 |
| $Pb_3S_2Cl_2$ | 211 | +2 | 0 | 5.0 | 377 | 3.48 | 67 |
| $Pb_3S_2Cl_2$ | -2-2-1 | 0 | 0 | 7.6 | 440 | 3.86 | 75 |
| $Pb_3S_2Cl_2$ | 111 | 0 | 0 | 9.5 | 126 | 3.28 | 88 |
| $Pb_3S_2Cl_2$ | 221 | -1 | 0 | 7.6 | 440 | 4.04 | 89 |
| $Pb_3S_2Cl_2$ | -1-1-1 | 0 | 0 | 9.5 | 126 | 3.47 | 93 |

## S9. $Bi_xPb_yS_z/CsPbBr_3$ interfaces

Considering *h,k,l* ≤ *2*, there are 105 possible non-equivalent epitaxial relations between the (100) surface of $CsPbBr_3$ and the $Bi_xPb_yS_z$ phases considered in this work (19 for heyrovskyite, 19 for lillianite, 29 for xilingolite, 19 for cosalite, and 19 for galenobismuthite). Among them, we identified 10 supercells (**Figure S26**) that fulfilled all criteria discussed in the Main Text. **Table S13** summarizes the outcome of *surface matching and ranking*, while **Figure S26** shows the best model for the three top-ranking interfaces.

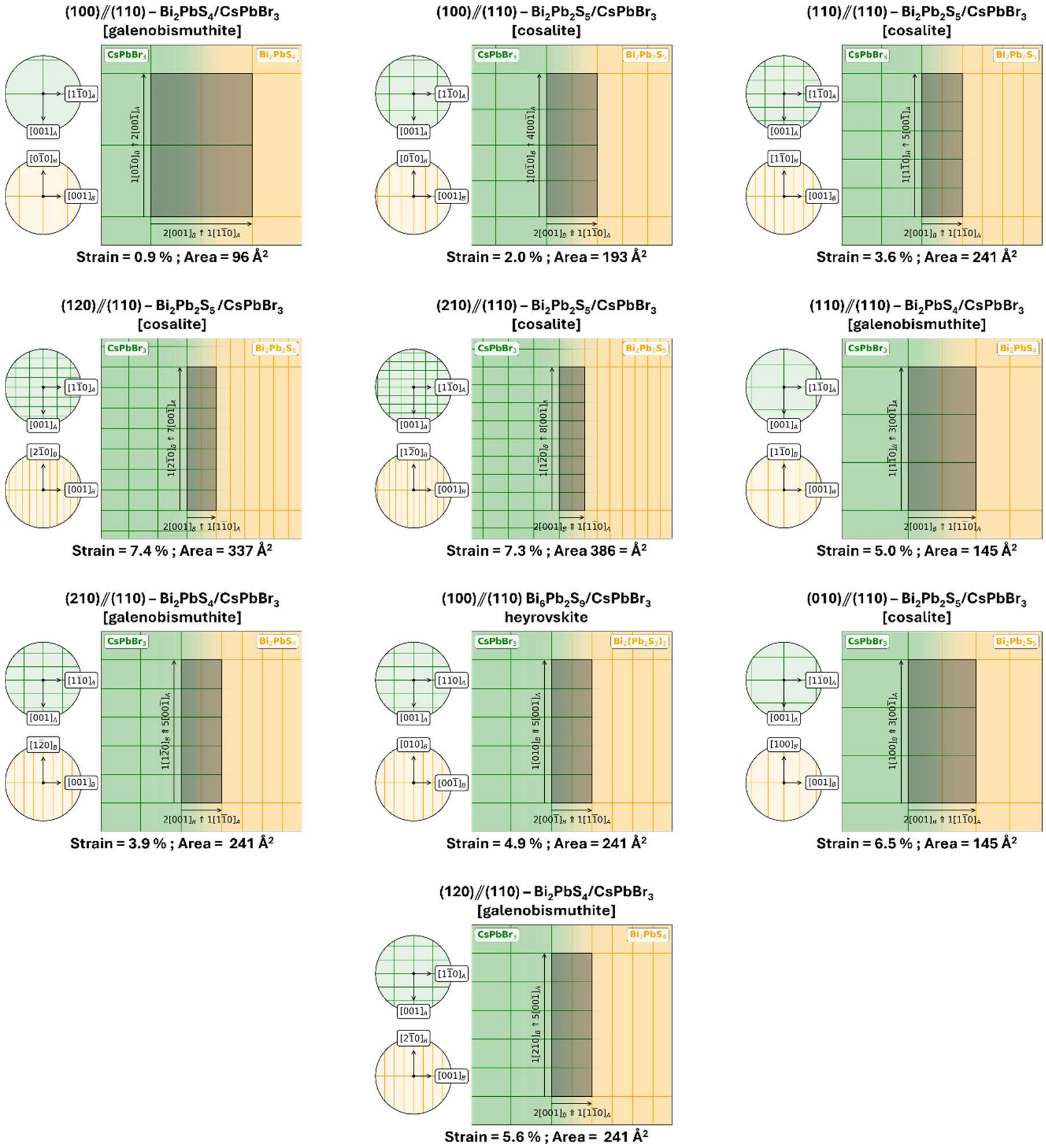

**Figure S26. 2D-supercells of the $Bi_xPb_yS_z/CsPbBr_3$ epitaxial relations considered.** All supercells share the same relative orientation between the lattices of $CsPbBr_3$ and $Bi_xPb_yS_z$, as shown by the circular dials on the left of each panel. The order of panels follows the stability of the related interfaces (see **Table S13**).

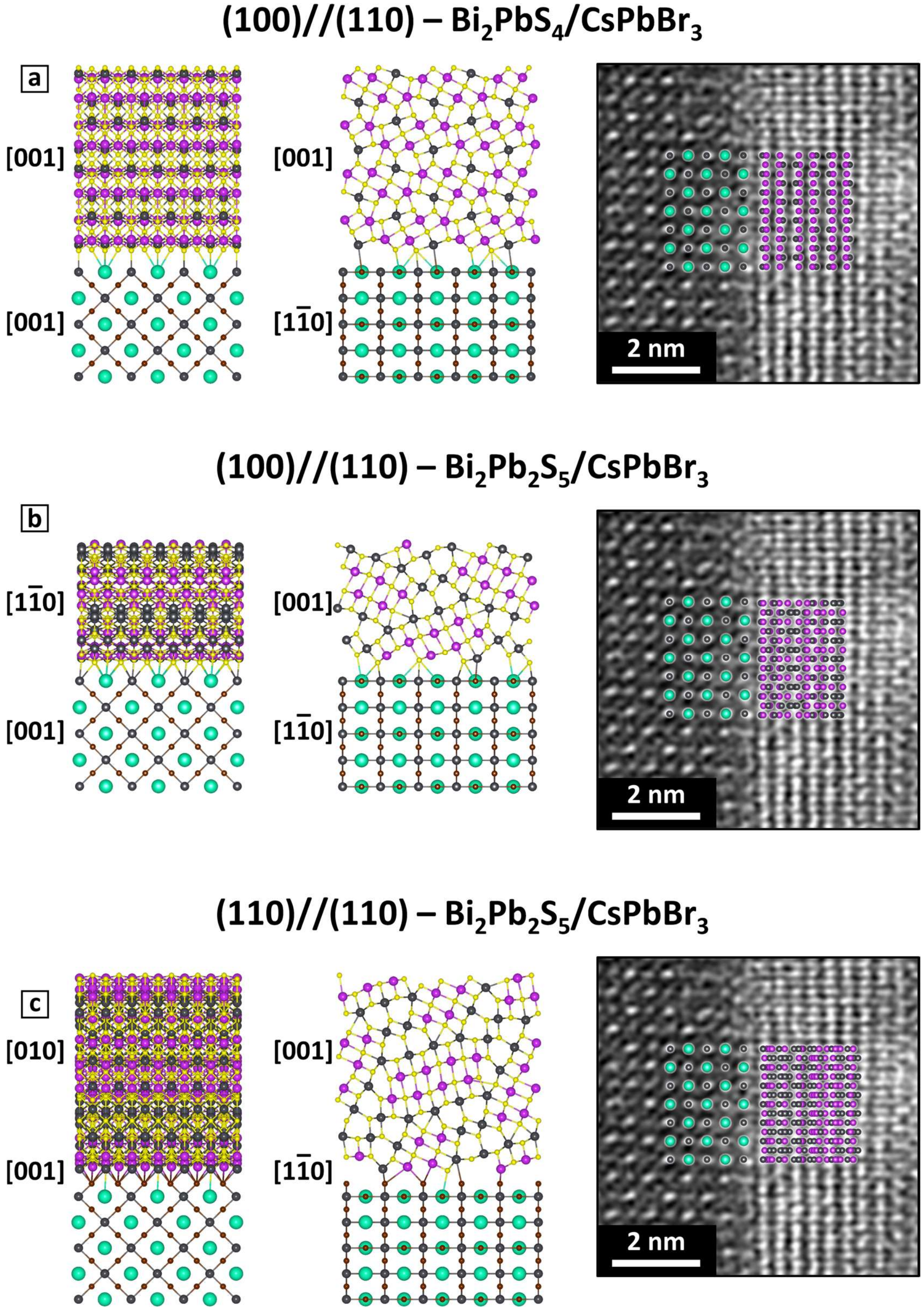

**Figure S27. Best $Bi_xPb_yS_z/CsPbBr_3$ interfaces by $E_{int}$.** From top to bottom: a) (100)⫽(110) – $Bi_2PbS_4/CsPbBr_3$ galenobismuthite. b) (100)⫽(110) – $Bi_2Pb_2S_5/CsPbBr_3$ cosalite. c) (110)⫽(110) – $Bi_2Pb_2S_5/CsPbBr_3$ cosalite. From left to right: Left) view of the interface along

the [001] $CsPbBr_3$ zone axis, as seen in the atomic-resolution TEM images. Center) Side-view of the interface (90° rotation), highlighting the partial formation of bonds between slabs. Right) Interface model overlaid to the atomic-resolution TEM image of a heterostructure. Only the heavy atoms (Pb and Bi) are shown to provide a better comparison with electron scattering contrast. Microscopy images reproduced with permission.[9] Copyright 2023, the Authors.

**Table S13. $Bi_xPb_yS_z$/$CsPbBr_3$ interfaces.** *Structure optimization* parameters for the interfaces identified between $CsPbBr_3$ and one of the $Bi_xPb_yS_z$ phases, based on the geometric and structural considerations illustrated in **Figure 7b** of the Main Text. The three best models are depicted and overlaid to atomic-resolution images of the heterostructure in **Figure S27**. Reference structures: ICSD-43657 (cosalite); ICSD-604473 (galenobismuthite); ICSD-60160 (heyrovskyite); ICSD-2737 (lillianite); ICSD-92981 (xilingolite).

| $Bi_xPb_yS_z$ material | $Bi_xPb_yS_z$ (hkl) | $CsPbBr_3$ (hkl) | BixPbySz slab charge | $CsPbCl_3$ slab charge | Strain [%] | Area [Å$^2$] | Int. dist. [Å] | $E_{int}$ [meV Å$^{-2}$] |
|---|---|---|---|---|---|---|---|---|
| Galenobismuthite $Bi_2PbS_4$ | 100 | 110 | 0 | +1 | 0.9 | 96 | 3.34 | 46 |
| Cosalite $Bi_2Pb_2S_5$ | 100 | 110 | -1 | +1 | 2.0 | 193 | 2.87 | 49 |
| Cosalite $Bi_2Pb_2S_5$ | 110 | 110 | +3 | -1 | 3.6 | 241 | 2.01 | 51 |
| Cosalite $Bi_2Pb_2S_5$ | 120 | 110 | +3 | -1 | 7.4 | 337 | 2.17 | 87 |
| Cosalite $Bi_2Pb_2S_5$ | 210 | 110 | -1 | +1 | 7.3 | 386 | 3.11 | 117 |
| Galenobismuthite $Bi_2PbS_4$ | 110 | 110 | +2 | -1 | 5.0 | 145 | 2.44 | 121 |
| Galenobismuthite $Bi_2PbS_4$ | 210 | 110 | -2 | +1 | 3.9 | 241 | 2.69 | 131 |
| Heyrovskyite A $Bi_6Pb_2S_9$ | 100 | 110 | +6 | -1 | 4.9 | 241 | 2.01 | 146 |
| Cosalite $Bi_2Pb_2S_5$ | 010 | 110 | +3 | -1 | 6.5 | 145 | 1.92 | 151 |
| Galenobismuthite $Bi_2PbS_4$ | 120 | 110 | +3 | -1 | 5.6 | 241 | 1.75 | 214 |
| Heyrovskyite B $Bi_6Pb_2S_9$ | 100 | 110 | -2 | +1 | 4.9 | 241 | 3.05 | 247 |

**Notes:** due to standards in the assignment of crystallographic axes, not all the bulk structures for $Bi_xPb_yS_z$ materials have their preferred growth axis labelled as *c*. However, to simplify the discussion we converted all structures into non-standard settings where the growth axis is *c*. We also note that the bulk structure of heyrovskyite features atomic sites with partial occupancy, which is not currently implemented in our software. To circumvent this issue, we constructed two stoichiometric bulk models with no occupational disorder (here denoted heyrovskyite A

and heyrovskyite B), and optimized both. Reference CIFs converted in the non-standard space groups and without occupational disorder are provided in the Supplementary Material.

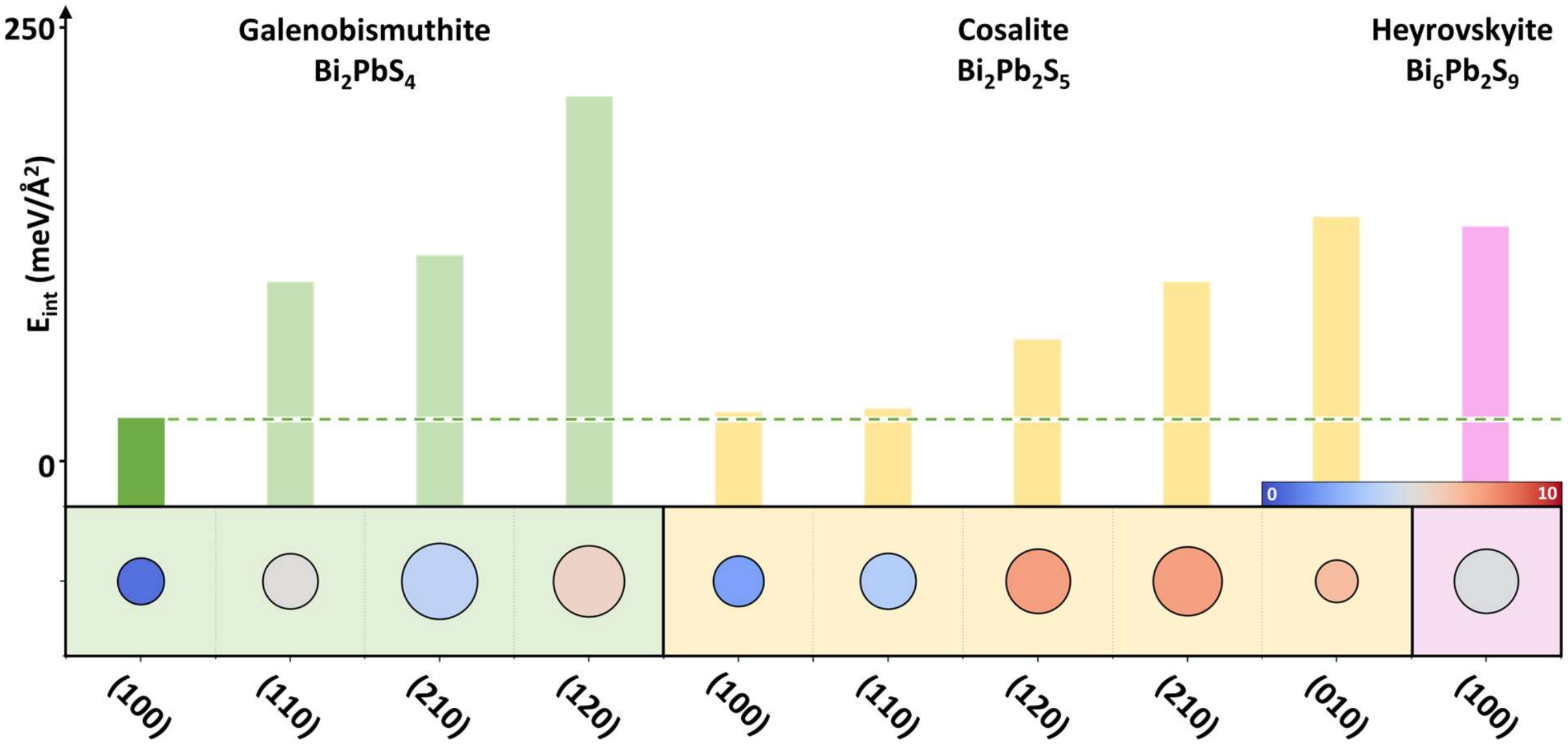

**Figure S28. $Bi_xPb_yS_z/CsPbBr_3$ interfaces.** *Lattice matching* and interface ranking results for the growth of galenobismuthite (green), cosalite (yellow) and heyrovskyite (purple) on the (110) surface of $CsPbBr_3$, summarized in **Table S13.** The circle size corresponds to the interface area and the color corresponds to the strain. The green dashed line marks the $E_{int}$ value of the experimental interface (100)∥(110) – $Bi_2PbS_4/CsPbBr_3$.

## S10. Other interfaces with $CsPbBr_3$

### S10.1. $CsPbBr_3/CsPb_2Br_5$

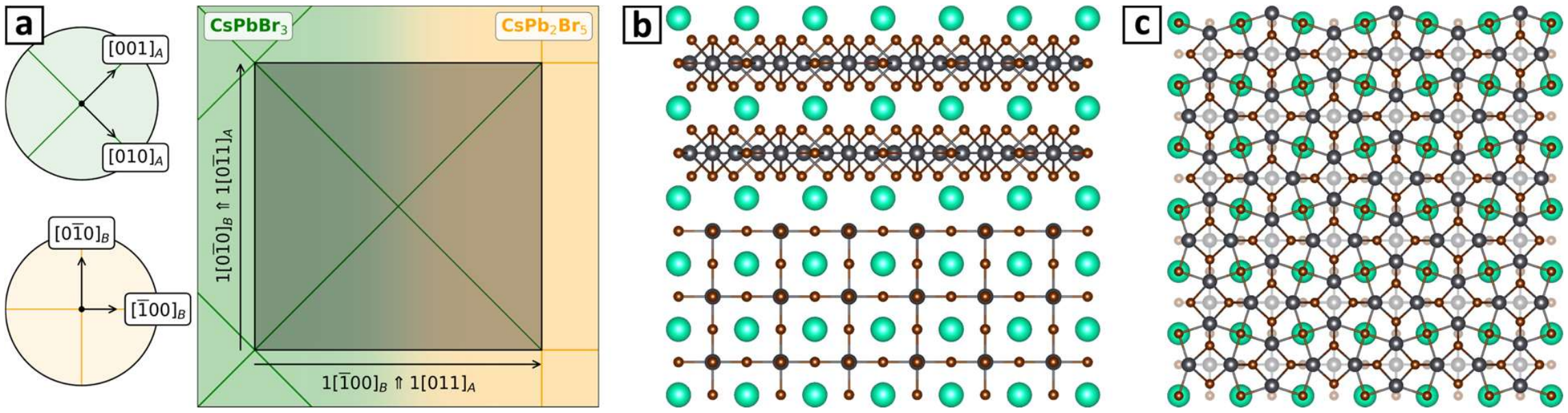

**Figure S29. (100)∥(001) – $CsPbBr_3/Pb_4S_3Br_2$ interface.** a) 2D-supercell of the interface (strain = 2.7%, area = 68 Å$^2$). b) Side view of the most stable model as ranked by Ogre (green in **Table S14**). c) Top-down view of the interface layer. The shaded atoms belong to $CsPbBr_3$. Reference structure: ICSD- 254290 ($CsPb_2Br_5$).

**Table S14. *Surface ranking* results**. Green indicates the most stable interface.

| $CsPb_2Br_5$ slab index | $CsPbBr_3$ slab index | Interfacial dist. [Å] | $CsPb_2Br_5$ slab charge | $CsPbBr_3$ slab charge | $E_{int}$ [meV Å$^{-2}$] |
|---|---|---|---|---|---|
| 1 | 0 | 2.87 | +1 | 0 | 10 |
| 1 | 1 | 3.88 | +1 | 0 | 15 |
| 2 | 1 | 4.31 | -1 | 0 | 19 |
| 2 | 0 | 4.31 | -1 | 0 | 20 |
| 3 | 0 | 3.28 | +3 | 0 | 133 |
| 3 | 1 | 4.07 | +3 | 0 | 155 |
| 0 | 1 | 4.02 | -3 | 0 | 160 |
| 0 | 0 | 4.31 | -3 | 0 | 161 |

## S10.2. $CsPbBr_3$/ZnS

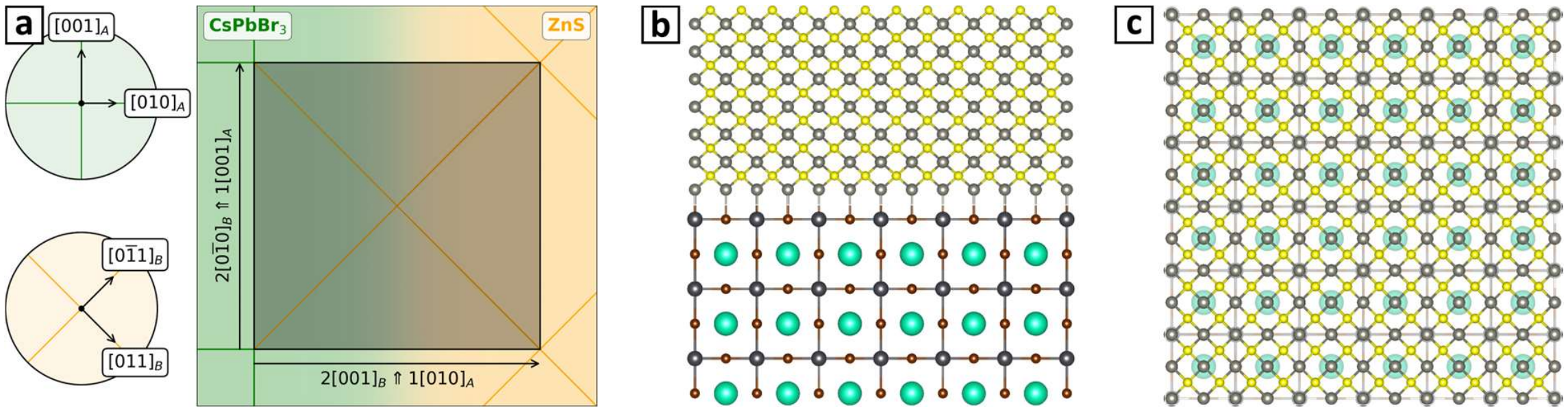

**Figure S30. (100)∥(100) – $CsPbBr_3$/ZnS interface.** a) 2D-supercell of the interface (strain = 8.0%, area = 34 Å$^2$). b) Side view of the most stable model as ranked by Ogre (green in **Table S15**). c) Top-down view of the interface layer. The shaded atoms belong to $CsPbBr_3$. Reference structure: ICSD- 254290 (ZnS sphalerite).

**Table S15.** ***Surface ranking* results**. Green indicates the most stable interface.

| ZnS slab index | $CsPbBr_3$ slab index | Interfacial dist. [Å] | ZnS slab charge | $CsPbBr_3$ slab charge | $E_{int}$ [meV Å$^{-2}$] |
|---|---|---|---|---|---|
| 0 | 0 | 2.68 | 1 | 0 | 412 |
| 1 | 0 | 3.89 | -1 | 0 | 439 |
| 0 | 1 | 4.22 | 1 | 0 | 441 |
| 1 | 1 | 4.22 | -1 | 0 | 444 |

## S10.3. $CsPbBr_3/Al_2O_3$

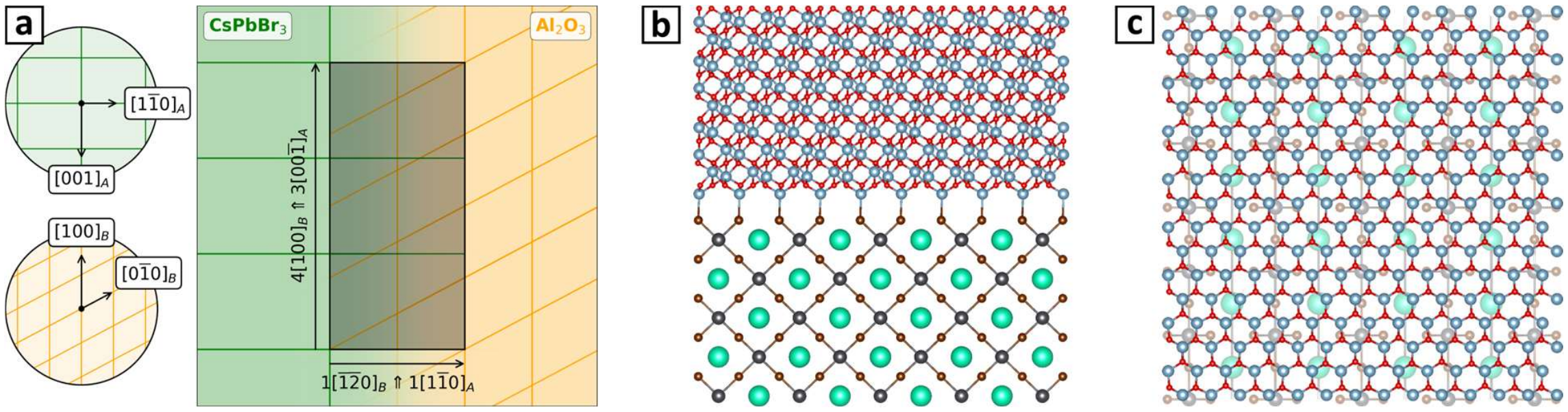

**Figure S31. (110)⫽(001) – $CsPbBr_3/Al_2O_3$ interface.** a) 2D-supercell of the interface (strain = 6.0%, area = 145 Å$^2$). b) Side view of the most stable model as ranked by Ogre (green in **Table S16**). c) Top-down view of the interface layer. The shaded atoms belong to $CsPbBr_3$. Reference structure: ICSD-111371 ($Al_2O_3$ sapphire).

**Table S16. *Surface ranking* results**. Green indicates the most stable interface.

| $Al_2O_3$ slab index | $CsPbBr_3$ slab index | Interfacial dist. [Å] | $Al_2O_3$ slab charge | $CsPbBr_3$ slab charge | $E_{int}$ [meV Å$^{-2}$] |
|---|---|---|---|---|---|
| 0 | 1 | 2.69 | 0 | -1 | 263 |
| 12 | 1 | 2.69 | 0 | -1 | 263 |
| 0 | 0 | 4.22 | 0 | +1 | 367 |
| 12 | 0 | 4.22 | 0 | +1 | 367 |
| 2 | 1 | 1.81 | +3 | -1 | 1807 |
| 1 | 0 | 3.02 | -3 | +1 | 2018 |

## S8.4. $CsPbBr_3/Bi_2WO_6$

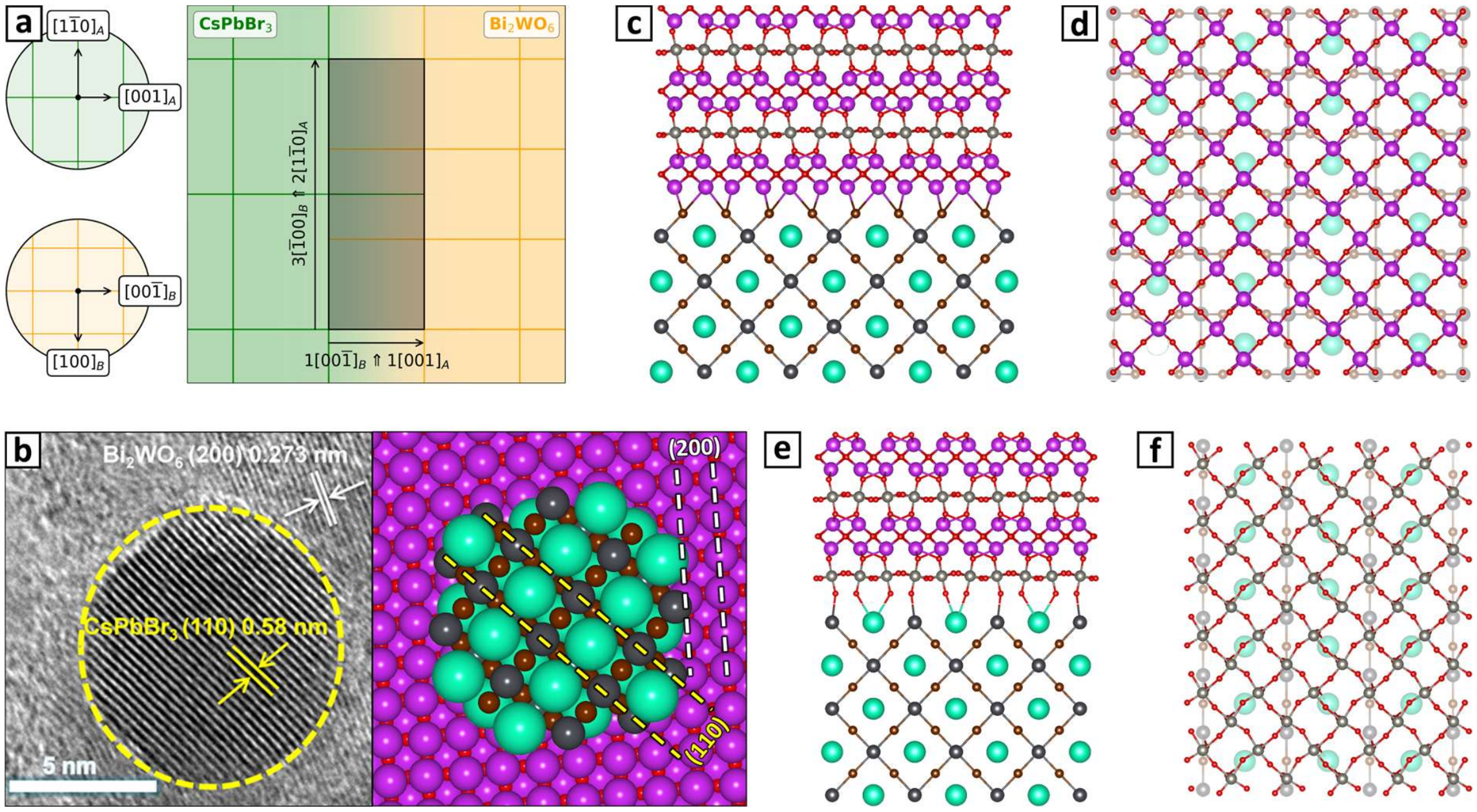

**Figure S32. (110)⫽(010) – $CsPbBr_3/Bi_2WO_6$ interface.** a) 2D-supercell of the interface (strain =5.0%, area = 96 Å$^2$). b) TEM image of a $CsPbBr_3/Bi_2WO_6$ heterostructure side-by-side with the proposed model. $CsPbBr_3$ is indexed in the *Pnma* setting, following the original publication. TEM reproduced with permission.[13] Copyright 2020, American Chemical Society. Side view of the 1$^{st}$-ranking model (green in **Table S17**). Its structure is similar to the 2$^{nd}$ ranking model, where $Bi_2WO_6$ is also bismuth-terminated (not shown). d) Top-down view of the same interface. e) Side view of the 3$^{rd}$-ranking model by Ogre (blue in **Table S17**). d) Top-down view of the interface layer. Shaded atoms belong to $CsPbBr_3$. Reference structure: ICSD- 67647 ($Bi_2WO_6$).

**Table S17. *Surface ranking* results**. Green = **Figure 32c-d** . Blue = **Figure 32b,e-f**.

| $Bi_2WO_6$ slab index | $CsPbBr_3$ slab index | Interfacial dist. [Å] | $Bi_2WO_6$ slab charge | $CsPbBr_3$ slab charge | $E_{int}$ [meV Å$^{-2}$] |
|---|---|---|---|---|---|
| 6 | 1 | 2.51 | +2 | -1 | 248 |
| 14 | 1 | 2.47 | +2 | -1 | 257 |
| 9 | 0 | 2.86 | -2 | +1 | 293 |
| 1 | 0 | 2.85 | -2 | +1 | 298 |
| 10 | 1 | 2.71 | +2 | -1 | 466 |
| 2 | 1 | 2.72 | +2 | -1 | 471 |
| 13 | 0 | 2.89 | -2 | +1 | 519 |
| 5 | 0 | 2.92 | -2 | +1 | 531 |
| 15 | 0 | 2.77 | -4 | +1 | 1383 |
| 7 | 0 | 2.71 | -4 | +1 | 1395 |
| 0 | 1 | 2.30 | +4 | -1 | 1407 |
| 8 | 1 | 2.34 | +4 | -1 | 1431 |
| 3 | 1 | 2.30 | +6 | -1 | 3546 |
| 11 | 1 | 2.34 | +6 | -1 | 3565 |
| 12 | 0 | 2.55 | -6 | +1 | 3818 |
| 4 | 0 | 2.55 | -6 | +1 | 3842 |

## S11. Validation on oxide interfaces

### S11.1. $ZnO/Zn_2GeO_4$ interface.

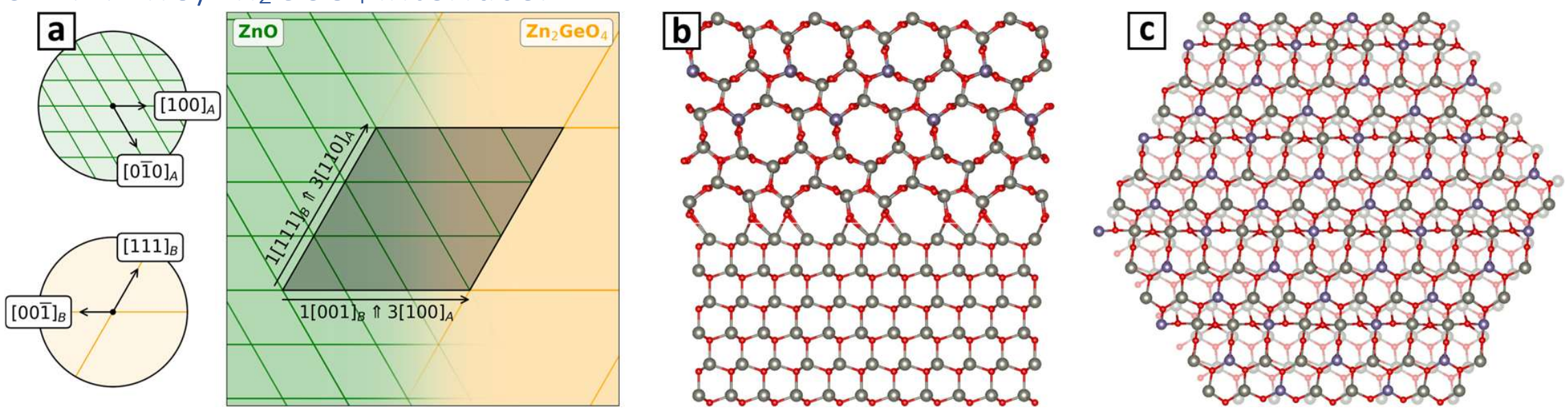

**Figure S33. (001)∥(-110) – $ZnO/Zn_2GeO_4$ interface.** a) 2D-supercell of the interface (strain = 11.9%, area = 82 Å$^2$). b) Side view of the most stable model as ranked by Ogre (green in **Table S18**). c) Top-down view of the interface layer. Shaded atoms belong to ZnO. Reference structures: ICSD-26170 (**ZnO**) ; ICSD-68382 (**$Zn_2GeO_4$**).

**Table S18.** ***Surface ranking*** **results**. Green indicates the most stable interface.

| $Zn_2GeO_4$ slab index | ZnO slab index | Interfacial dist. [Å] | $Zn_2GeO_4$ slab charge | ZnO slab charge | $E_{int}$ [meV Å$^{-2}$] |
|---|---|---|---|---|---|
| 8 | 0 | 1.70 | -6 | 0 | 333 |
| 6 | 1 | 0.83 | 10 | -2 | 518 |
| 4 | 0 | 1.74 | 0 | 0 | 528 |
| 1 | 0 | 1.69 | -2 | 0 | 570 |
| 5 | 0 | 1.96 | -8 | 0 | 577 |
| 3 | 1 | 0.73 | 8 | -2 | 817 |
| 2 | 0 | 1.70 | -10 | 0 | 973 |
| 0 | 1 | 1.41 | 6 | -2 | 1198 |
| 7 | 1 | 1.50 | 2 | -2 | 2087 |
| 4 | 1 | 2.17 | 0 | -2 | 2790 |

## S11.2. $LaAlO_3/TiO_2$ (anatase) interface.

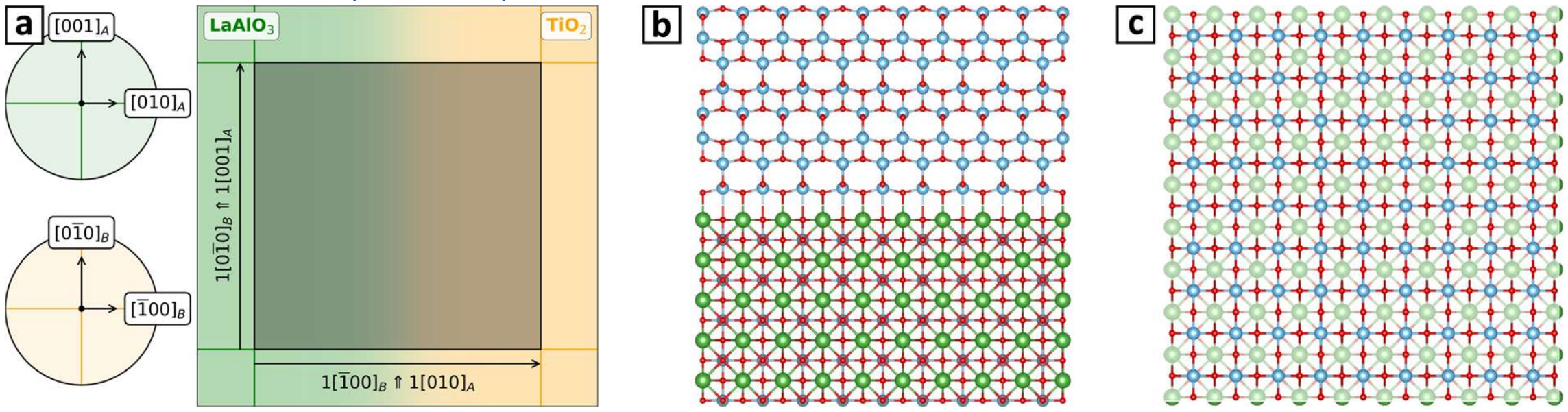

**Figure S34. (100)⫽(001) – $LaAlO_3/TiO_2$ (anatase) interface.** a) 2D-supercell of the interface (strain = 0.2%, area = 14 Å$^2$). b) Side view of the most stable model as ranked by Ogre (green in **Table S19**). c) Top-down view of the interface layer. Shaded atoms belong to $LaAlO_3$. Reference structures: ICSD-170772 ($LaAlO_3$) ; ICSD-9852 ($TiO_2$ anatase).

**Table S19.** ***Surface ranking*** **results**. Green indicates the most stable interface.

| $TiO_2$ slab index | $LaAlO_3$ slab index | Interfacial dist. [Å] | $TiO_2$ slab charge | $LaAlO_3$ slab charge | $E_{int}$ [meV Å$^{-2}$] |
|---|---|---|---|---|---|
| 1 | 0 | 2.57 | 0 | +0.5 | 102 |
| 1 | 1 | 2.58 | 0 | -0.5 | 167 |
| 2 | 1 | 2.09 | +2 | -0.5 | 1002 |
| 0 | 0 | 2.46 | -2 | +0.5 | 1369 |

## S11.4. $LaAlO_3$/ZnO interface.

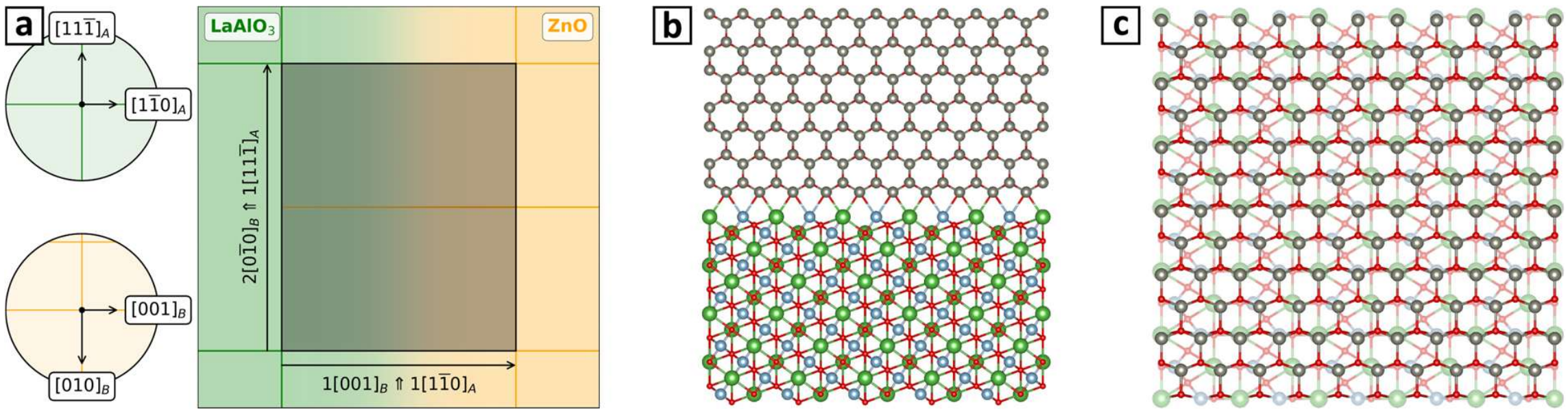

**Figure S35. (112)⫽(100) – $LaAlO_3$/ZnO interface.** a) 2D-supercell of the interface (strain = 2.2%, area = 35 Å$^2$). b) Side view of the most stable model as ranked by Ogre (green in **Table S20**). c) Top-down view of the interface layer. Shaded atoms belong to $LaAlO_3$. Reference structures: ICSD-170772 ($LaAlO_3$) ; ICSD-26170 (ZnO).

**Table S20.** ***Geometry optimization*** **results**. Green indicates the most stable interface.

| ZnO slab index | $LaAlO_3$ slab index | Interfacial dist. [Å] | ZnO slab charge | $LaAlO_3$ slab charge | $E_{int}$ [meV Å$^{-2}$] |
|---|---|---|---|---|---|
| 0 | 0 | 2.43 | 0 | 2 | 462 |
| 0 | 1 | 2.39 | 0 | -2 | 544 |

## S11.5. $Fe_3O_4/SrTiO_3$ interface.

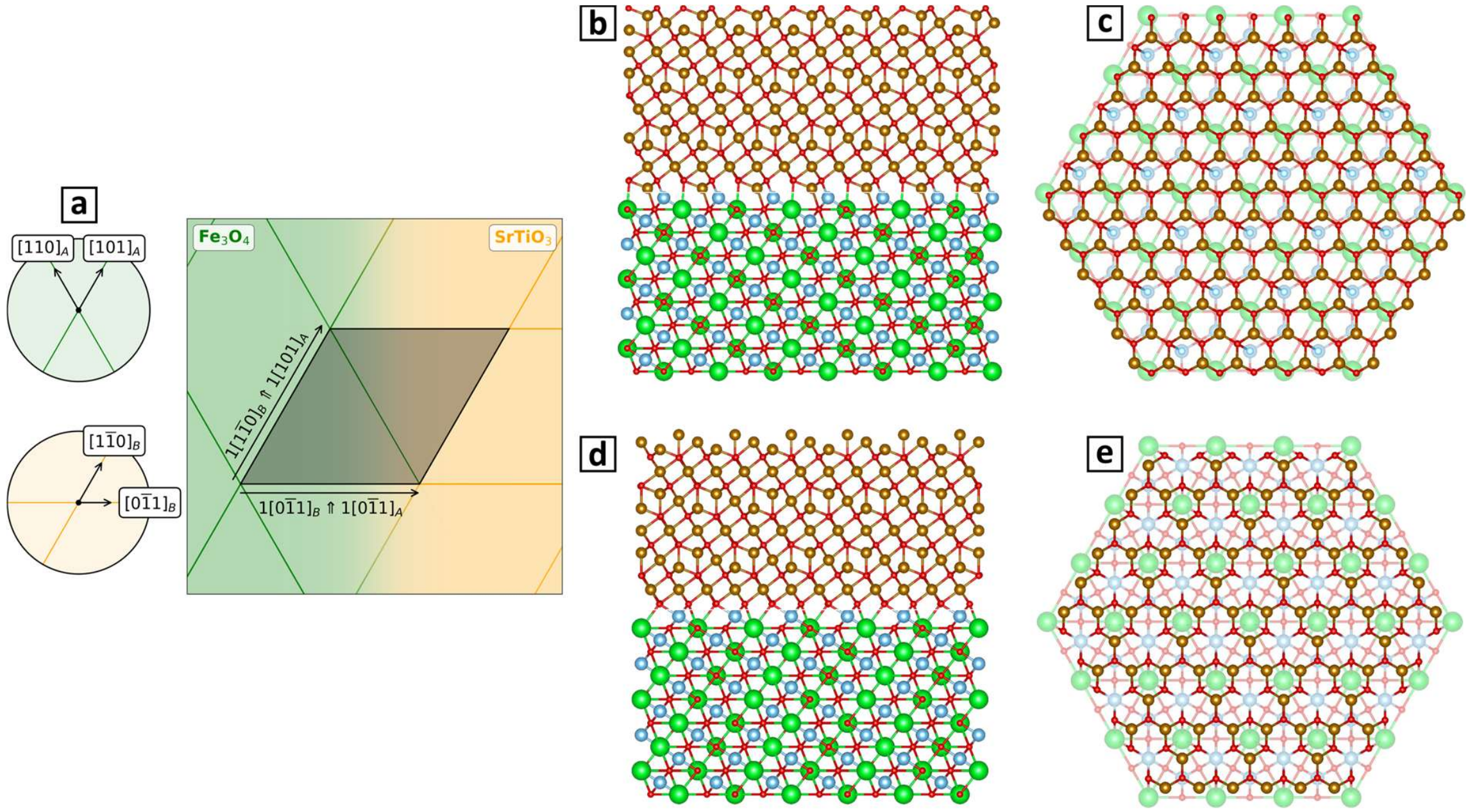

**Figure S36. (-111)∥(111) – $Fe_3O_4/SrTiO_3$ interface.** a) 2D-supercell of the interface (strain = 7.5%, area = 31 $Å^2$). b) Side view of the 1st-ranking model (green in **Table S21**). c) Top-down view of the same interface. d) Side view of the 2rd-ranking model by Ogre (blue in **Table S21**). d) Top-down view of the interface layer. Shaded atoms belong to $SrTiO_3$. Reference structures: ICSD- 26410 ($Fe_3O_4$) ; ICSD- 23076 ($SrTiO_3$).

**Table S21.** ***Geometry optimization*** **results**. Green indicates the most stable interface.

| $SrTiO_3$ slab index | $Fe_3O_4$ slab index | Interfacial dist. [Å] | $SrTiO_3$ slab charge | $Fe_3O_4$ slab charge | $E_{int}$ [meV $Å^{-2}$] |
|---|---|---|---|---|---|
| 0 | 5 | 0.84 | +2 | -1.3 | 140 |
| 0 | 4 | 1.07 | +2 | -4.0 | 407 |
| 1 | 0 | 1.92 | -2 | +1.3 | 461 |
| 0 | 2 | 1.26 | +2 | -4.0 | 522 |
| 1 | 1 | 1.63 | -2 | +4.0 | 957 |
| 1 | 3 | 2.12 | -2 | +4.0 | 1080 |

## S12. Introduction to the OgreInterface app

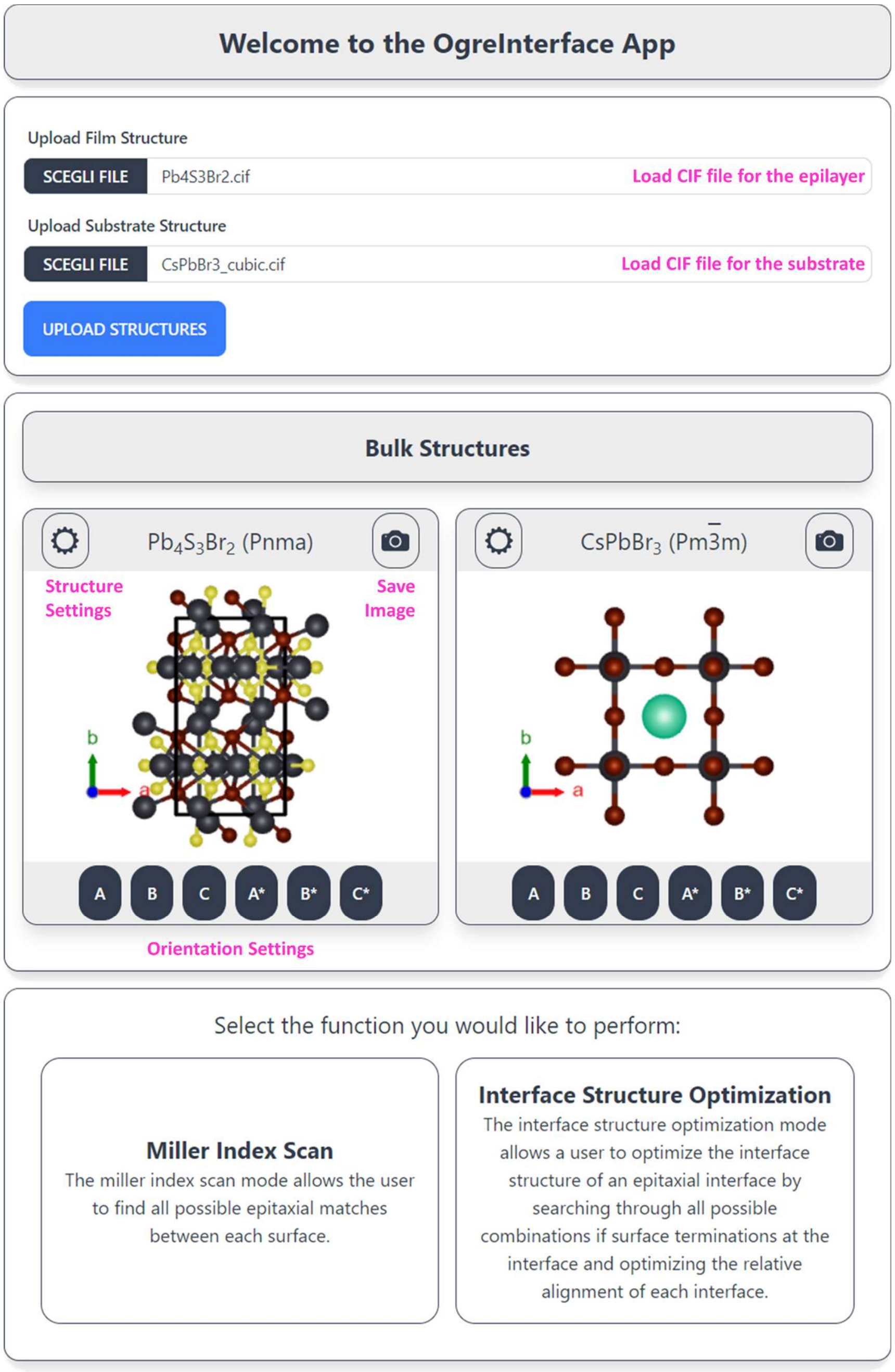

**Figure S37. Welcome and structure upload panel of the OgreInterface app.** The version shown here is the 0.0.14, and might not reflect the current version available online.

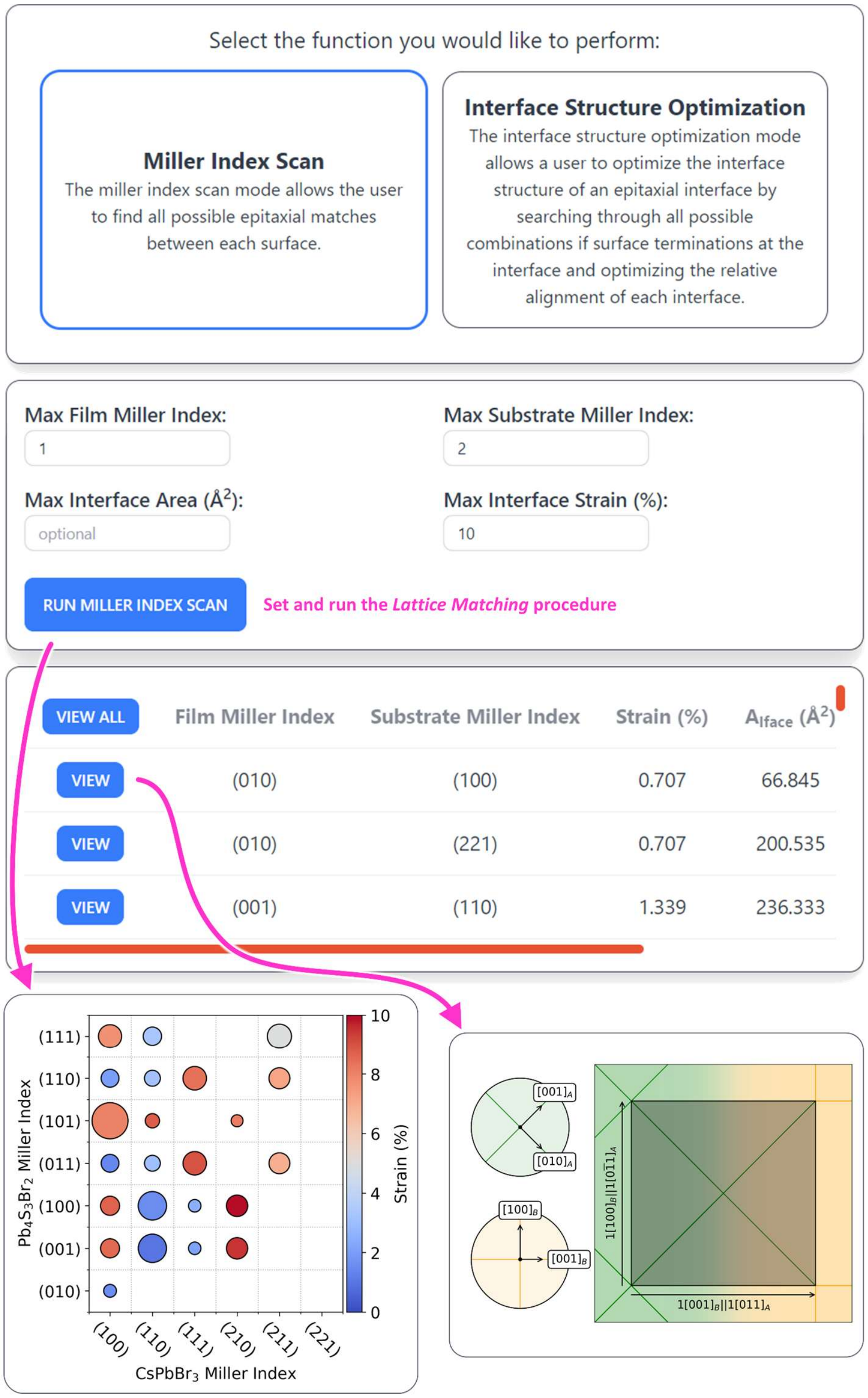

**Figure S38.** ***Lattice Matching*** **panel of the OgreInterface app.** The purple arrows indicate clickable elements of the panel and their respective outcome when selected.

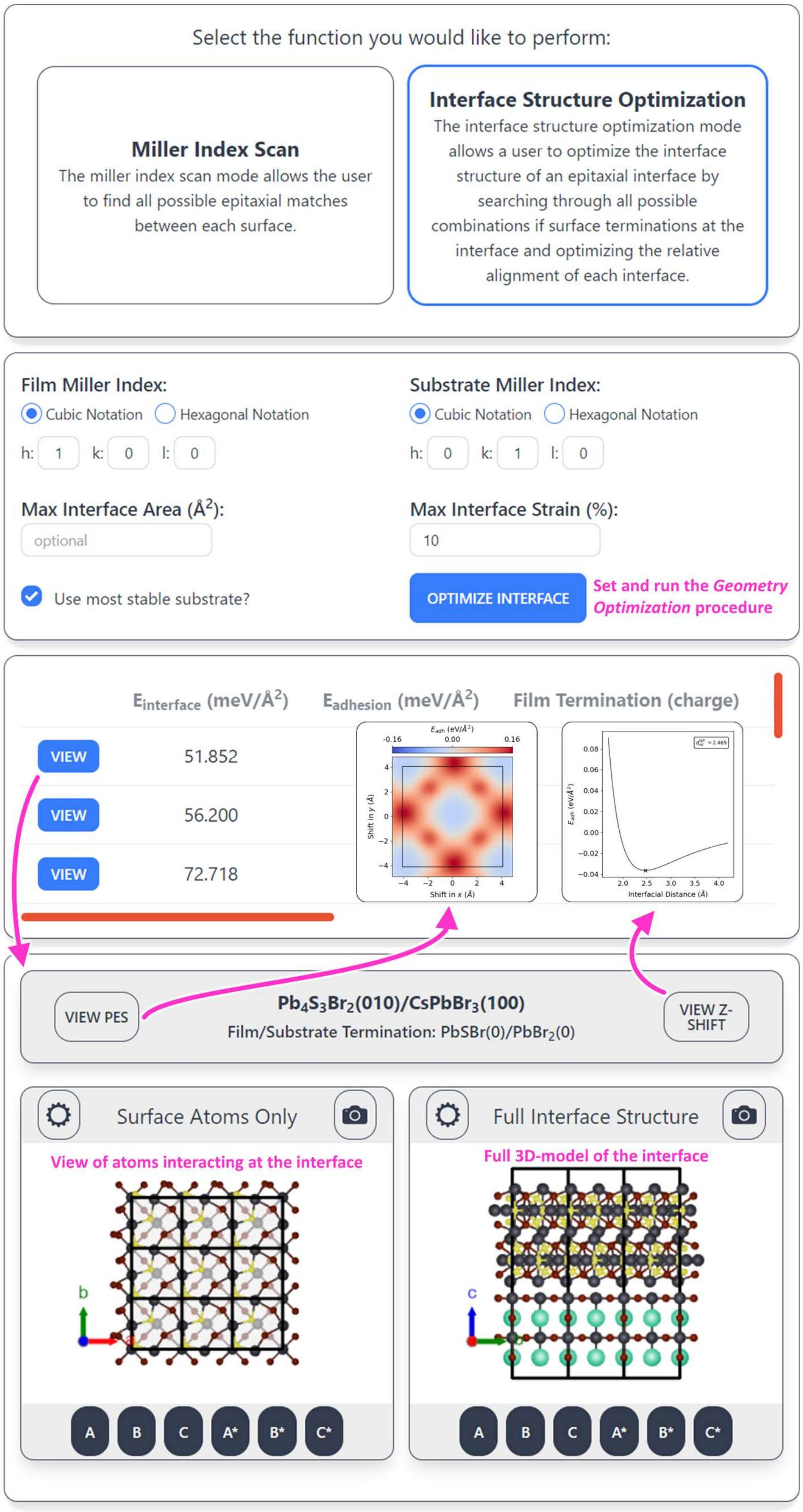
Select the function you would like to perform:
Miller Index Scan
The miller index scan mode allows the user to find all possible epitaxial matches between each surface.
Interface Structure Optimization
The interface structure optimization mode allows a user to optimize the interface structure of an epitaxial interface by searching through all possible combinations if surface terminations at the interface and optimizing the relative alignment of each interface.
Film Miller Index:
Cubic Notation
Hexagonal Notation
Substrate Miller Index:
Max Interface Area (Å²):
optional
Max Interface Strain (%):
10
Use most stable substrate?
OPTIMIZE INTERFACE
Set and run the Geometry Optimization procedure
Einterface (meV/Å²)
Eadhesion (meV/Å²)
Film Termination (charge)
VIEW
51.852
56.200
72.718
VIEW PES
Pb4S3Br2(010)/CsPbBr3(100)
Film/Substrate Termination: PbSBr(0)/PbBr2(0)
VIEW Z-SHIFT
Surface Atoms Only
View of atoms interacting at the interface
Full Interface Structure
Full 3D-model of the interface
A B C A* B* C*

**Figure S39.** ***Structure Optimization*** **panel of the OgreInterface app.** The purple arrows indicate clickable elements of the panel and their respective outcome when selected.

### S13. Ogre performances on a consumer-grade laptop

All simulations based on the Ogre classical potential work were executed on a consumer-grade laptop (ASUS-N552V) equipped with an Intel® Core™ i7-6700HQ CPU @ 2.60GHz processor, 16 GB of RAM and a Samsung SSD 970 EVO Plus hard drive. On this configuration, simulations for the (100)⫽(010) – $CsPbBr_3/Pb_4S_3Br_2$ test interface required ∼100 s. Factors influencing the duration of calculations are the area of the supercell and the complexity of crystal structures involved, which might significantly increase computation times (up to ∼1500 s for the most complex interfaces considered in this work).

The algorithm was executed and is made available in the form of a Jupyter notebook script, in the same form as uploaded to GitHub as a part of the Supplementary Material of this work. An additional script is made available to reproduce all simulations in this work. Note that the execution might require several hours depending on the computer performances.

### S14. Supplementary References